\documentclass[12pt]{iopart}
\usepackage[utf8]{inputenc}
\pdfoutput=1
\usepackage[english]{babel}
\usepackage{csquotes}
\usepackage{amsmath}
\usepackage{amssymb}
\usepackage{siunitx}
\usepackage{graphicx}
\usepackage{subcaption}
\usepackage{booktabs}
\usepackage{xcolor}
\usepackage{multirow}
\usepackage{comment}
\usepackage{setspace}
\usepackage{ulem}
\usepackage[colorinlistoftodos,textsize=tiny]{todonotes}
\usepackage[maxnames=1, doi=false, url=false, eprint=false, related=false, isbn=false, style=numeric, sorting=none]{biblatex}
\bibliography{biblio}

\begin{document}
\newcommand{\el}{\text{e}}
\newcommand{\Dp}{\text{D}^+}
\newcommand{\Dtp}{\text{D}_2^+}
\newcommand{\D}{\text{D}}
\newcommand{\Dt}{\text{D}_2}

\title[Simulations of detached plasma and neutrals]{Self-consistent multi-component simulation of plasma turbulence and neutrals in detached conditions}

\author{D Mancini$^1$, P Ricci$^1$, N Vianello$^{2,3}$, G Van Parys$^{1}$, D S Oliveira$^{1}$}
 \ead{davide.mancini@epfl.ch}

\address{$^1$ Ecole Polytechnique Fédérale de Lausanne (EPFL), Swiss Plasma Center (SPC), CH-1015 Lausanne, Switzerland}
\address{$^2$ Consorzio RFX (CNR, ENEA, INFN, Università di Padova, Acciaierie Venete SpA), C.so Stati Uniti 4, 35127
Padova, Italy}
\address{$^3$  Istituto per la Scienza e la Tecnologia dei Plasmi, CNR, Padova, Italy}

\begin{abstract}
Simulations of high-density deuterium plasmas in a lower single-null magnetic configuration based on a TCV discharge are presented. We evolve the dynamics of three charged species (electrons, $\Dp$ and $\Dtp$), interacting with two neutrals species ($\D$ and $\Dt$) through ionization, charge-exchange, recombination and molecular dissociation processes. The plasma is modelled by using the drift-reduced fluid Braginskii equations, while the neutral dynamics is described by a kinetic model. To control the divertor conditions, a $\Dt$ puffing is used and the effect of increasing the puffing strength is investigated. The increase in fuelling leads to an increase of density in the scrape-off layer and a decrease of the plasma temperature. At the same time, the particle and heat fluxes to the divertor target decrease and the detachment of the inner target is observed. The analysis of particle and transport balance in the divertor volume shows that the decrease of the particle flux is caused by a decrease of the local neutral ionization together with a decrease of the parallel velocity, caused by the lower plasma temperature and the increase in momentum losses. The relative importance of the different collision terms is assessed, showing the crucial role of molecular interactions, as they are responsible for increasing the atomic neutral density and temperature, since most of the $\D$ neutrals are produced by molecular activated recombination and $\Dt$ dissociation. The presence of strong electric fields in high-density plasmas is also shown, revealing the role of the $E \times B$ drift in setting the asymmetry between the divertor targets. Simulation results are in agreement with experimental observations of increased density decay length, attributed to a decrease of parallel transport, together with an increase of plasma blob size and radial velocity. 
\end{abstract}

\maketitle
\section{Introduction}
In order to operate within the constraint imposed by the materials used for the plasma facing components, future fusion reactors will need to work in regimes where a large fraction of the power is dissipated via radiation \cite{Reimerdes_2017,Loarte_2007, Pitts_2019}. This can be achieved by operating the divertor in detached conditions, reached in present devices for example by increasing the core density, where a reduction of target temperature, heat and particle fluxes to the walls is observed \cite{Stangeby_2000, Loarte_1998, Reimerdes_2017}. This reduction is largely determined by the plasma-neutral interactions present at low temperatures, $T \lesssim 5$~eV, with an important role played by molecules as sink of particles, momentum and energy \cite{Stangeby_2000, Verhaegh_2019, Verhaegh_2021_2}. Indeed, neutral atoms and molecules can be ionized at these temperatures, generating atomic and molecular ions at the cost of the ionisation energy, or participate in recombination and charge-exchange reactions, which act as a particle and momentum sink. Molecules can also undergo dissociative processes, increasing the channels for ionization and recombination, e.g. through molecular activated recombination (MAR) reactions.

Even if the overall importance of MAR reactions in high density discharges is still debated \cite{Kukushkin_2017}, their role as ion sink is shown to be dominant compared to the atomic ion recombination \cite{Hollmann_2006,Fil_2018, Verhaegh_2021_2}, producing excited atoms that contribute to the total radiative losses \cite{Verhaegh_2021_1, Groth_2019}.
Moreover, molecular interactions contribute to momentum losses and are expected to play an important role in the transport dynamics and, as a consequence, in the asymmetries observed between the inner and outer divertor targets \cite{Park_2018,Potzel_2014} and in the dependence of the detachment threshold on the divertor leg length \cite{Reimerdes_2017}.
The importance of molecules in detachment calls for multi-component simulations that include molecular species.

The multi-component simulations of a tokamak plasma is usually based on fluid-diffusive models that consider a version of the Braginskii fluid equations simplified by modelling cross-field transport through empirical anomalous diffusion coefficients. The plasma dynamics is coupled with a kinetic Monte-Carlo model for the neutral dynamics. This approach is used in several modelling studies of detachment. For example, the SOLPS-ITER code \cite{Bonnin_2016} is used to model a TCV density ramp in Ref. \cite{Fil_2018} and the ASDEX detachment and X-point radiator regimes in Ref. \cite{Wu_2021, Pan_2023, Senichenkov_2022}, while deuterium molecular emissions in DIII-D ohmic discharges are studied by using the EDGE2D-EIRENE \cite{Groth_2019}. Despite the significant progress obtained by using fluid-diffusive models, simulating the plasma-neutral reactions self-consistently with turbulence is crucial to improve our predictive capabilities and, ultimately, the control of detachment \cite{Reimerdes_2017, Verhaegh_2021_2}.

Different models are able to capture the turbulent plasma dynamics by using fluid and gyrofluid models. These models are implemented in codes such as BOUT++ \cite{Dudson_2009}, FELTOR \cite{Wiesenberger_2019}, GRILLIX \cite{Stegmeir_2018}, GDB \cite{Zhu_2018}, GBS \cite{Ricci_2012,Coroado_2022_1,Giacomin_2022} and TOKAM3X \cite{Tamain_2016}. However, multi-component turbulent plasma simulations are very recent. They are used, for example, in the analysis of carbon impurities dynamics with SOLEDGE3X (combination of SOLEDGE2D and TOKAM3X) \cite{Bufferand_2019} or in the simulation of a gas puff imaging diagnostics in a limited magnetic configuration with GBS \cite{Coroado_2022_2}. Single-seeded blobs are studied by using the multi-species version of FELTOR \cite{Poulsen_2020}, the mult-species model implemented in the Hermes-3 module of the BOUT++ code \cite{Dudson_2023}, and the nHesel code, that simulates a single-species plasma with multiple species of neutrals modelled with a fluid approach \cite{Thrysoe_2018,Thrysoe2023}.

In this work, we present the first turbulence simulations of a deuterium plasma including molecules in a diverted tokamak geometry. The plasma we consider is composed of electrons and two ion species, $\Dp$ and $\Dtp$, coupled with a kinetic neutral model that include the dynamics of two deuterium neutral species, $\D$ and $\Dt$. The plasma and neutral models are described in Ref. \cite{Coroado_2022_1}. The simulations are carried out with the GBS code generalized here to perform multi-component simulations of the full tokamak plasma volume, considering a diverted magnetic configuration, retaining the SOL-edge-core interplay \cite{Giacomin_2022}. The solution of a kinetic model for the neutrals allows us to simulate self-consistently the neutral dynamics, without introducing ad-hoc diffusion coefficients, which are required by fluid approaches.

The interplay between molecular interactions, plasma target profiles and turbulent transport is investigated in a lower single-null L-mode discharge, with increasing plasma core density. Understanding these processes in the L-mode confinement regime is a first essential step, since it simplifies both the experimental and the numerical effort, mitigating the need to understand the transient phenomena induced, e.g., by edge localized modes.

The outcome of two different simulations is presented, where the electron density at the separatrix is increased by a factor of two by varying the intensity of the $\Dt$ gas puff. With higher density, we find a steady-state scenario where the inner strike point (ISP) presents a reduction of particle and heat fluxes, with large plasma pressure gradient along the magnetic field lines, which recalls one of the most important features of detached conditions \cite{Loarte_1998}. Our simulations show that molecular interactions affect the plasma dynamics increasing the $\D$ density in the divertor volume through MAR, modifying the average $\D$ temperature and ultimately decreasing the plasma temperature via ionization and charge-exchange reactions. The increase in plasma collisionality due to lower temperature establishes strong electric fields in the SOL, with an associated $E \times B$ drift, which increases the plasma asymmetry between the two targets. In addition, we observe the formation of a density shoulder at the outer mid-plane (OMP) \cite{Mancini_2021}, due to the increase of turbulent transport observed in high resistivity scenarios \cite{Mancini_2021, Beadle_2020, Vianello_2020, Offeddu_2022}.

The present paper is organised as follows. After the Introduction, in Sec. \ref{sec:model} we introduce the model used to self-consistently simulate plasma turbulence and neutral dynamics, as well as its implementation in the GBS code and the simulation setup. Sec. \ref{sec:results} provides an overview of the results obtained from our simulations, focusing on the analysis of the density, temperature and pressure profiles, with particular attention to the role played by neutral-plasma interaction terms and the importance of molecules. The analysis of the fluxes to the target and the assessment of the detachment conditions are described in Sec. \ref{sec:transport}. The conclusions follow.

\section{Simulation model and set-up}
\label{sec:model}
The simulations presented in this study are carried out with the GBS code, a three-dimensional, flux-driven code used to study plasma turbulence in the tokamak boundary \cite{Ricci_2012, Giacomin_2022}. GBS was initially developed to simulate basic plasma physics experiments \cite{Ricci_2013} and then ported to the geometry of the tokamak boundary, first in limited \cite{Halpern_2016} and later in diverted configurations \cite{Paruta_2018}. GBS can now perform simulations of three dimensional magnetic equilibrium configurations such as stellarators \cite{Coelho_2022}. In GBS the plasma description is provided by the drift-reduced Braginskii equations \cite{Zeiler_1997} coupled to a self-consistent kinetic neutral model \cite{Wersal_2015}. Thanks to recent efforts, both plasma and neutral models are now extended to simulate multiple species \cite{Coroado_2022_1}. The results we discuss in the present paper are based on simulations of the dynamics of five species ($\Dp$, $\Dtp$, electrons, $\D$ and $\Dt$) in a diverted configuration. In the following, we first describe the plasma and then the neutral model. Finally, we turn to the setup of the simulations presented in this work.

\subsection{The plasma model}
The model of the three plasma species ($\Dp$, $\Dtp$ and electrons) is based on the Braginskii fluid equations \cite{Braginskii_1965}, with the multi-species closure proposed by Zhdanov \cite{Zhdanov_2002} that include plasma-neutral collision terms in the form of Krook operators \cite{Coroado_2022_1}. In our model, we consider the drift-reduced approximation \cite{Zeiler_1997}, i.e. the limit of turbulent time scales slower than the ion cyclotron time scale, $\Omega_{c i} \tau_{\text{turb}} \gg 1 $, and turbulent scale lengths larger than the ion Larmor radius, $k_{\perp} \rho_{i} \ll 1$, with $\Omega_{c i} = eB/m_i $ and $\rho_{s i} = c_{s i}/\Omega_{c i}$ the cyclotron frequency and Larmor radius are defined for each ion species $i = \Dp , \Dtp$. Within these hypotheses, the component of the velocity perpendicular to the magnetic field is written as $\mathbf{v}_{\perp i} = \mathbf{v}_{E \times B} + \mathbf{v}_{d i} + \mathbf{v}_{\text{pol,}i} + \mathbf{v}_{\text{fric,}i}$, where $\mathbf{v}_{E \times B} = (\mathbf{E} \times \mathbf{B})/B^2$ is the $E \times B$ drift, $\mathbf{v}_{di} = (\mathbf{B} \times \nabla p_i)/(e n_i B^2)$ the diamagnetic drift, $\mathbf{v}_{\text{pol,}i}$ the polarization drift and $\mathbf{v}_{\text{fric,}i}$ the drift due to friction between different ion species and neutrals. The detailed expressions of the velocities are given in Refs. \cite{Giacomin_2022, Coroado_2022_1}. The electron perpendicular velocity is approximated by its leading order component $\mathbf{v}_{\perp e} = \mathbf{v}_{E \times B} + \mathbf{v}_{de}$. Exploiting the collisional Zdhanov closure proposed in Refs. \cite{Bufferand_2019} and \cite{Zhdanov_2002}, with the approximation of $n_{\Dtp}/n_{\Dp} \ll 1$ proposed in Ref. \cite{Coroado_2022_1}, the plasma equations implemented in GBS take the form:
\begingroup
\allowdisplaybreaks
\begin{align}
    \begin{split}\label{eq:density_e}
        \frac{\partial n_{\el}}{\partial t} =& -\frac{\rho_*^{-1}}{B}[\phi,n_{\el}] + \frac{2}{B} \left[C(p_{\el}) - n_{\el} C(\phi)\right] - \nabla_{\|}(n_{\el} v_{\|\el}) + \mathcal{D}_{n_{\el}}\nabla_{\perp}^2 n_{\el} \\
        &+ n_{D} \nu_{\text{iz,D}} - n_{\Dp} \nu_{\text{rec,D}^+} + n_{\Dt} \nu_{\text{iz,D}_2} - n_{\Dtp} \nu_{\text{rec,D}_2^+} \\
        &+ n_{\Dt} \nu_{\text{diss-iz,}\Dt} + n_{\Dtp} \nu_{\text{diss-iz,}\Dtp} - n_{\Dtp} \nu_{\text{diss-rec,D}_2^+} \quad ,
    \end{split}
    \\
    \begin{split}\label{eq:density_D2p}
        \frac{\partial n_{\Dtp}}{\partial t} =& -\frac{\rho_*^{-1}}{B}[\phi,n_{\Dtp}] - \nabla_{\|}(n_{\Dtp} v_{\|\Dtp}) - \frac{2}{B} \left[C(p_{\Dtp})+n_{\Dtp}C(\phi)\right] \\
        & + \mathcal{D}_{n_{\Dtp}}\nabla_{\perp}^2 n_{\Dtp} + n_{\Dt}\nu_{\text{iz,}\Dt}-n_{\Dtp}\nu_{\text{rec,D}_2^+}+n_{\Dp_2}\nu_{\text{cx,D}_2-\Dp} \\
        & - n_{\Dp}\nu_{\text{cx,D-D}_2^+}-n_{\Dtp}\left(\nu_{\text{diss-iz,D}_2^+}+\nu_{\text{diss,D}_2^+}+\nu_{\text{diss-rec,D}_2^+}\right) \quad ,
    \end{split}
    \\
    \begin{split}\label{eq:vorticity}
        \frac{\partial \Omega}{\partial t} = &-\frac{\rho_*^{-1}}{B} \nabla \cdot \left( \left[\phi,\mathbf{\omega}_{\Dp}\right] + 2 \left[\phi,\mathbf{\omega}_{\Dtp} \right] \right) - \nabla \cdot \left( v_{\|\Dp} \nabla_{\|} \mathbf{\omega}_{\Dp} + v_{\|\Dtp} \nabla_{\|} \mathbf{\omega}_{\Dtp} \right)\\
        &+ 2 B C(p_{\el}+p_{\Dp}+p_{\Dtp})+B^2 \nabla_{\|} j_{\|} + \frac{B}{3} C(G_{\Dp} + G_{\Dtp}) \\
        &+ \eta_{0 \Omega}\nabla_{\|}^2 \Omega + \mathcal{D}_{\perp \Omega}\nabla_{\perp}^2 \Omega - \nabla \cdot \left[\frac{2 n_{\Dt}}{n_{\Dtp}}\left(\nu_{\text{cx,}\Dt} + \nu_{\text{iz,}\Dt} + \nu_{\text{cx,}\Dt\text{-}\Dp}\right)\mathbf{\omega}_{\Dtp}\right]\\
        &- \nabla \cdot \left[\frac{n_{\Dp}}{n_{\Dp}}\left(\nu_{\text{cx,}\D} + \nu_{\text{iz,}\D} + \nu_{\text{cx,}\D\text{-}\Dtp}\right)\mathbf{\omega}_{\Dp}\right] - \nabla \cdot \left[\frac{n_{\Dt}}{n_{\Dp}}\nu_{\text{di-iz,}\Dt}\mathbf{\omega}_{\Dp}\right]\\
        &+ \nabla \cdot \left[\frac{n_{\Dtp}}{n_{\Dp}}\left(2\nu_{\text{di-iz},\Dtp}+\nu_{\text{di,}\Dtp}\right)\left(\mathbf{\omega}_{\Dtp}-\mathbf{\omega}_{\Dp}\right)\right] \quad ,
    \end{split}
    \\
    \begin{split}\label{eq:vpare}
        \frac{\partial U_{\|\el}}{\partial t} =& -\frac{\rho_*^{-1}}{B}[\phi,v_{\|\el}]+\frac{m_{\Dp}}{m_{\el}}\left[\nu j_{\|} + \nabla_{\|}\phi-\frac{\nabla_\| p_{\el}}{n_{\el}} -\frac{2}{3n_{\el}}\nabla_\|G_{\el}-0.71\nabla_{\|}T_{\el}\right] \\
        &-v_{\|\el}\nabla_\| v_{\|\el}+\mathcal{D}_{v_{\|\el}}\nabla_{\perp}^2 v_{\|\el} + \frac{1}{n_{\el}} [ n_{\Dp}\left( 2 \nu_{\text{iz,}\D}+\nu_{\text{e-}\D} \right) \left( v_{\|\Dp}-v_{\|\el} \right) \\ 
        &+ n_{\Dt} \left( 2 \nu_{\text{iz,}\Dt} + \nu_{\text{e-}\Dt} + 2 \nu_{\text{diss-iz,}\Dt} + \nu_{\text{diss,}\Dt} \right) \left( v_{\|\Dt}-v_{\|\el} \right) \\
        &+ n_{\Dtp} \left( 2\nu_{\text{diss-iz,}\Dtp}+\nu_{\text{diss,}\Dtp} \right) \left( v_{\|\Dtp}-v_{\| \el} \right)] \quad ,
    \end{split}
    \\
    \begin{split} \label{eq:vparD} 
        \frac{\partial v_{\|\Dp}}{\partial t} =& -\frac{\rho_*^{-1}}{B}[\phi,v_{\|\Dp}]-v_{\|\Dp}\nabla_\| v_{\|\Dp}-\nabla_{\|}\phi-\frac{\nabla_\| p_\Dp}{n_{\Dp}}\\
        &- \frac{2}{3n_{\Dp}}\nabla_\|G_{\Dp}+0.71\frac{n_{\el}}{n_{\Dp}}\nabla_{\|}T_{\el}-\nu\frac{n_{\el}}{n_{\Dp}}j_{\|} + \mathcal{D}_{v_{\|\Dp}}\nabla_{\perp}^2 v_{\|\Dp} \\
        &+ \frac{1}{n_{\Dp}} [ n_{\D} \left(\nu_{\text{iz,}\D}+\nu_{\text{cx,}\D}+\nu_{\text{cx,}\D\text{-}\Dtp} \right) \left( v_{\|\D}-v_{\|\Dp} \right) \\
        &+ n_{\Dtp}  \left( 2\nu_{\text{diss-iz,}\Dtp}+\nu_{\text{diss,}\Dtp} \right) \left( v_{\|\Dtp}-v_{\|\Dp} \right) \\
        &+ n_{\Dt} \nu_{\text{diss-iz},\Dt} \left( v_{\|\Dt}-v_{\|\Dp} \right) ] \quad ,
    \end{split}
    \\
    \begin{split}\label{eq:vparD2}
        \frac{\partial v_{\|\Dtp}}{\partial t} =& -\frac{\rho_*^{-1}}{B}[\phi,v_{\|\Dtp}]-v_{\|\Dtp}\nabla_\| v_{\|\Dtp} + \frac{1}{2}\left[-\nabla_{\|}\phi -\frac{\nabla p_{\Dtp}}{n_{\Dtp}} -\frac{2}{3 n_{\Dtp}}\nabla_{\|}G_{\Dtp}\right] \\
        &+ \mathcal{D}_{v_{\|\Dtp}}\nabla_{\perp}^2 v_{\|\Dtp} +\frac{n_{\Dt}}{n_{\Dtp}}(\nu_{\text{iz,D}_2}+\nu_{\text{cx,}\Dt}+\nu_{\text{cx,}\Dt\text{-}\Dp})(v_{\|\Dtp}-v_{\|\Dtp}) \quad ,
    \end{split}
    \\
    \begin{split}\label{eq:temperature_e}
        \frac{\partial T_{\el}}{\partial t} =& -\frac{\rho_*^{-1}}{B}[\phi,T_{\el}] - v_{\|\el}\nabla_\| T_{\el} + \frac{4 T_{\el}}{3 B} \left[ \frac{7}{2} C(T_{\el}) + \frac{T_{\el}}{n_{\el}} C(n_{\el}) - C(\phi) \right] - \frac{2 T_{\el}}{3} \nabla_{\|} v_{\| \el} \\
        &+\frac{2}{3 n_{\el}} \left[ \frac{1.62}{\nu} \nabla_{\|}\left(n_{\el} T_{\el} \nabla_{\|} T_{\el}\right) - 0.71 \nabla_{\|} \left( T_{\el} j_{\|} \right) \right] \\
        &+\mathcal{\chi}_{\perp \el}\nabla_{\perp}^2 T_{\el} + \nabla_{\|}\left(\mathcal{\chi}_{\|\el}\nabla_{\|} T_{\el}\right) - \frac{4}{3} \frac{m_e}{m_{\Dp}} \frac{1}{\tau_{\el}} (T_{\el} - T_{\Dp}) + S_{T_{\el}} \\
        &+\frac{1}{n_{\el}} \{ n_{\D} \nu_{\text{iz,D}}\left[-\frac{2}{3}E_{\text{iz,D}}-T_{\el}+\frac{m_{\el}}{m_{\Dp}} v_{\| \el}\left(v_{\| \el}-\frac{4}{3}v_{\|\Dp}\right)\right]\\
        &+n_{\Dt}\nu_{\text{iz,D}_2}\left[-\frac{2}{3} E_{\text{iz,D}_2} - T_{\el} + \frac{m_{\el}}{m_{\Dp}} v_{\| \el}\left(v_{\| \el}-\frac{4}{3}v_{\|\Dp_2}\right)\right]\\
        &+n_{\Dt}\nu_{\text{diss,D}_2}\left[-\frac{2}{3}E_{\text{diss,D}_2}+\frac{2}{3}\frac{m_{\el}}{m_{\Dp}} v_{\| \el}\left(v_{\| \el}-v_{\|\Dp_2}\right)\right] \\
        &+n_{\Dt}\nu_{\text{diss-iz,D}_2}\left[-\frac{2}{3} E_{\text{diss-iz,D}_2} - T_{\el} + \frac{m_{\el}}{m_{\Dp}} v_{\| \el}\left(v_{\| \el}-\frac{4}{3}v_{\|\Dp_2}\right)\right] \\
        &+n_{\Dtp}\nu_{\text{diss,D}_2^+}\left[-\frac{2}{3}E_{\text{diss,D}_2^+}+\frac{2}{3}\frac{m_{\el}}{m_{\Dp}} v_{\| \el}\left(v_{\| \el}-v_{\|\Dtp}\right)\right] \\
        &+n_{\Dtp}\nu_{\text{diss-iz,D}_2^+}\left[-\frac{2}{3} E_{\text{diss-iz,D}_2^+} - T_{\el} + \frac{m_{\el}}{m_{\Dp}} v_{\| \el}\left(v_{\| \el}-\frac{4}{3}v_{\|\Dtp}\right)\right] \\
        &-n_{\Dp}\nu_{\text{e-D}}\frac{m_{\el}}{m_{\Dp}}\frac{2}{3}v_{\| \el}(v_{\|\Dp}-v_{\| \el}) - n_{\Dp_2}\nu_{\text{e-D}_2}\frac{m_{\el}}{m_{\Dp}}\frac{2}{3}v_{\| \el}(v_{\|\Dp_2}-v_{\| \el}) \} \quad ,
    \end{split}
    \\
    \begin{split}\label{eq:temperature_D}
        \frac{\partial T_{\Dp}}{\partial t} =& -\frac{\rho_*^{-1}}{B}[\phi,T_{\Dp}] - v_{\|\Dp}\nabla_\| T_{\Dp} +\frac{4}{3} \frac{T_{\Dp}}{B} \left[\frac{1}{n_{\Dp}}C(p_{\el} + p_{\Dtp})-C(\phi)\right] \\
        &-\frac{2 T_{\Dp}}{3 n_{\Dp}} \left[ \nabla_{\|} \left( n_{\el} v_{\|\el} \right) - \nabla_{\|} \left( n_{\Dtp} v_{\|\Dtp} \right) - v_{\|\Dp} \nabla_{\|} \left( n_{\Dtp} \right) \right] \\
        &-\frac{10}{3} \frac{T_{\Dp}}{B} C(T_{\Dp})
        +\frac{2}{3 n_{\Dp}}\frac{2.32}{\sqrt{2} \nu} \sqrt{\frac{m_{\el}}{m_{\Dp}}}\nabla_{\|}\left(n_{\el} T_{\Dp}\nabla_\|T_{\Dp}\right) + \frac{4}{3} \frac{m_e}{m_{\Dp}} \frac{1}{\tau_{\el}} (T_{\el} - T_{\Dp}) \\
        &+ \mathcal{\chi}_{\perp \Dp}\nabla_{\perp}^2 T_{\Dp} + \nabla_{\|}\left(\mathcal{\chi}_{\|\Dp}\nabla_{\|} T_{\Dp}\right)\\
        &+ \frac{1}{n_{\Dp}}\left\{n_{\D}\left(\nu_{\text{iz,}\D}+\nu_{\text{cx,}\D}+\nu_{\text{cx,}\D\text{-}\Dtp}\right)\left[T_{\D}-T_{\Dp}+\frac{1}{3}\left(v_{\|\D}-v_{\|\Dp}\right)^2\right]\right. \\
        &+n_{\Dt}\nu_{\text{diss-iz,}\Dt}\left[T_{\D\text{,diss-iz}\left(\Dt\right)}-T_{\Dp}+\frac{1}{3}\left(v_{\|\Dt}-v_{\|\Dp}\right)^2\right] \\
        &+ 2 n_{\Dtp}\nu_{\text{diss-iz,}\Dtp}\left[T_{\D\text{,diss-iz}\left(\Dtp\right)}-T_{\Dp}+\frac{1}{3}\left(v_{\|\Dtp}-v_{\|\Dp}\right)^2\right] \\
        &\left. + n_{\Dtp}\nu_{\text{diss,}\Dtp}\left[T_{\D\text{,diss}\left(\Dtp\right)}-T_{\Dp}+\frac{1}{3}\left(v_{\|\Dtp}-v_{\|\Dp}\right)^2\right]\right\} \quad ,
    \end{split}
    \\
    \begin{split}\label{eq:temperature_D2}
        \frac{\partial T_{\Dtp}}{\partial t} =& -\frac{\rho_*^{-1}}{B}[\phi,T_{\Dtp}] - v_{\|\Dtp}\nabla_\| T_{\Dtp} - \frac{4}{3} \frac{T_{\Dtp}}{B} \left[\frac{1}{n_{\Dtp}}C(p_{\Dtp}) + C(\phi) \right] \\
        &- \frac{10}{3} \frac{T_{\Dtp}}{B} C(T_{\Dtp}) - \frac{2 T_{\Dtp}}{3} \nabla_{\|} v_{\|\Dtp} + \frac{2}{3 n_{\Dtp}}\frac{0.92}{\sqrt{2} \nu}\sqrt{\frac{m_{\el}}{m_{\Dp}}}\nabla_{\|} \left(n_{\el} T_{\Dp}\nabla_\|T_{\Dp} \right) \\
        &+ \mathcal{\chi}_{\perp \Dtp}\nabla_{\perp}^2 T_{\Dtp} + \nabla_{\|}\left(\mathcal{\chi}_{\|\Dtp}\nabla_{\|} T_{\Dtp}\right) + S_{T_{\Dtp}} \\
        &+ \frac{n_{\Dt}}{n_{\Dtp}}(\nu_{\text{cx,D}_2}+\nu_{\text{iz,D}_2}+\nu_{\text{cx,D}_2-\Dp})\left[T_{\Dt}-T_{\Dtp}+\frac{2}{3}(v_{\|\Dt}-v_{\|\Dtp})^2\right] \quad ,
    \end{split}
\end{align}
\endgroup

\noindent solved with the Poisson and Ampère equations
\begin{align}
    \label{eq:poisson}
    & \nabla \cdot \left[ \left(n_{\Dp} + 2 n_{\Dtp} \right) \nabla_{\perp} \phi \right] = \Omega - \tau \nabla_{\perp}^2 \left(p_{\Dp} + 2 p_{\Dtp} \right)
\end{align}
and
\begin{align}
    \label{eq:ampere}
    & \left( \nabla^2_{\perp} - \frac{\beta_{e 0}}{2}\frac{m_{\Dp}}{m_{\el}} n_{\el} \right) = \nabla_{\perp}^2 U_{\| \el} - \frac{\beta_{e 0}}{2}\frac{m_{\Dp}}{m_{\el}} n_{\Dp} v_{\| \Dp} + \frac{\beta_{e0}}{2}\frac{m_{\Dp}}{m_{\el}}\overline{j}_{\|} \quad ,
\end{align}
while the atomic ions density is evaluated imposing quasi-neutrality $n_{\Dp} = n_{\el} - n_{\Dtp}$.

In Eqs. (\ref{eq:density_e}-\ref{eq:ampere}), $U_{\| \el} = V_{\| \el} + e \psi/m_{\el}$ is the sum of electron inertia and electromagnetic induction, $p_a = n_a T_a$ is the pressure for the species $a$, $a = {\el, \Dp, \Dtp}$, and $\Omega = \Omega_{\Dp} + 2 \Omega_{\Dtp}$ is the plasma vorticity, with $\Omega_i = \nabla \cdot \omega_i = \nabla \cdot \left(n_i \nabla_{\perp} \phi + \nabla_{\perp} p_i \right)$ the contribution of each ion species. The operator $[\phi, f] = \mathbf{b} \cdot \left(\nabla \phi \times \nabla f \right)$ is the $\mathbf{E} \times \mathbf{B}$ convective operator, $C(f) = B/2[\nabla \times (\mathbf{b}/B)] \cdot \nabla f$ is the curvature operator, $\nabla_{\parallel} f = \mathbf{b} \cdot \nabla f$ is the parallel gradient, and $\nabla^2_{\perp} f = \nabla \cdot [(\mathbf{b} \times \nabla f) \times \mathbf{b}]$ is the perpendicular Laplacian, with $\mathbf{b} = \mathbf{B}/B$ the unit vector in the direction of the magnetic field.  The electron gyroviscous term is given by $G_{\el} = -\eta_{0 \el} \left[ 2 \nabla_{\|} v_{\| \el} + C(\phi)/B - C(p_{\el})/(e n_{\el} B) \right]$, while for the ion species $G_i = \eta_{0 i} \left[ 2 \nabla_{\|} v_{\| i} + C(\phi)/B + C(p_{i})/(e n_{i} B) \right]$. ompared to the GBS model with one single ion species, implemented in diverted configuration in \cite{Giacomin_2022}, the model considered here provides the evolution of the molecular ion profiles, taking into account the contribution of both ion species into the vorticity evolution and the contributions of several new plasma-neutrals interaction terms.

The plasma-neutrals interaction terms considered in this work are ionization, recombination, dissociation, charge-exchange and electron-neutral elastic collisions, all listed in Table \ref{tab:reactions}. We consider the collisional processes that have larger cross sections in the deuterium plasma in typical conditions of the tokamak boundary \cite{Wensing_2019}, where the reaction rates $\langle v \sigma \rangle$ are obtained from the AMJUEL \cite{Reiter_2011} and HYDEL \cite{Janev_1987} databases. The reaction frequencies for ionization, recombination, elastic collisions and dissociative processes are averaged over the electron velocity distribution function, assumed Maxwellian, while the one for charge exchange processes are averaged over the ion velocity distribution function. Velocities and energies of the particles that results for the reactions are evaluated by using momentum and energy considerations, resulting in the values listed in Table \ref{tab:products} \cite{Coroado_2022_1}. In particular, for an elastic collision between an electron and an atomic or molecular neutral, it is assumed that the neutral velocity is not affected by the reaction, while the electron is emitted isotropically according to a Maxwellian distribution function centered at the velocity of the incoming electron. Regarding the ionization processes, the electrons and ions are generated according to a Maxwellian distribution function centered at the fluid velocity of the incoming neutral, with the electron temperature taking into account the loss of the ionization energy, $\langle E_{\text{iz,}\D} \rangle$ or $\langle E_{\text{iz,}\Dt} \rangle$. Ionization processes in fusion plasma involve radiation emission \cite{Stangeby_2000}, not taken into account in our model equations. To include the related additional losses, we follow the procedure suggested in \cite{Stangeby_2000} (chapter 3.5) and consider an effective ionization energy of $E_{\text{iz,}\D}^{\text{eff}}=30.0 \text{ eV}$. For dissociation processes, we follow a similar procedure, with the reaction-specific electron energy loss assumed to be the energy necessary to excite the molecule and incur in a Franck-Condon dissociation. The D atoms generated by dissociative-recombination reactions, namely MAR processes, considered in this work, are described by a Maxwellian distribution function, with average temperature $T_{D\text{, diss-rec(}\Dtp\text{)}}$.

\begin{table}[ht]
\centering
\caption{\label{tab:reactions} Collisional processes considered and their respective reaction rates, source: \cite{Coroado_2022_1}}
\begin{tabular}{@{}lll} 
\toprule
\textbf{Collisional process} & \textbf{Equation} & \textbf{Reaction Frequency} \\
\midrule
Ionization of $\D$ & $\el^- +\D \rightarrow 2\el^- + \Dp$ & $\nu_{\text{iz,}\D} = n_{\el}\left\langle v_{\el}\sigma_{\text{iz,}\D}(v_{\el})\right\rangle$ \\ 
Recombination of $\Dp$ and $\el^-$ & $\el^- + \Dp \rightarrow \D$ & $\nu_{\text{rec,}\Dp} = n_{\el}\left\langle v_{\el}\sigma_{\text{rec,}\Dp}(v_{\el})\right\rangle$ \\ 
$\el^- -\Dp$ elastic collisions & $\el^- +\Dp \rightarrow \el^- +\Dp$ & $\nu_{\el\text{-}\D} = n_{\el}\left\langle v_{\el}\sigma_{\el\text{-}\D}(v_{\el})\right\rangle$ \\ 
Ionization of $\Dt$ & $\el^- + \Dt \rightarrow 2\el^- + \Dtp$ & $\nu_{\text{iz,}\Dt} = n_{\el}\left\langle v_{\el}\sigma_{\text{iz,}\Dt}(v_{\el})\right\rangle$ \\ 
Recombination of $\Dtp$ and $\el^-$ & $\el^- + \Dtp \rightarrow \Dt$ & $\nu_{\text{rec,}\Dtp} = n_{\el}\left\langle v_{\el}\sigma_{\text{rec,}\Dtp}(v_{\el})\right\rangle$ \\ 
$\el^- -\Dp_2$ elastic collisions & $\el^- + \Dp_2 \rightarrow \el^- + \Dp_2$ & $\nu_{\el\text{-}\Dt} = n_{\el}\left\langle v_{\el}\sigma_{\el\text{-}\Dt}(v_{\el})\right\rangle$ \\ 
Dissociation of $\Dt$ & $\el^- + \Dt \rightarrow \el^- + \D + \D$ & $\nu_{\text{diss,}\Dt} = n_{\el}\left\langle v_{\el}\sigma_{\text{diss,}\Dt}(v_{\el})\right\rangle$ \\ 
Dissociative ionization of $\Dt$ & $\el^- + \Dt \rightarrow 2\el^- + \Dp +  \D$ & $\nu_{\text{diss-iz,}\Dt} = n_{\el}\left\langle v_{\el}\sigma_{\text{diss-iz,}\Dt}(v_{\el})\right\rangle$ \\ 
Dissociation of $\Dtp$ & $\el^- + \Dtp \rightarrow \el^- + \Dp +  \D$ & $\nu_{\text{diss,}\Dtp} = n_{\el}\left\langle v_{\el}\sigma_{\text{diss,}\Dtp}(v_{\el})\right\rangle$ \\ 
Dissociative ionization of $\Dtp$ & $\el^- + \Dtp \rightarrow 2\el^- + 2\Dp$ & $\nu_{\text{diss-iz,}\Dtp} = n_{\el}\left\langle v_{\el}\sigma_{\text{diss-iz,}\Dtp}(v_{\el})\right\rangle$ \\ 
Dissociative recombination of $\Dtp$ & $\el^- + \Dtp \rightarrow 2\D$ & $\nu_{\text{diss-rec,}\Dtp} = n_{\el}\left\langle v_{\el}\sigma_{\text{diss-rec,}\Dtp}(v_{\el})\right\rangle$ \\ 
Charge-exchange of $\Dp,\D$ & $\Dp +\D \rightarrow \D +  \Dp$ & $\nu_{\text{cx,}\D} = n_{\Dp}\left\langle v_{\Dp}\sigma_{\text{cx,}\Dp}(v_{\Dp})\right\rangle$ \\ 
Charge-exchange of $\Dtp,\Dt$ & $\Dtp + \Dt \rightarrow \Dt + \Dtp$ & $\nu_{\text{cx,}\Dt} = n_{\Dtp}\left\langle v_{\Dtp}\sigma_{\text{cx,}\Dtp}(v_{\Dtp})\right\rangle$ \\ 
Charge-exchange of $\Dtp,\D$ & $\Dtp +\D \rightarrow \Dt + \Dp$ & $\nu_{\text{cx,}\D\text{-}\Dtp} = n_{\Dtp}\left\langle v_{\Dtp}\sigma_{\text{cx,}\D\text{-}\Dtp}(v_{\Dtp})\right\rangle$ \\ 
Charge-exchange of $\Dt,\Dp$ & $\Dt + \Dp \rightarrow \Dtp +\D$ & $\nu_{\text{cx,}\Dt\text{-}\Dp} = n_{\Dp}\left\langle v_{\Dp}\sigma_{\text{cx,}\Dt\text{-}\Dp}(v_{\Dp})\right\rangle$ \\ 
\bottomrule
\end{tabular}
\end{table}

\begin{table}[h]
\centering
\caption{\label{tab:products}Average electron energy loss and average energy of reaction products for the ionization and dissociative processes included in the model, source: \cite{Coroado_2022_1}.}
\begin{tabular}{@{}lll} 
\toprule
\textbf{Collisional process} & \textbf{$\el^-$ Energy loss} & \textbf{Temperature of products} \\
\midrule
Ionization of $\D$ & $\left\langle E_{\text{iz,}\D}\right\rangle = 13.60 \text{eV}$ & ---------------------------\\ 
Ionization of $\Dt$ & $\left\langle E_{\text{iz,}\Dt}\right\rangle = 15.43 \text{eV}$ & ---------------------------\\ 
Dissociation of $\Dt$ & $\left\langle E_{\text{diss,}\Dt}\right\rangle \simeq 14.3 \text{eV}$ & $T_{\D\text{,diss}\left(\Dt\right)} \simeq 1.95 \text{eV}$\\ 
Dissociative ionization \\ of $\Dt$ ($E_{\el} < 26 \text{eV}$) & $\left\langle E_{\text{diss-iz,}\Dt}\right\rangle \simeq 18.25 \text{eV}$ & $T_{\D\text{,diss-iz}\left(\Dt\right)} \simeq 0.25 \text{eV}$\\
Dissociative ionization \\ of $\Dt$ ($E_{\el} > 26 \text{eV}$) & $\left\langle E_{\text{diss-iz,}\Dt}\right\rangle \simeq 33.6 \text{eV}$ & $T_{\D\text{,diss-iz}\left(\Dt\right)} \simeq 7.8 \text{eV}$\\
Dissociation of $\Dtp$ & $\left\langle E_{\text{diss,}\Dtp}\right\rangle \simeq 13.7 \text{eV}$ & $T_{\D\text{,diss}\left(\Dtp\right)} \simeq 3.0 \text{eV}$\\
Dissociative ionization of $\Dtp$ & $\left\langle E_{\text{diss-iz,}\Dtp}\right\rangle \simeq 15.5 \text{eV}$ & $T_{\D\text{,diss-iz}\left(\Dtp\right)} \simeq 0.4 \text{eV}$\\
Dissociative recombination of $\Dtp$ & ----------------------------- & $T_{\D\text{,diss-rec}\left(\Dtp\right)} \simeq 11.7 \text{eV}$\\
\bottomrule
\end{tabular}
\end{table}

The simulation domain encompasses the whole tokamak plasma volume, with a rectangular poloidal cross section of vertical extension $L_Z$ and radial extension $L_R$, leading to a natural choice of a cylindrical coordinate system $(R, \varphi, Z)$, where $R$ is the radial distance from the tokamak axis of symmetry, $\varphi$ the toroidal angle and $Z$ the vertical coordinate. The magnetic field is expressed in terms of the flux function $\psi$, $\mathbf{B} = R B_{\varphi} \nabla \varphi + \nabla \psi \times \nabla \varphi$, where $\nabla \psi$ is the direction orthogonal to the flux surface, defining a flux-aligned coordinate system, $(\nabla \psi, \nabla \chi, \nabla \varphi)$ where $\nabla \chi = \nabla \varphi \times \nabla \psi$, used in the analysis of the simulation results.

In the following of the present paper, all quantities in Eqs. (\ref{eq:density_e}-\ref{eq:ampere}) are normalized to their reference value. Densities are normalized to the reference density $n_0$, $T_{\el}$ to $T_{\el 0}$, both $T_{\Dp}$ and $T_{\Dtp}$ to $T_{\Dp 0}$ and parallel velocities to the sound speed $c_{s0} = \sqrt{T_{\el 0}/m_{\Dp}}$. The magnetic field strenght $B$ is normalized to the field value on the magnetic axis $B_0$, perpendicular lengths to the ion sound Larmor radius $\rho_{s0} = c_{s0}/\Omega_{c \Dp}$, parallel lengths to the tokamak major radius $R_0$, and time to $t_0 = R_{0}/c_{s0}$. The dynamics is then set by the following dimensionless parameters: the normalized ion Larmor radius, $\rho_{*} = \rho_{s0}/R_0$, the ion to electron temperature ratio, $\tau = T_{\Dp 0}/T_{\el 0}$, and the normalized Spitzer resistivity $\nu = e^2 n_0 R_0/(m_{\Dp} c_{s0} \sigma_{\parallel}) = \nu_0 T_{\el}^{-3/2}$, with
\begin{equation}\label{eq:sigma}
    \sigma_{\parallel} =  \left(1.96 \frac{n_0 e^2 \tau_e}{m_e}\right) n = \left(\frac{5.88}{4\sqrt{2\pi}}\frac{(4\pi\epsilon_0)^2}{e^2}\frac{T^{3/2}_{\el 0}}{\lambda \sqrt{m_e}}\right)(T_{\el})^{3/2}
\end{equation}
and, as a consequence,
\begin{equation}\label{eq:param}
        \nu_0 = \frac{4\sqrt{2 \pi}}{5.88} \frac{e^4}{(4 \pi \epsilon_0)^2} \frac{\sqrt{m_{\el}} R_0 n_0 \lambda}{m_{\Dp} c_{s0} T_{\el 0}^{3/2}} \, .
\end{equation}

The expression for the normalized viscosities, $\eta_{0 e}$, $\eta_{0 \Dp}$ and $\eta_{0 \Dtp}$ in Eqs. (\ref{eq:vpare}-\ref{eq:vparD2}), and thermal conductivities, $\chi_{0 e}$, $\chi_{0 \Dp}$ and $\chi_{0 \Dtp}$ in Eqs. (\ref{eq:temperature_e}-\ref{eq:temperature_D2}), can be found in Ref. \cite{Giacomin_2022} and are all assumed constant in this work. The normalized diffusion coefficients $D_f$, for each field $f$, are introduced for numerical stability.

In our simulations, fuelling is entirely the result of self-consistent neutral ionization processes, while external electron heating is added in Eqs. \eqref{eq:temperature_e} and \eqref{eq:temperature_D2} through the $s_{T_{\el}}$ and $s_{T_{\Dtp}}$ source terms. Both temperature sources are toroidally uniform and expressed as an analytical function of the flux function
\begin{equation}
        s_{T} = \frac{s_{T0}}{2} \left[\tanh\left(-\frac{\psi(R,Z)-\psi_T}{\Delta_T}\right) + 1 \right] \, ,
    \label{eq:source_cost}    
\end{equation}
where $\psi_T$ is a flux surface localized inside the LCFS, as shown in Fig. \ref{fig:sim_setup}. The heating source is therefore the sum of the contributions given by the external heating and by the neutral interactions present in Eqs. (\ref{eq:density_e}-\ref{eq:temperature_D2}), i.e.
\begin{align}
    \label{eq:sptot}
    s_{P_{\text{tot}}} = n_{\el} (s_{T_{\el}} + s^{\text{neu}}_{T_{\el}}) + n_{\Dp} s^{\text{neu}}_{T_{\Dp}} + n_{\Dtp} (s_{T_{\Dtp}} + s^{\text{neu}}_{T_{\Dtp}}) + T_{\el} s^{\text{neu}}_{n_{\el}} + T_{\Dp} s^{\text{neu}}_{n_{\Dp}} + T_{\Dtp} s^{\text{neu}}_{n_{\Dtp}} \quad .
\end{align}

We implement a pre-sheath set of magnetic boundary conditions at the walls where the strike points are located, i.e. the lower and the left walls, as detailed in Ref. \cite{Loizu_2012} and extended in Ref. \cite{Coroado_2022_1} to include molecular deuterium, that is
\begin{align}
    v_{\| \el} =& \pm c_{s}  \, \text{max}\{\exp\left(\Lambda - \frac{\phi}{T_e}\right),\exp(\Lambda)\}  \label{eq:bounplasma_start}\\
    v_{\| \Dp} =& \pm c_s \sqrt{1 + \frac{T_{\Dp}}{T_{\el}}}\\
    v_{\| \Dtp} =& \, \frac{v_{\| \Dp}}{\sqrt{2}}\\
    \partial_{s} \phi =& \mp \frac{c_s}{\sqrt{1 + \frac{T_{\Dp}}{T_{e}}}} \partial_n v_{\parallel \Dp} \\
    \partial_{s} n_{\el} =& \, \partial_{s} n_{\Dp} = \mp \frac{n_{\el}}{c_s \sqrt{1 + \frac{T_{\Dp}}{T_{e}}}} \partial_n v_{\parallel \Dp} \\
    \partial_{s} n_{\Dtp} =& \mp \frac{n_{\Dtp}}{c_s \sqrt{1 + \frac{T_{\Dp}}{T_{e}}}} \partial_n v_{\parallel \Dtp} \\
    \partial_{s} T_e =& \, \partial_{s} T_{\Dp} = \partial_{n} T_{\Dtp} = 0 \\
    \Omega =& \mp (n_{\el} + n_{\Dtp})\sqrt{1 + \frac{T_{i}}{T_{e}}} \partial^2_{nn} v_{\parallel \Dp}
    \label{eq:boundplasma_end}
\end{align}
where $s$ is the direction perpendicular to the vessel wall, the plus (minus) sign refers to the magnetic field pointing toward (away from) the wall, the dimensionless ion sound speed is $c_s = \sqrt{T_{\el}}$ and $\Lambda = \log \sqrt{m_{\Dp}/(2 \pi m_e)} \simeq 3$. A set of simplified boundary conditions is also used at the top and right walls that do not present strike points. With respect to the set of boundary conditions in Eqs. (\ref{eq:bounplasma_start}-\ref{eq:boundplasma_end}), in these cases the electrostatic potential is set to $\phi = \Lambda T_e$, implying $v_{\| e} = \pm c_s $.

\subsection{The multi-species kinetic neutral model}
The neutral model used in this work is based on the kinetic description introduced in a limited tokamak configuration in Ref. \cite{Wersal_2015} for the case of a mono-atomic neutral species, then extended in Ref. \cite{Coroado_2022_1} to take into account molecular deuterium. The same approach to study the neutral dynamics was used in Ref. \cite{Mancini_2021} to carry out in a diverted configuration, with a single species model. Here, we extend Ref. \cite{Mancini_2021} to include the molecular dynamics. We underline that our model can be extended to include, in principle, an arbitrary number of species.

The kinetic equation we consider to evolve the distribution function $f_{\D}$ is
\begin{align}
    \begin{split}\label{eq:fD}
        \frac{\partial f_{\D}}{\partial t} + \mathbf{v}\cdot\frac{\partial f_{\D}}{\partial \mathbf{x}} =& - \nu_{\text{iz,D}} f_{\D} - \nu_{\text{cx,D}}\left(f_{\D}-\frac{n_{\D}}{n_{\Dp}}f_{\Dp}\right) + \nu_{\text{rec,}\Dp} f_{\Dp} \\
        &+ \nu_{\text{cx,}\Dt\text{-}\Dp} \left(\frac{n_{\Dt}}{n_{\Dp}} f_{\Dp}\right) - \nu_{\text{cx,}\D\text{-}\Dtp} f_{\D} + 2 \nu_{\text{diss,}\Dt} f_{\Dt} + \nu_{\text{diss-iz,}\Dt} f_{\Dt} \\
        &+ \nu_{\text{diss,}\Dtp} f_{\Dtp} + 2 \nu_{\text{diss-rec,}\Dtp} f_{\Dtp} \quad ,
    \end{split}
\end{align}
and a similar one is used for $f_{\Dt}$,
\begin{align}
    \begin{split}\label{eq:fD2}
    \frac{\partial f_{\Dt}}{\partial t} + \mathbf{v}\cdot\frac{\partial f_{\Dt}}{\partial \mathbf{x}} =& - \nu_{\text{iz,}\Dt} f_{\Dt} - \nu_{\text{cx,}\Dt}\left(f_{\Dt}-\frac{n_{\Dt}}{n_{\Dtp}}f_{\Dtp}\right) \\
    &+ \nu_{\text{rec,}\Dtp} f_{\Dtp} - \nu_{\text{cx,}\Dt\text{-}\Dp} f_{\Dt} + \nu_{\text{cx,}\D\text{-}\Dtp} \left(\frac{n_{\D}}{n_{\Dtp}} f_{\Dtp}\right) \\
    &- \nu_{\text{diss,}\Dt} f_{\Dt} - \nu_{\text{diss-iz,}\Dt} f_{\Dt} \quad ,
    \end{split}
\end{align}
where $f_{\Dp}$ and $f_{\Dtp}$ are the velocity distribution functions of the $\Dp$ and $\Dtp$ ions and all the reaction frequencies are defined in Table \ref{tab:reactions}.

The formal solution of Eqs. (\ref{eq:fD}-\ref{eq:fD2}) can be found by applying the method of characteristics, yielding:
\begin{align}
    \begin{split} \label{eq:intfD}
        f_{n}(\mathbf{x},\mathbf{v},t) =& \int_0^{r'_{\text{b}}} \left[\frac{S_n (\mathbf{x}',\mathbf{v},t')}{v}+\delta \left(r'-r'_{\text{b}}\right) f_n (\mathbf{x}'_{\text{b}},\mathbf{v},t'_{\text{b}})\right] \\
            &\times\exp\left[-\frac{1}{v}\int_0^{r'} \nu_{\text{eff},n}(\mathbf{x}'',t'') dr''\right] dr' \quad ,
    \end{split}
\end{align}
for the two neutral species, $n = \D$ or $\Dt$. Equation \eqref{eq:intfD} describes the distribution function of neutrals at position $\mathbf{x}$, velocity $\mathbf{v}$ and time $t$, as the result of neutrals generated at position $\mathbf{x}' = \mathbf{x} - r' \mathbf{v}/v$, and time $t' = t - r'/v$, where $r'$ is the coordinate along the characteristic connecting $\mathbf{x}'$ and $\mathbf{x}$, and $r'_{\text{b}}$ denotes the distance between the position $\mathbf{x}$ and the intersection of the characteristic with the boundary. The term $S_n$ is the volumetric source of $\D$ or $\Dt$, generated by charge-exchange, recombination and dissociation reactions. The exponential term in Eq. \eqref{eq:intfD} takes into account all processes that lead to a loss of neutrals on the way from $\mathbf{x}'$ to $\mathbf{x}$.

We now turn to the boundary condition for the distribution function, $f_{n}(\mathbf{x}_b,\mathbf{v},t'_b)$ in Eq. \eqref{eq:intfD}. In typical experimental conditions, $\D$ or $\Dt$ are emitted from the wall as a result of the reciclying of the $\Dp$ or $\Dtp$ ions impacting the wall. A fraction of the outflowing ions, $\alpha_{\text{refl}}$, is reflected as fast neutrals, with the same temperature of the impacting ions, while the rest of the ions are absorbed by the wall and emitted with the boundary temperature $T_b$. Similarly, the reflection or re-emission of the outflowing neutrals contribute to the neutral emission of the same neutral species flux from the wall. In addition, a small fraction, $\beta_{\text{assoc}}$, of the absorbed $\Dp$ and $\D$ goes through association processes, contributing to the $\Dt$ emission. The resulting boundary condition for the $\D$ species is therefore
\begin{align}
    \begin{split}\label{eq:fDb}
        f_{\D}(\mathbf{x}'_{\text{b}},\mathbf{v},t') =& (1-\alpha_{\text{refl}}) \Gamma_{\text{emiss,}\D}(\mathbf{x'_b},t') \chi_{\text{in,}\D}(\mathbf{x'_b},\mathbf{v},T_b) \\
            &+ \alpha_{\text{refl}} \left[ f_{\D}(\mathbf{x'_b}, \mathbf{v}-2 \mathbf{v_p},t') + f_{\Dp}(\mathbf{x'_b}, \mathbf{v}-2 \mathbf{v_p},t')\right] \quad ,
    \end{split}
\end{align}
where $\chi_{\text{in,}\D}$ is the velocity distribution of the emitted neutrals and $\mathbf{v}_p$ is the velocity in the direction perpendicular to the wall. For the $\Dt$ species we impose a similar boundary condition,
\begin{align}
    \begin{split}\label{eq:fD2b}
        f_{\Dt}(\mathbf{x}'_{\text{b}},\mathbf{v},t') =& (1-\alpha_{\text{refl}}) \Gamma_{\text{emiss,}\Dt}(\mathbf{x'_b},t') \chi_{\text{in,}\Dt}(\mathbf{x'_b},\mathbf{v},T_b) \\
            &+ \alpha_{\text{refl}} \left[ f_{\Dt}(\mathbf{x'_b}, \mathbf{v}-2 \mathbf{v_p},t') + f_{\Dtp}(\mathbf{x'_b}, \mathbf{v}-2 \mathbf{v_p},t')\right] \quad .
    \end{split}
\end{align}
The fluxes of the emitted neutrals takes into account the probability of association, $\beta_{\text{assoc}}$
\begin{align}
    \begin{split}\label{eq:GammaD}
        \Gamma_{\text{emiss,}\D} =& \, (1-\beta_{\text{assoc}}) \left( \Gamma_{\text{out,}\D} + \Gamma_{\text{out,}\Dp} \right)
    \end{split}
    \\
    \begin{split}\label{eq:GammaD2}
        \Gamma_{\text{emiss,}\Dt} =& \, \Gamma_{\text{out,}\Dt} + \Gamma_{\text{out,}\Dtp} + \frac{\beta_{\text{assoc}}}{2} \left( \Gamma_{\text{out,}\D} + \Gamma_{\text{out,}\Dp} \right) \quad ,
    \end{split}
\end{align}
and include the fluxes of ions to the walls due to their parallel motion, the diamagnetic and $\mathbf{E} \times \mathbf{B}$ drifts, $\Gamma_{\text{out},\Dp}$ and $\Gamma_{\text{out},\Dtp}$, as well as the outflowing fluxes of neutrals, $\Gamma_{\text{out},\D}$ and $\Gamma_{\text{out},\Dt}$.

As detailed in Refs. \cite{Coroado_2022_1, Wersal_2015}, Eq. \eqref{eq:intfD} can be integrated in velocity space, obtaining an integral equation for the neutral densities, $n_{\D}$ and $n_{\Dt}$. The resulting equations can be simplified under the assumptions that the time of flight of neutrals is lower than the turbulence timescale and that the mean free path of neutrals is shorter than the typical turbulence scale lengths in the parallel direction, obtaining a set of two-dimensional equations for the variables $n_{\D}$ and $n_{\Dt}$, that is
\begingroup
\allowdisplaybreaks
\begin{align}
    \begin{split}\label{eq:density_D}
        n_{\text{D}}(\mathbf{x}_\perp) =& \int_{\text{S}} n_{\D}(\mathbf{x}'_\perp) \nu_{\text{cx,}\D}(\mathbf{x}'_\perp) K_{p\rightarrow p}^{\D\text{,}\Dp}(\mathbf{x}_\perp,\mathbf{x}'_\perp) dA' \\
        &+ \int_{\text{S}} n_{\Dt}(\mathbf{x}'_\perp) \nu_{\text{cx,}\D2\text{-}\Dp}(\mathbf{x}'_\perp) K_{p\rightarrow p}^{\D\text{,}\Dp}(\mathbf{x}_\perp,\mathbf{x}'_\perp) dA' \\
        &+ \int_{\text{S}} 2 n_{\Dt}(\mathbf{x}'_\perp) \nu_{\text{diss,}\Dt}(\mathbf{x}'_\perp) K_{p\rightarrow p}^{\D\text{,diss}\left(\ \Dt \right)}(\mathbf{x}_\perp,\mathbf{x}'_\perp) dA' \\
        &+ \int_{\text{S}} n_{\Dt}(\mathbf{x}'_\perp) \nu_{\text{diss-iz,}\Dt}(\mathbf{x}'_\perp) K_{p\rightarrow p}^{\D\text{,diss-iz}\left( \Dt \right)}(\mathbf{x}_\perp,\mathbf{x}'_\perp) dA' \\
        &+\int_{\partial \text{S}} (1-\alpha_{\text{refl}}(\mathbf{x}'_{\perp,\text{b}})) (1-\beta_{\text{assoc}}) \Gamma_{\text{out,}\D}(\mathbf{x}'_{\perp,\text{b}})K_{b\rightarrow p}^{\D\text{,reem}} (\mathbf{x}_\perp,\mathbf{x}'_{\perp,\text{b}}) da'_{\text{b}} \\
        &+n_{\D[\text{rec}(\Dp)]}(\mathbf{x}_\perp) + n_{\D[\text{out}(\Dp)]}(\mathbf{x}_\perp) + n_{\D[\text{diss}(\Dtp)]}(\mathbf{x}_\perp) \quad ,
    \end{split}
\end{align}
and
\begin{align}
    \begin{split}\label{eq:density_D2}
        n_{\Dt}(\mathbf{x}_\perp) =& \int_{\text{S}} n_{\Dt}(\mathbf{x}'_\perp) \nu_{\text{cx,}\Dt}(\mathbf{x}'_\perp) K_{p\rightarrow p}^{\Dt,\Dtp}(\mathbf{x}_\perp,\mathbf{x}'_\perp) dA' \\
        &+ \int_{\partial \text{S}} (1-\alpha_{\text{refl}}(\mathbf{x}'_{\perp,\text{b}})) \Gamma_{\text{out,}\Dt}(\mathbf{x}'_{\perp,\text{b}})K_{b\rightarrow p}^{\Dt}(\mathbf{x}_\perp,\mathbf{x}'_{\perp,\text{b}}) da'_{\text{b}} \\
        &+ \int_{\partial \text{S}} (1-\alpha_{\text{refl}}(\mathbf{x}'_{\perp,\text{b}})) \frac{\beta_{\text{assoc}}}{2} \Gamma_{\text{out,}\D}(\mathbf{x}'_{\perp,\text{b}})K_{b\rightarrow p}^{\Dt}(\mathbf{x}_\perp,\mathbf{x}'_{\perp,\text{b}}) da'_{\text{b}} \\
        &+ \int_{\text{S}} n_{\D}(\mathbf{x}'_\perp) \nu_{\text{cx,}\D\text{-}\Dtp}(\mathbf{x'_\perp}) K_{p\rightarrow p}^{\Dt\text{,}\Dtp}(\mathbf{x}_\perp,\mathbf{x}'_\perp) dA' \\ 
        &+ n_{\Dt[\text{rec}(\Dtp)]}(\mathbf{x}_\perp) + n_{\Dt[\text{out}(\Dtp)]}(\mathbf{x}_\perp) + n_{\Dt[\text{out}(\Dp)]}(\mathbf{x}_\perp) \quad ,
      \end{split}
\end{align}
\endgroup
which are coupled with two equations for the outgoing neutral fluxes, $\Gamma_{\text{out,}\D}$ and $\Gamma_{\text{out,}\Dt}$,
\begingroup
\allowdisplaybreaks
\begin{align}
    \begin{split} \label{eq:Gammaout_D}
        \Gamma_{\text{out,}\D}(\mathbf{x}_{\perp,\text{b}}) =& \int_{\text{S}} n_{\D}(\mathbf{x}'_\perp) \nu_{\text{cx,}\D}(\mathbf{x}'_\perp) K_{p\rightarrow b}^{\D\text{,}\Dp}(\mathbf{x}_\perp,\mathbf{x}'_\perp) dA' \\
        &+ \int_{\text{S}} n_{\Dt}(\mathbf{x}'_\perp) \nu_{\text{cx,}\Dt\text{-}\Dp}(\mathbf{x}'_\perp) K_{p\rightarrow b}^{\D\text{,}\Dp}(\mathbf{x}_\perp,\mathbf{x}'_\perp) dA' \\
        &+ \int_{\text{S}} 2 n_{\Dt}(\mathbf{x}'_\perp) \nu_{\text{diss,}\Dt}(\mathbf{x}'_\perp) K_{p\rightarrow b}^{\D\text{,diss}\left( \Dt \right)}(\mathbf{x}_\perp,\mathbf{x}'_\perp) dA' \\
        &+ \int_{\text{S}} n_{\Dt}(\mathbf{x}'_\perp) \nu_{\text{diss-iz,}\Dt}(\mathbf{x}'_\perp) K_{p\rightarrow b}^{\D\text{,diss-iz}\left( \Dt \right)}(\mathbf{x}_\perp,\mathbf{x}'_\perp) dA' \\
        &+ \int_{\partial \text{S}} (1-\alpha_{\text{refl}}(\mathbf{x}'_{\perp,\text{b}})) (1-\beta_{\text{assoc}}) \Gamma_{\text{out,}\D}(\mathbf{x}'_{\perp,\text{b}})K_{b\rightarrow b}^{\D\text{,reem}} (\mathbf{x}_\perp,\mathbf{x}'_{\perp,\text{b}}) da'_{\text{b}} \\
        &+ \Gamma_{\D[\text{rec}(\Dp)]}(\mathbf{x}_\perp) + \Gamma_{\D[\text{out}(\Dp)]}(\mathbf{x}_\perp) + \Gamma_{\D[\text{diss}(\Dtp)]}(\mathbf{x}_\perp) \quad ,
    \end{split}
\end{align}
and
\begin{align}
      \begin{split}\label{eq:Gammaout_D2}
        \Gamma_{\text{out,}\Dt}(\mathbf{x}_{\perp,\text{b}}) =& \int_{\text{S}} n_{\Dt}(\mathbf{x}'_\perp) \nu_{\text{cx,}\Dt}(\mathbf{x}'_\perp) K_{p\rightarrow b}^{\Dt\text{,}\Dtp}(\mathbf{x}_\perp,\mathbf{x}'_\perp) dA' \\
            &+ \int_{\partial \text{S}} (1-\alpha_{\text{refl}}(\mathbf{x}'_{\perp,\text{b}})) \Gamma_{\text{out,}\Dt}(\mathbf{x}'_{\perp,\text{b}})K_{b\rightarrow b}^{\Dt}(\mathbf{x}_\perp,\mathbf{x}'_{\perp,\text{b}}) da'_{\text{b}} \\
            &+ \int_{\partial \text{S}} (1-\alpha_{\text{refl}}(\mathbf{x}'_{\perp,\text{b}})) \frac{\beta_{\text{assoc}}}{2} \Gamma_{\text{out,}\D}(\mathbf{x}'_{\perp,\text{b}})K_{b\rightarrow b}^{\Dt}(\mathbf{x}_\perp,\mathbf{x}'_{\perp,\text{b}}) da'_{\text{b}} \\
            &+ \int_{\text{S}} n_{\D}(\mathbf{x}'_\perp) \nu_{\text{cx,}\D\text{-}\Dtp}(\mathbf{x'_\perp}) K_{p\rightarrow b}^{\Dt\text{,}\Dtp}(\mathbf{x}_\perp,\mathbf{x}'_\perp) dA' \\
            &+ \Gamma_{\text{out,}\Dt[\text{rec}(\Dtp)]}(\mathbf{x}_\perp) + \Gamma_{\text{out,}\Dt[\text{out}(\Dtp)]}(\mathbf{x}_\perp) + \Gamma_{\text{out,}\Dt[\text{out}(\Dp)]}(\mathbf{x}_\perp) \quad ,
      \end{split}
\end{align}
\endgroup
where the integrals appearing in Eqs. (\ref{eq:density_D}-\ref{eq:Gammaout_D2}) are carried out over the area, $S$, of each poloidal plane or over its boundary, $\partial S$.
In Eqs. (\ref{eq:density_D}-\ref{eq:Gammaout_D2}) the terms $K_{i\rightarrow j}$ are the kernel functions presented in Ref. \cite{Coroado_2022_1}, where the integrals in the velocity space are performed. The contribution to the neutral densities and fluxes at the wall, which are proportional to the ion densities and fluxes, include the contribution to $n_{\D}$ coming from volumetric $\Dp$ recombination
\begin{align}
    \begin{split} \label{eq:nD_rec}
        n_{\D \text{[rec(}\Dp\text{]}}(\mathbf{x}_\perp) =& \int_{\text{S}} n_{\Dp}(\mathbf{x}'_\perp) \nu_{\text{rec,}\Dp}(\mathbf{x}'_\perp) K_{p\rightarrow p}^{\D\text{,}\Dp}(\mathbf{x}_\perp,\mathbf{x}'_\perp) dA' \quad ,
    \end{split}
\end{align}
from $\Dp$ recombination on the boundary, 
\begin{align}
    \begin{split} \label{eq:nD_out}
        n_{\D \text{[out(}\Dp\text{]}} =& \int_{\partial \text{S}} \Gamma_{\text{out,}\Dp}(\mathbf{x}'_{\perp,\text{b}}) [ (1-\alpha_{\text{refl}}(\mathbf{x}'_{\perp,\text{b}})) (1-\beta_{\text{assoc}}) K_{b\rightarrow p}^{\D\text{,reem}} (\mathbf{x}_\perp,\mathbf{x}'_{\perp,\text{b}}) \\
            &+ \alpha_{\text{refl}} K_{b\rightarrow p}^{\D\text{,refl}} (\mathbf{x}_\perp,\mathbf{x}'_{\perp,\text{b}}) ] da'_{\text{b}} \quad ,
    \end{split}
\end{align}
and from $\Dtp$ dissociation,
\begin{align}
    \begin{split} \label{eq:nD_MAR}
        n_{\D \text{[diss(}\Dtp\text{]}} =& \int_{\text{S}} n_{\Dtp}(\mathbf{x}'_\perp) [ \nu_{\text{diss,}\Dtp}(\mathbf{x}'_\perp) K_{p\rightarrow p}^{\D\text{,diss(}\Dtp\text{)}}(\mathbf{x}_\perp,\mathbf{x}'_\perp) \\
            &+ 2 \nu_{\text{diss-rec,}\Dtp}(\mathbf{x}'_\perp) K_{p\rightarrow p}^{\D\text{,diss-rec(}\Dtp\text{)}}(\mathbf{x}_\perp,\mathbf{x}'_\perp) ] dA' \quad .
    \end{split}
\end{align}

The set of Eqs. (\ref{eq:density_D}-\ref{eq:Gammaout_D2}) is discretized on a Cartesian grid, $(R,Z)$, written in matrix form and then numerically solved for $n_{\D}$ and $n_{\Dt}$. Once the densities are known, all moments of the neutral distribution function can be evaluated (see Ref. \cite{Coroado_2022_1}).

\subsection{Simulation setup}
\label{subsec:setup}
\begin{figure*}
    \centering
    \includegraphics[width=0.4 \textwidth]{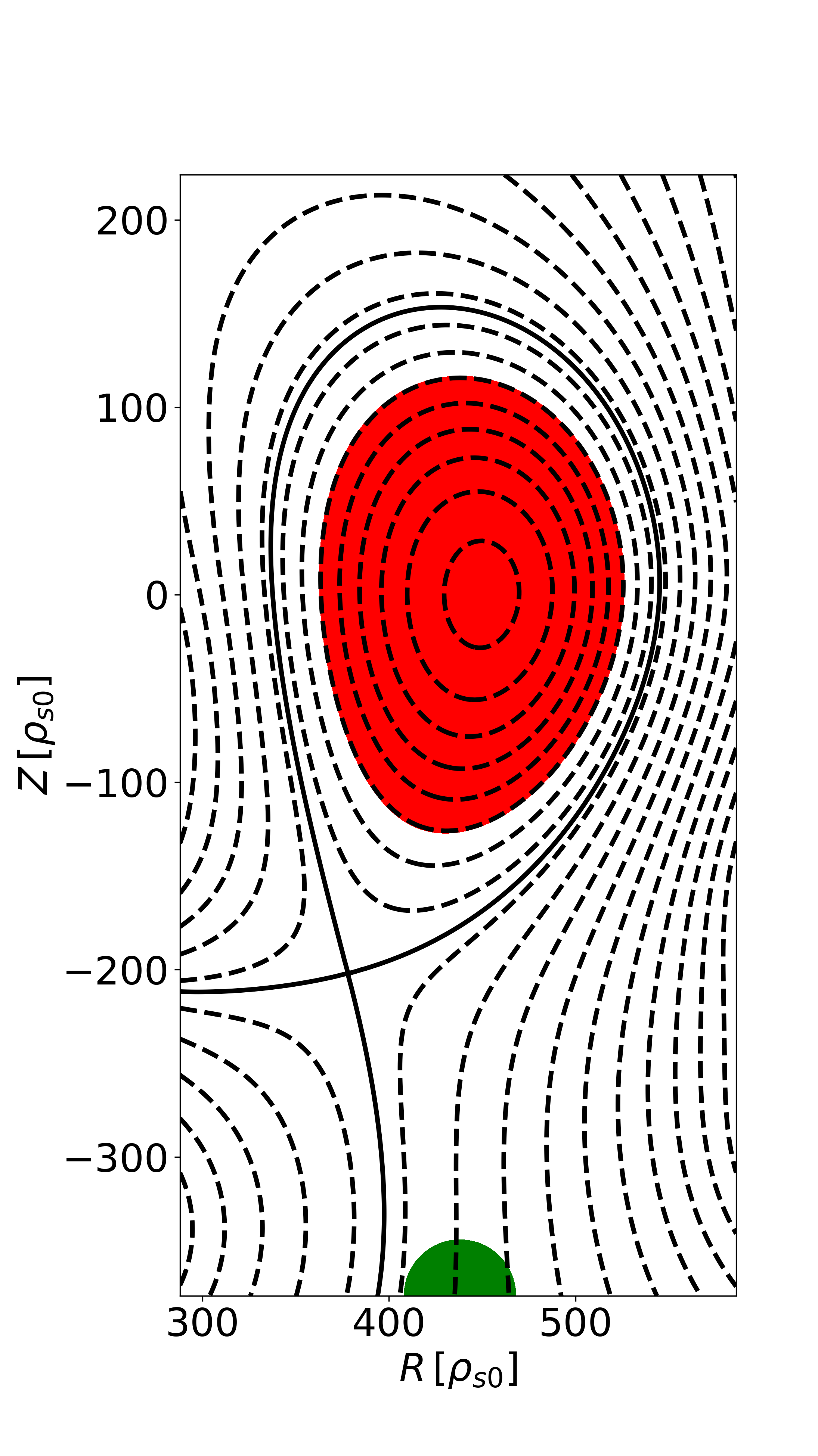}
    \caption{Magnetic flux surfaces in the poloidal plane considered for the simulations. The area covered by the electron temperature source is in red, inside the core, the position of the neutral gas puff is in green, on the bottom boundary.}
    \label{fig:sim_setup}
\end{figure*}

In this work, we consider two simulations carried out with the GBS code implementing the model described in Sec. \ref{sec:model}. The simulation parameters are based on an experimental dataset developed for validation studies, TCV-X21 \cite{Oliveira_2022}. TCV-X21 is a lower single-null L-mode discharge performed at low toroidal magnetic field, with value at magnetic axis $B_0=0.95$~T, in forward field direction (ion-$\nabla B$ drift direction pointing from the core toward the X-point), with plasma current $I_p$=165~kA. The upstream experimental density and electron temperature at the separatrix, taken as the reference density and temperature for the simulations, are $n_0=0.6\times 10^{19}$~m$^{-3}$ and $T_{e0}=35$~eV. This corresponds to ion sound Larmor radius $\rho_{s0}\simeq 1$~mm, sound speed $c_{s0}\simeq 4.1\times 10^4$~m/s and reference time $t_0 = 0.02$~ms.
Given the explorative nature of the present study, the computational cost of the simulation is reduced by considering a domain corresponding to, approximately, half the size of the TCV tokamak ($R/\rho_{s0} = 450$), i.e. $L_R = 300\,\rho_{s0}$, $L_Z= 600\,\rho_{s0}$ and ${L_\varphi = 2\pi R_0 \simeq 2800\,\rho_{s0}}$. We note that simulations carried out with a realistic TCV size, performed with a previous version of GBS, were compared with experimental results \cite{Oliveira_2022}. At the same time, single species simulations with a realistic TCV size that take into account also the neutrals evolution are currently under analysis. The simulation setup used in this work is shown in Fig. \ref{fig:sim_setup}, where the chosen magnetic configuration, the position of the temperature source and the position of the neutral gas puffing are indicated.

The dimensionless simulation parameters are $\rho_*^{-1}=450$, $\tau=1$, $\eta_{0e}=3\times10^{-4}$, $\eta_{0\Dp}=\eta_{0\Dtp}=2\times10^{-2}$, $\chi_{\parallel e}=20$, $\chi_{\parallel \Dp}=\chi_{\parallel \Dtp}=1$, $m_{\Dp}/m_e=2500$, $\beta_{e0}=2\times 10^{-6}$, and $\nu_0=0.05$. The value of the ion to electron mass ratio is chosen to keep the inertial effects subdominant with respect to resistive effects, while reducing the computational cost of the simulations. The diffusion coefficients for numerical stability are set to $D_f=15$, with the field $f=\{n, T_e, T_{D^+}, T_{D_{2}^{+}}, \Omega, U_{\parallel e},v_{\parallel \Dp},v_{\parallel \Dtp}\}$, and the cross-field transport associated to those terms are verified to be at least one order of magnitude lower than the effective transport coefficients, as evaluated from the analysis of our results (see Sec. \ref{sec:transport}). The amplitude of the temperature source is chosen so that the power source, integrated over the core region, is close to the estimated experimental value of the power crossing the separatrix in the TCV-X21 case, $P_\text{sep} = 150$~kW \cite{Oliveira_2022}. These parameters are chosen to mimic the typical conditions found in L-mode diverted discharges, as described in Ref. \cite{Giacomin_2020}, where turbulent transport is mostly interchange driven.

Recycling is not considered on the top and right walls, where no strike points are present. A constant reflection coefficient, $\alpha_{\text{refl}} = 0.2$, and an association coefficient $\beta_{\text{ass}} = 0.1$ are considered on the left and bottom walls \cite{Stangeby_2000}. A gas puff is located on the bottom wall, with a narrow gaussian profile centered at the coordinate $R = 450 \rho_{s0}$, corresponding, approximately, to one of the gas puff positions present in TCV. The neutrals are puffed from the wall at room temperature $T_{\text{wall}} = T_{\text{GP,} \Dt} = 0.03\text{eV}$.

The two simulations presented in this work have the same setup, except for the strength of the $\Dt$ gas puff on the bottom wall. In the first simulation we introduce no puffing. Therefore, the presence of neutrals results only from plasma recycling and recombination processes. We label this simulation as \textit{low density}. In the second simulation, we increase the neutrals and plasma density by introducing a gas puff of $\Dt$, labelling it as \textit{high density} simulation. The two simulations allow us to explore the dynamics at two different separatrix densities. The low-density simulation is characterized by $n_{e,\text{sep}} = 1.62 \times 10^{19} \text{m}^{-3}$ and the high density by $n_{e,\text{sep}} = 3.42 \times 10^{19} \text{m}^{-3}$ at $Z = Z_{\text{axis}} = 0$.

Regarding the numerical parameters of our simulations, we use a plasma grid of $N_R \times N_Z \times N_\phi = 150 \times 300 \times 64$ points, while the neutral grid is $N^{n}_R \times N^{n}_Z \times N^{n}_\phi = 50 \times 100 \times 64$. The time step for the plasma evolution is $\Delta t \simeq 3 \times 10^{-5} t_0 $, while the solution for the neutral model is evaluated every $\Delta t \simeq 3 \times 10^{-2} t_0 $ \cite{Giacomin_2022}.
The initial conditions of the low-density simulation are provided by a quasi-steady state simulation with only atomic neutrals interactions \cite{Giacomin_2022}. A turbulent quasi-steady state in the low-density simulation is reached after approximately $20\,t_0$, when losses at the vessel balance the particle sources and the plasma and neutral quantities oscillate around constant values. The high-density simulation is then obtained introducing the $\Dt$ gas puff in the quasi-steady state of the low-density simulation. The time-averaged profiles are evaluated over an interval $\Delta t = 10 \, t_0$ during the quasi-steady states of the simulations. In the analysis, we also present flux tube averages that leverage the magnetic field aligned coordinate system introduced in Sec. \ref{sec:model}. The poloidal coordinate $\chi$ goes from $\chi = 0$ at the inner strike point (ISP) to $\chi = 1$ at the outer strike point (OSP). The radial coordinate is expressed as $\rho_{\psi} = \sqrt{(\psi - \psi_{\text{axis}})/(\psi_{\text{LCFS}} - \psi_{\text{axis}})}$, where $\psi_{\text{LCFS}}$ and $\psi_{\text{axis}}$ are the poloidal flux function values at the last closed flux surface and at the magnetic axis, having $\rho_{\psi} = 1$ at the last closed flux surface.

\section{The plasma and neutral turbulent dynamics}
\label{sec:results}
In this section, we provide an overview of the simulation results presented in this work, focusing on the turbulent dynamics of the plasma and neutral species, and their interactions. The density profiles of the plasma and neutral species are detailed in Sec. \ref{subsec:densities}. In Sec. \ref{subsec:pressures} we discuss the pressure profile, together with the analysis of the energy sink due to neutral interactions. The temperature profile is the subject of Sec. \ref{subsec:temperatures}. We present the electric field appearing in our simulations in Sec. \ref{subsec:electric_fields}. Finally, the formation of a density shoulder is the subject of Sec. \ref{subsec:turbulence}.

\subsection{Plasma and neutrals density profiles}
\label{subsec:densities}
\begin{figure*}
    \centering
    \begin{subfigure}[b]{0.45\textwidth}
        \includegraphics[width=\textwidth]{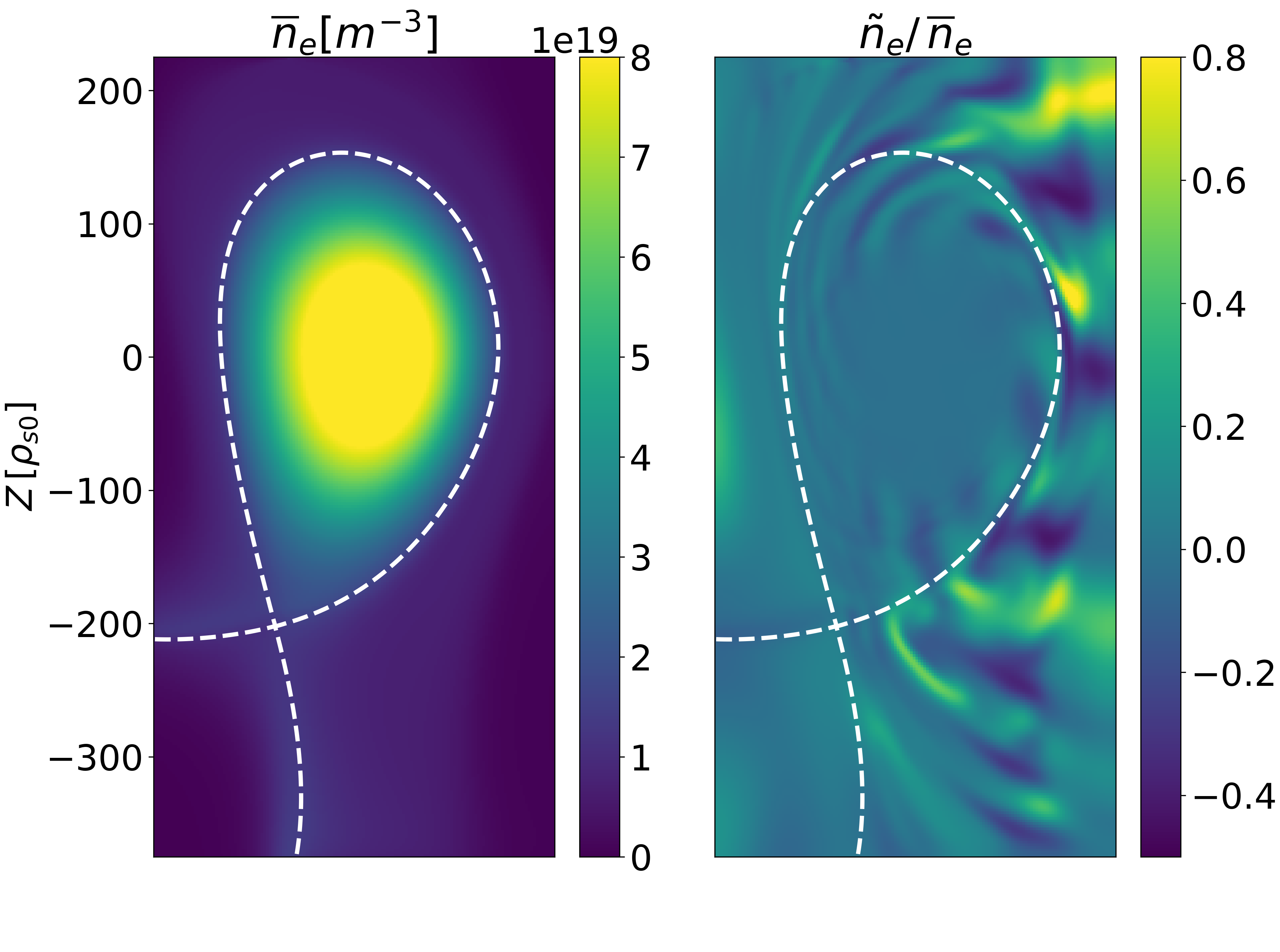}
    \end{subfigure}
    \begin{subfigure}[b]{0.45\textwidth}
        \includegraphics[width=\textwidth]{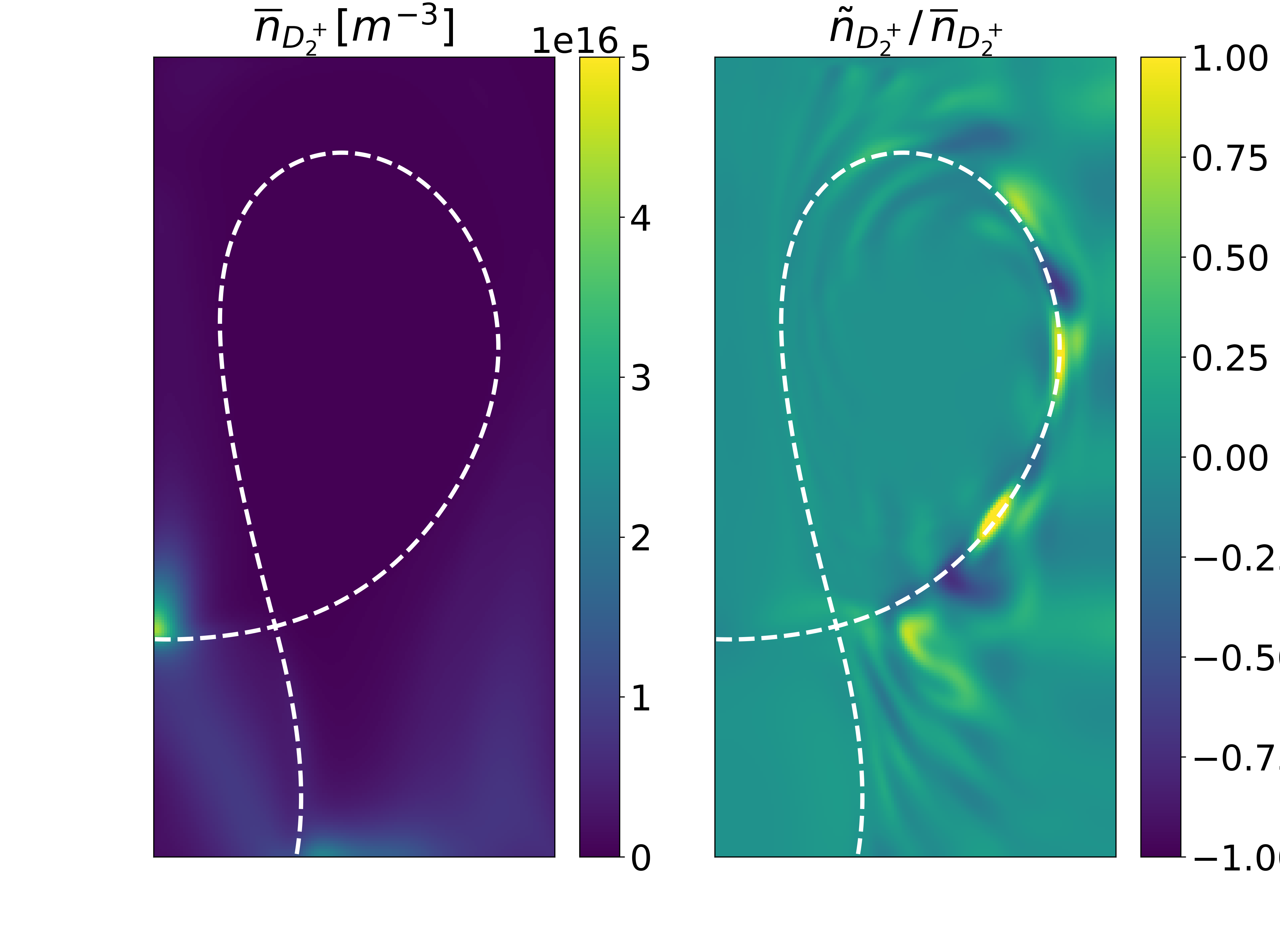}
    \end{subfigure}
    \\
    \begin{subfigure}[b]{0.45\textwidth}
        \includegraphics[width=\textwidth]{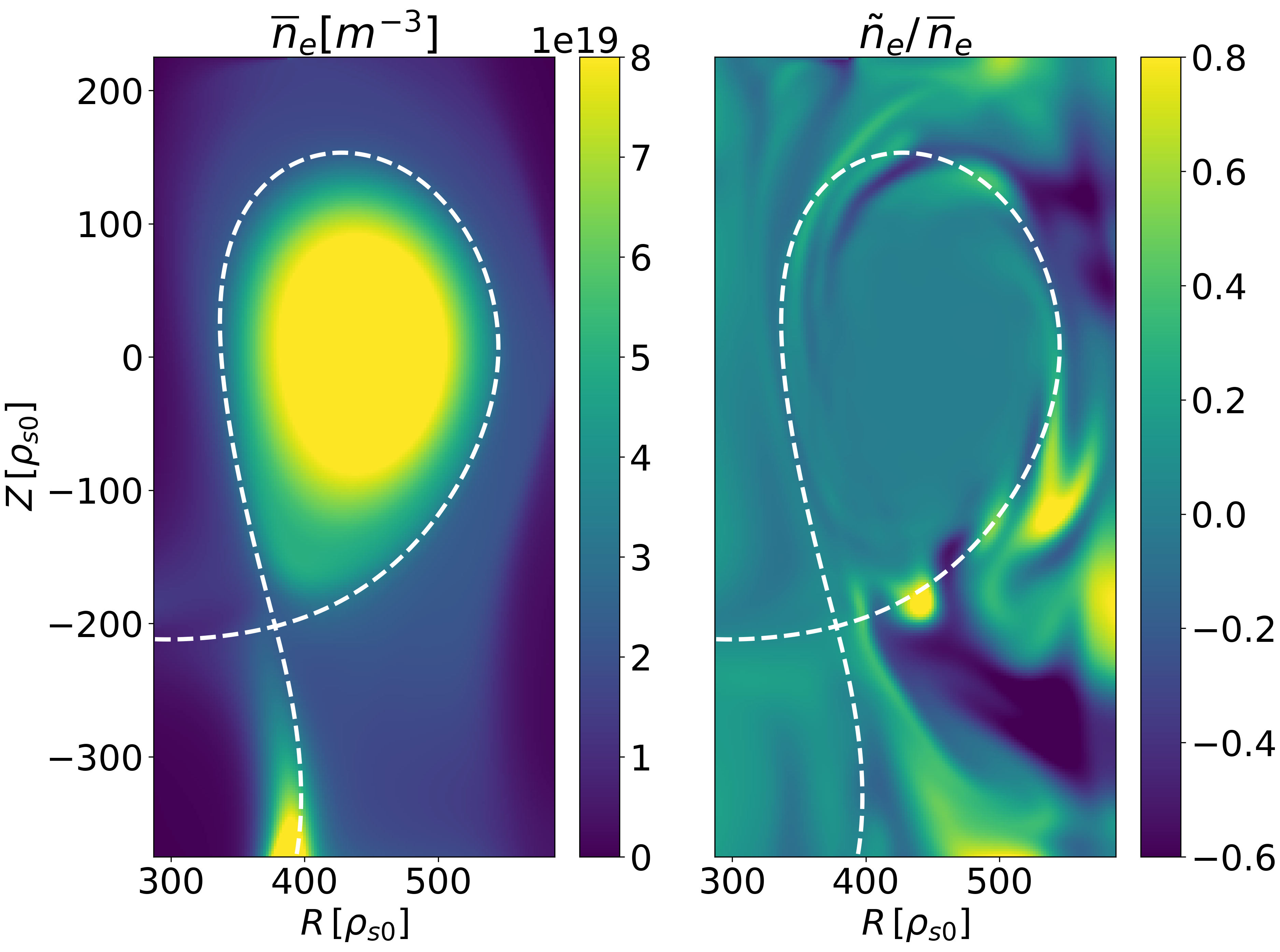}
    \end{subfigure}
    \begin{subfigure}[b]{0.45\textwidth}
        \includegraphics[width=\textwidth]{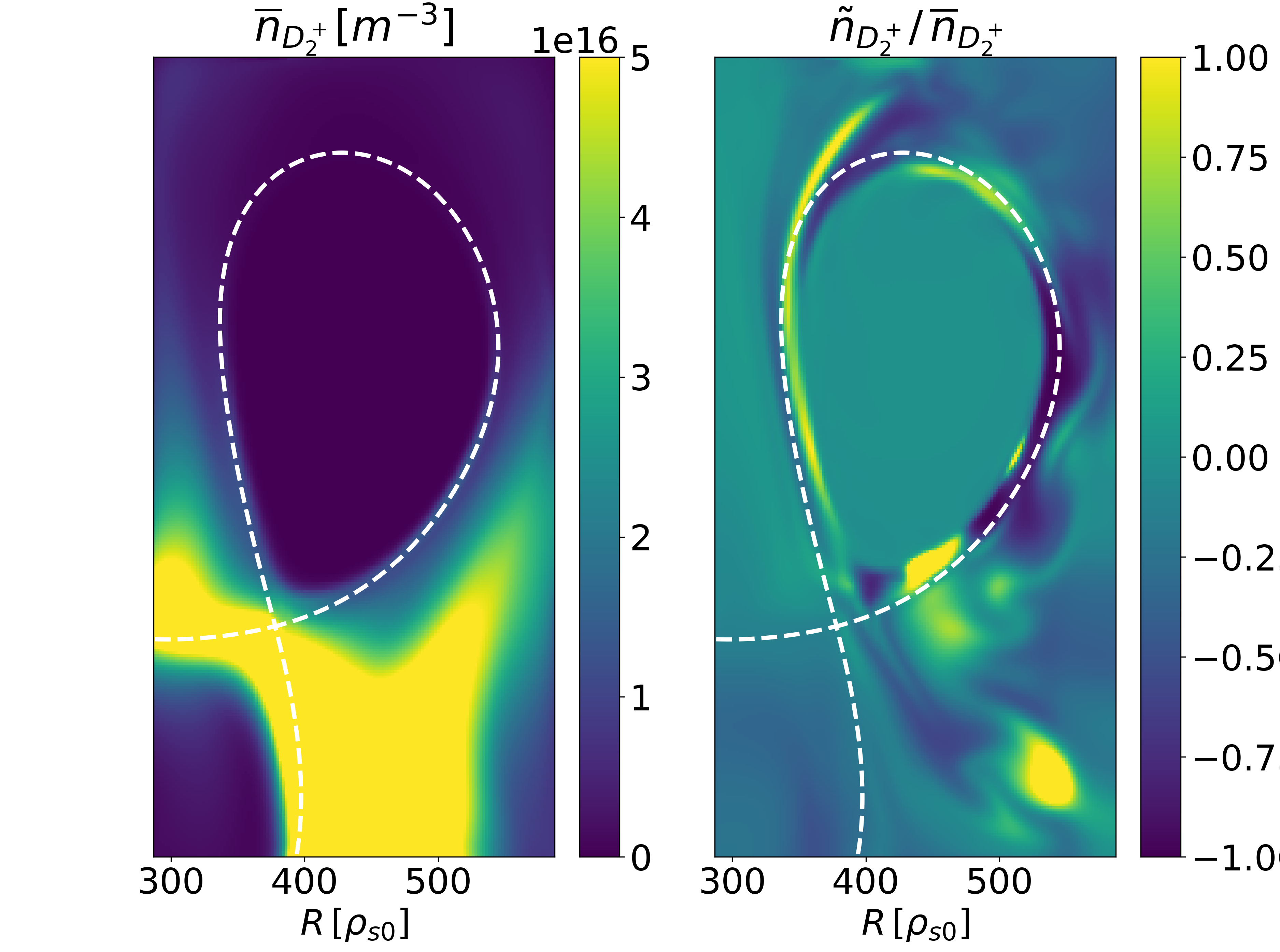}
    \end{subfigure}
    \caption{Time and toroidally-averaged profiles and typical snapshot of the normalized fluctuations of electron density, $n_{\el}$, and molecular ion density, $n_{\Dtp}$, for the low-density (top row) and high-density (bottom row) simulations.}
    \label{fig:plasma_density}
\end{figure*}

In Fig. \ref{fig:plasma_density}, time- and toroidally-averaged profiles of the electron and molecular ion densities on the poloidal plane are shown for the low- and high-density simulations, together with a typical snapshot of their fluctuations, normalized to their average values (we denote with tilde the fluctuating quantities and with overline their time- and toroidal-average values, e.g. $n_{\el} = \tilde{n}_{\el} + \overline{n}_{\el}$). 
The high-density simulation presents not only increased core density, but also increased density in the SOL region, with higher level of turbulence fluctuations. A similar increase of turbulent fluctuation amplitude with density are reported in Refs. \cite{Beadle_2020,Giacomin_2020}. The $\Dtp$ density is, at least, two orders of magnitude lower than the electron one and, as a consequence, $n_{\Dp} \simeq n_{\el}$, as assumed by our model. The high-density simulation exhibits strong enhancement of the plasma density in the private flux region close to the OSP, not observed at the ISP, resulting from the balance of the fluxes and the colder target existing at higher plasma density, as discussed in Sec. \ref{sec:transport}. The density of molecular ions is large in the region close to the targets, with a negligible value inside the last-closed flux surface.

\begin{figure*}
    \centering
    \begin{subfigure}[b]{0.495\textwidth}
        \includegraphics[width=\textwidth]{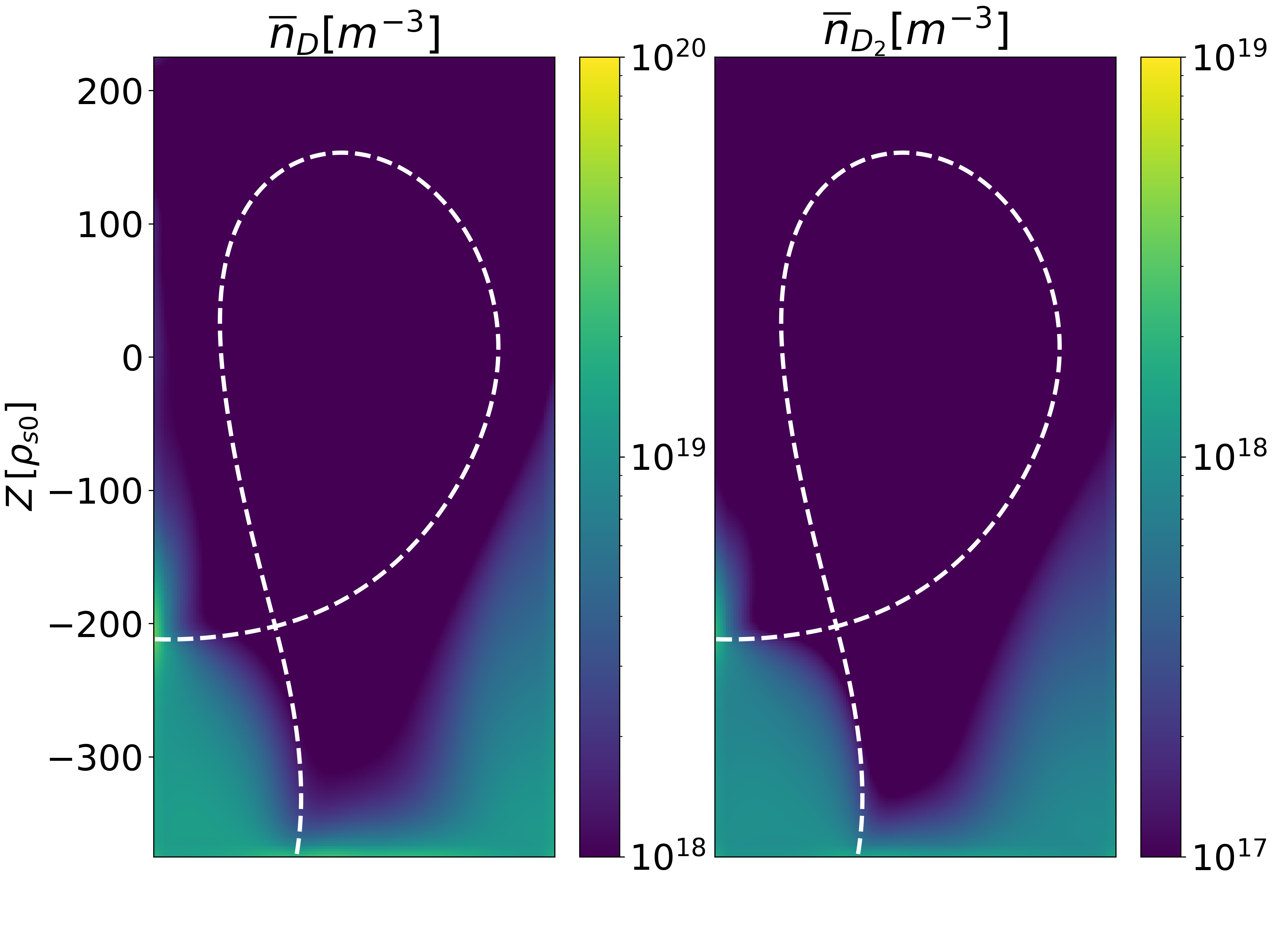}
    \end{subfigure}
    \begin{subfigure}[b]{0.495\textwidth}
        \includegraphics[width=\textwidth]{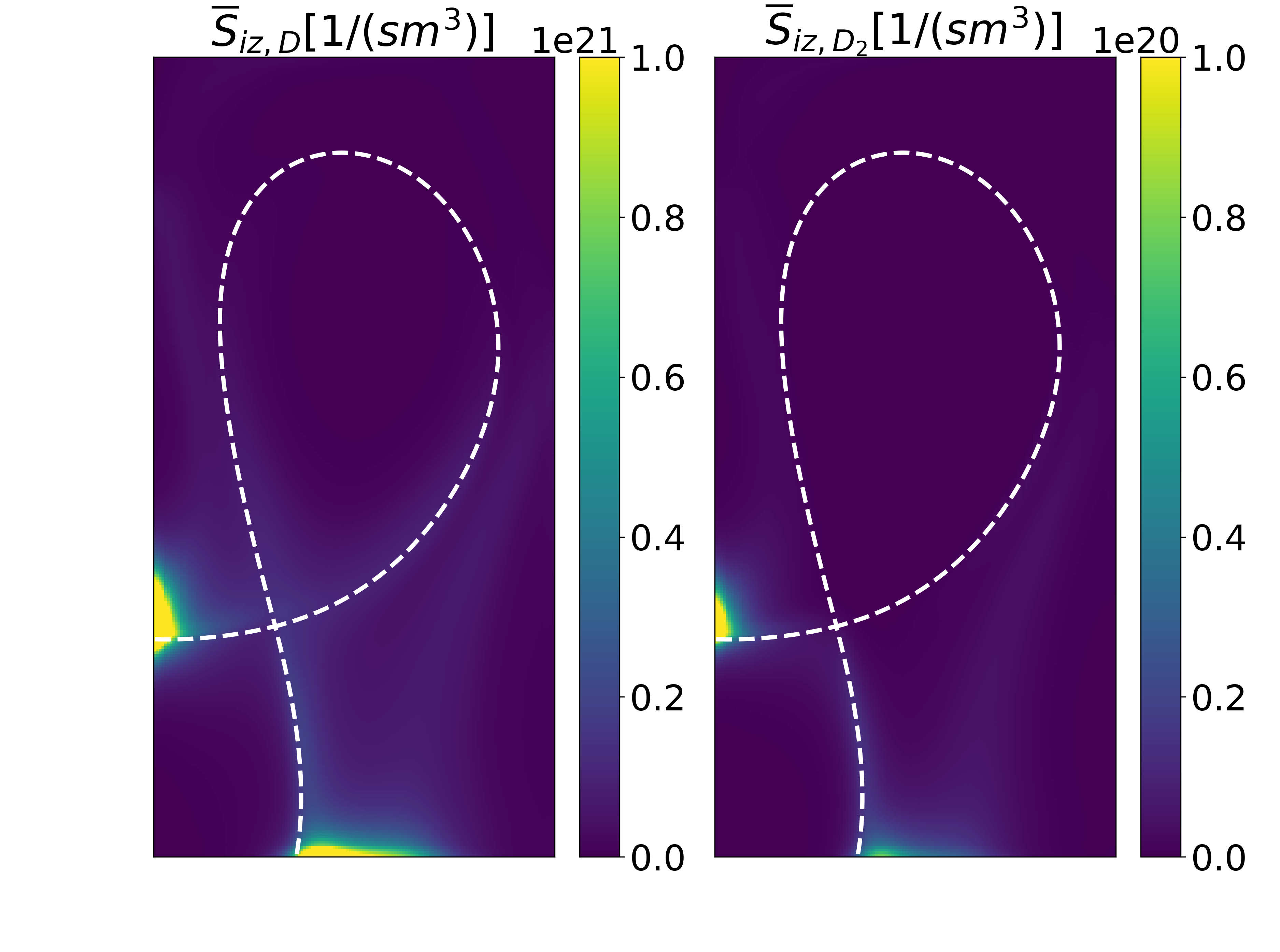}
    \end{subfigure}
    \\
    \begin{subfigure}[b]{0.495\textwidth}
        \includegraphics[width=\textwidth]{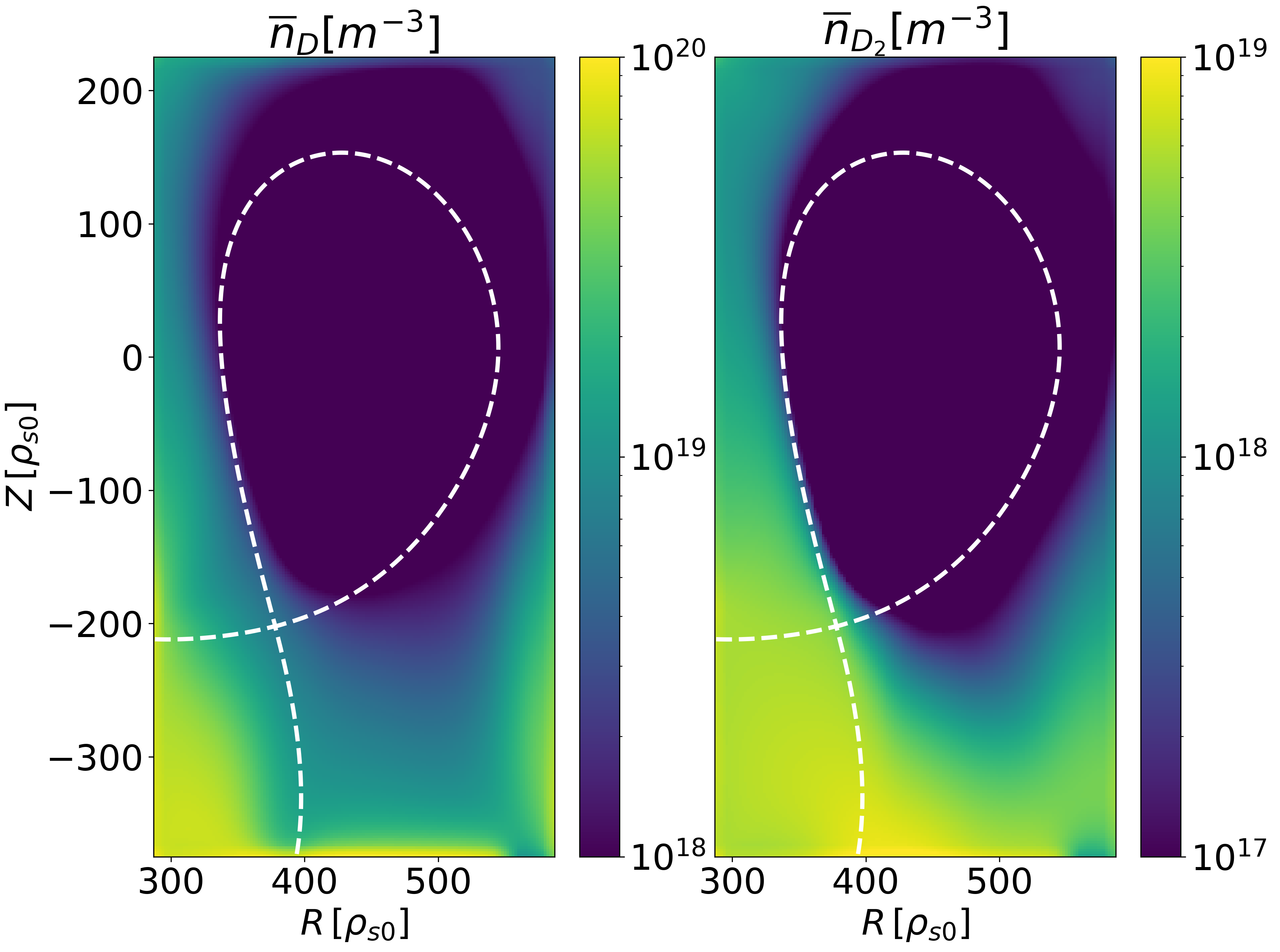}
    \end{subfigure}
    \begin{subfigure}[b]{0.495\textwidth}
        \includegraphics[width=\textwidth]{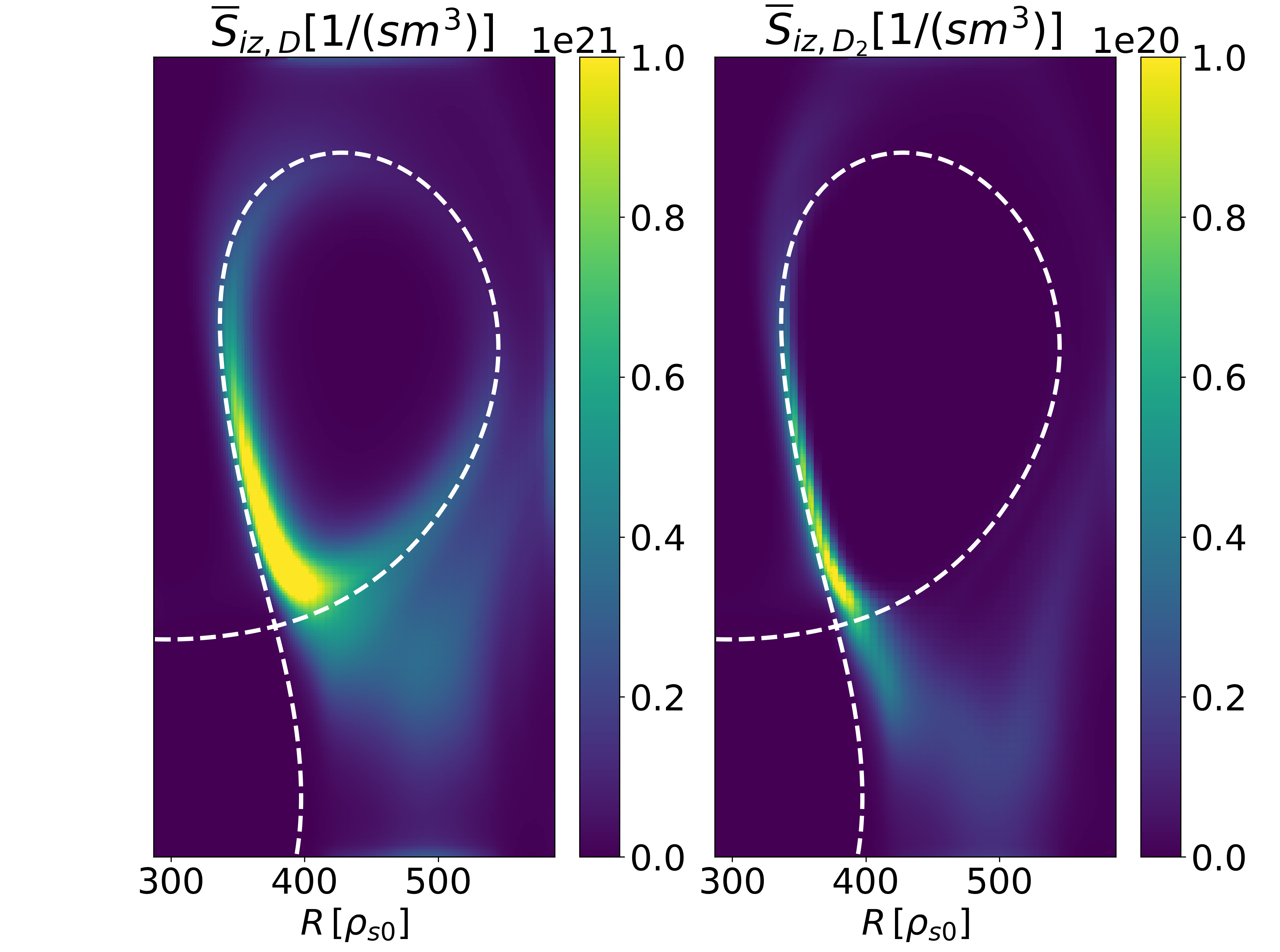}
    \end{subfigure}
    \caption{Time- and toroidally-averaged poloidal atomic neutrals density, $n_{D}$, molecular neutrals density, $n_{\Dt}$, $\Dp$ ion density source, $S_{n_{\Dp},\text{iz}}$, and $\Dtp$ ion density source, $S_{n_{\Dtp},\text{iz}}$, for the low-density (top row) and high-density (bottom row) simulations.}
    \label{fig:neutrals_density}
\end{figure*}

In Fig. \ref{fig:neutrals_density} we show the time- and toroidally-averaged profile of the neutral densities and of the ion density sources, $S_{\text{iz},\Dp}$ and $S_{\text{iz},\Dtp}$, where only direct ionizations of $\D$ and $\Dt$, respectively, are taken into account.
At low density, neutrals result from recombination processes at the wall and are recycled at the target, most of the ionizations occurring close to the targets. With the introduction of the gas puff, molecular neutrals penetrate deeper in the tokamak volume and the ionization front enters the edge and core regions.
In the high-density simulation, most of the ionization occurs inside the core, leading to a power radiated in this region due to ionization reactions that is up to $80 \%$ of the total power radiated in the entire tokamak volume, describing a scenario compatible with an X-point radiator \cite{Pan_2023}. We note that this is different from the experimental observations in the same magnetic configuration. We believe the differences with experiments is due to neutrals penetration in the core, which in our simulations is larger due to the reduced domain size.
At the same time, the atomic neutral density increases in the high-density simulation in all the SOL volume. This is due, at the same time, to the increase of recombination and the decrease of the ionization processes they undergo in the SOL. Focusing on recombination processes, we observe that atomic neutrals are produced through recombination, dissociation of $\Dt$ or $\Dtp$ molecules and MAR processes, as Table \ref{tab:reactions} shows. By performing a series of simulations where we artificially remove one of these reactions at a time, we identify MAR reactions as the main source of $n_{\D}$ in our high-density simulation. Indeed, they account for $40 \%$ of the produced neutrals. This result is in agreement with experimental findings, where MAR in high-density discharges are estimated dominant compared to other recombination channels \cite{Fil_2018, Verhaegh_2021_2}. Regarding the ionization processes, we note that their decrease in the target regions is a consequence of the strong decrease of the local temperature to values smaller than $3$~eV (see Sec. \ref{subsec:temperatures}).


\subsection{Power losses and pressure drop}
\label{subsec:pressures}
A detached scenario is characterized by a significant pressure drop between the upstream (OMP) region and the target \cite{Loarte_1998}, often observed with the increase of radiative losses due to a set of interactions with neutrals, which are important in the SOL region up to the X-point \cite{Reimerdes_2017,Potzel_2014}. The pressure drop is also the result of momentum loss mechanisms, e.g. due to ion-neutral charge-exchange reactions \cite{Stangeby_2000, krasheninnikov_2017}.

\begin{figure*}
    \centering
    \begin{subfigure}[b]{0.47\textwidth}
        \includegraphics[width=\textwidth]{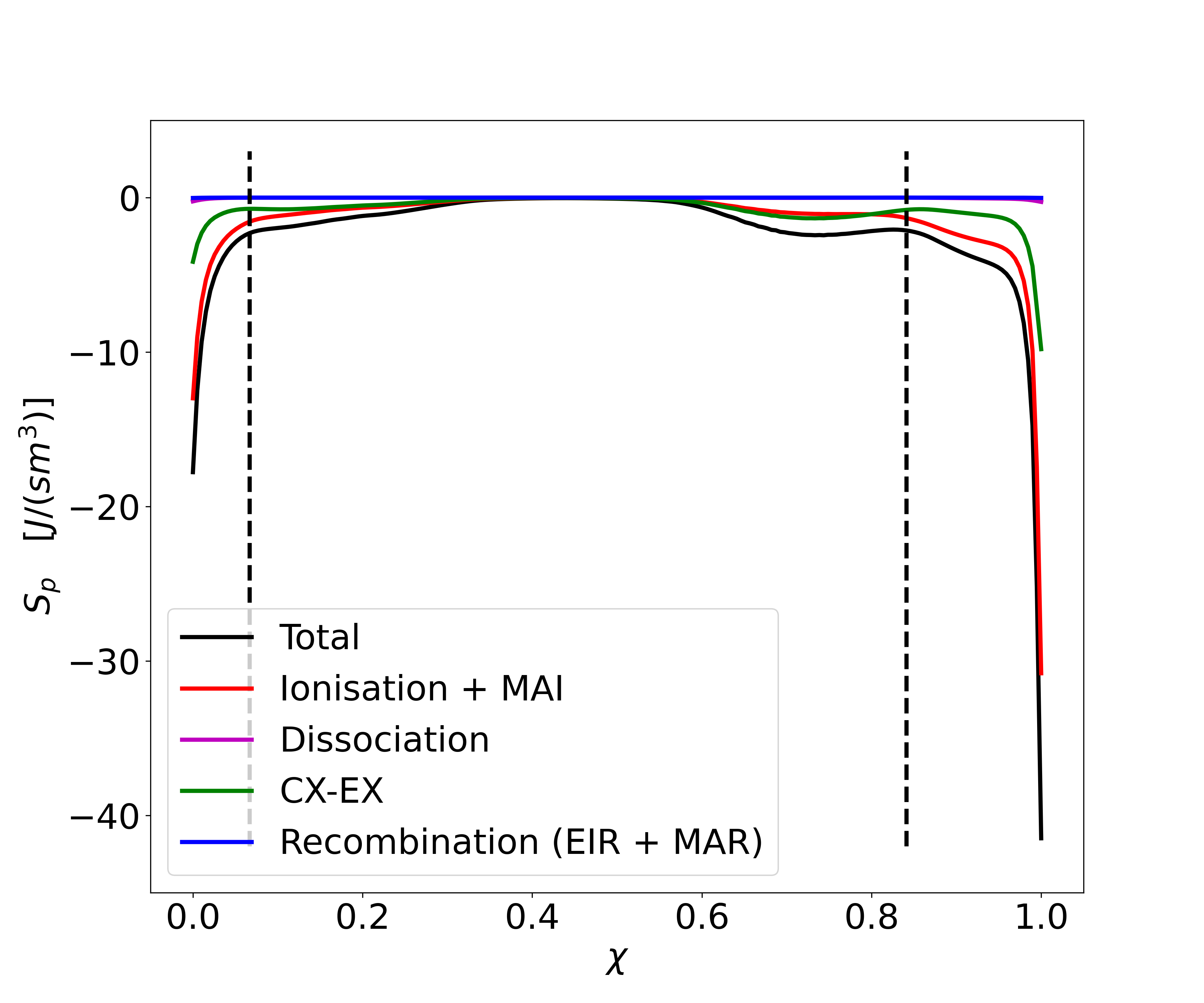}
    \end{subfigure}
    \begin{subfigure}[b]{0.47\textwidth}
        \includegraphics[width=\textwidth]{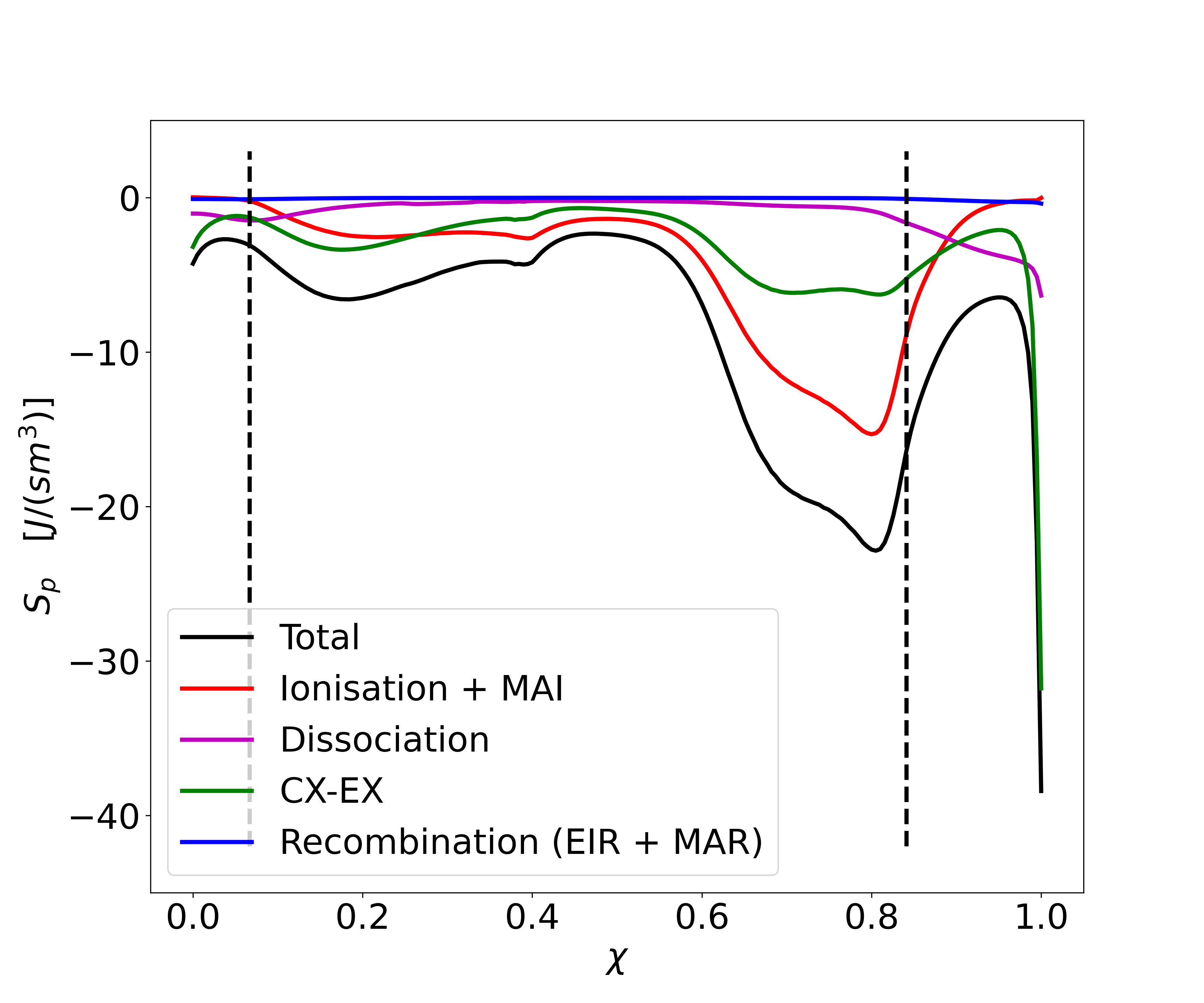}
    \end{subfigure}
    \caption{Time- and toroidally-averaged energy sink due to plasma-neutral interactions for the low-density (left) and high-density (right) simulations, along a flux tube close to the separatrix $1 \leq \rho_{\psi} \leq 1.1$, as a function of the poloidal coordinate $\chi$. The black dashed lines denote the X-point coordinates. MAI stands for Molecular Activated Ionization, CX-EX stands for the sum of charge-exchange reactions, EIR stands for Electron-Ion Recombination.}
    \label{fig:Sptot_sol}
\end{figure*}

In order to investigate the pressure drop in our simulations, we first consider the energy losses due to plasma-neutral interactions. In our model they are obtained by combining the density and temperature sources in Eqs. (\ref{eq:density_e}-\ref{eq:temperature_D2}), and they appear in Eq. \eqref{eq:sptot}. The losses associated with the neutral-plasma reactions considered in our model, evaluated separately in order to estimate their relative importance, are shown in Fig. \ref{fig:Sptot_sol} along a flux tube close to the separatrix, $1 \leq \rho_{\psi} \leq 1.1$, as a function of the poloidal coordinate $\chi$.
In the low-density simulation, ionization processes are relevant only in the target region and energy losses are present only below the X-point ($\chi_{\text{Xpt, HFS}} = 0.05$ and $\chi_{\text{Xpt, LFS}} = 0.86$), where both ionization and charge-exchange losses are important due to the significant neutral density.
On the other hand, the high-density simulation presents strong energy losses also above the X-point, where the ionization sink peaks. We point out that the integral of the energy losses above the X-point in the high-density simulation is twenty times as high as the low-density one, mainly due to the high $n_{\D}$ in the SOL resulting, as already discussed, from the molecular interactions \cite{Verhaegh_2021_2, Groth_2019}.
Focusing on the divertor legs, we note that both atomic and molecular ionization losses are practically absent close to both targets, a feature already observed in detachment experiments \cite{Reimerdes_2017}. On the outer divertor leg, $0.85 \leq \chi \leq 0.95$, the main energy sink is due to radiative losses caused by $\Dt$ and $\Dtp$ dissociation, together with charge-exchange reactions, which dominates closer to the target. On the other hand, along the inner divertor leg, for $\chi < 0.08$, we observe that charge-exchange reactions are the main loss mechanism in a wider region than for the outer leg.



\begin{figure*}
    \centering
    \begin{subfigure}[b]{0.45\textwidth}
        \includegraphics[width=\textwidth]{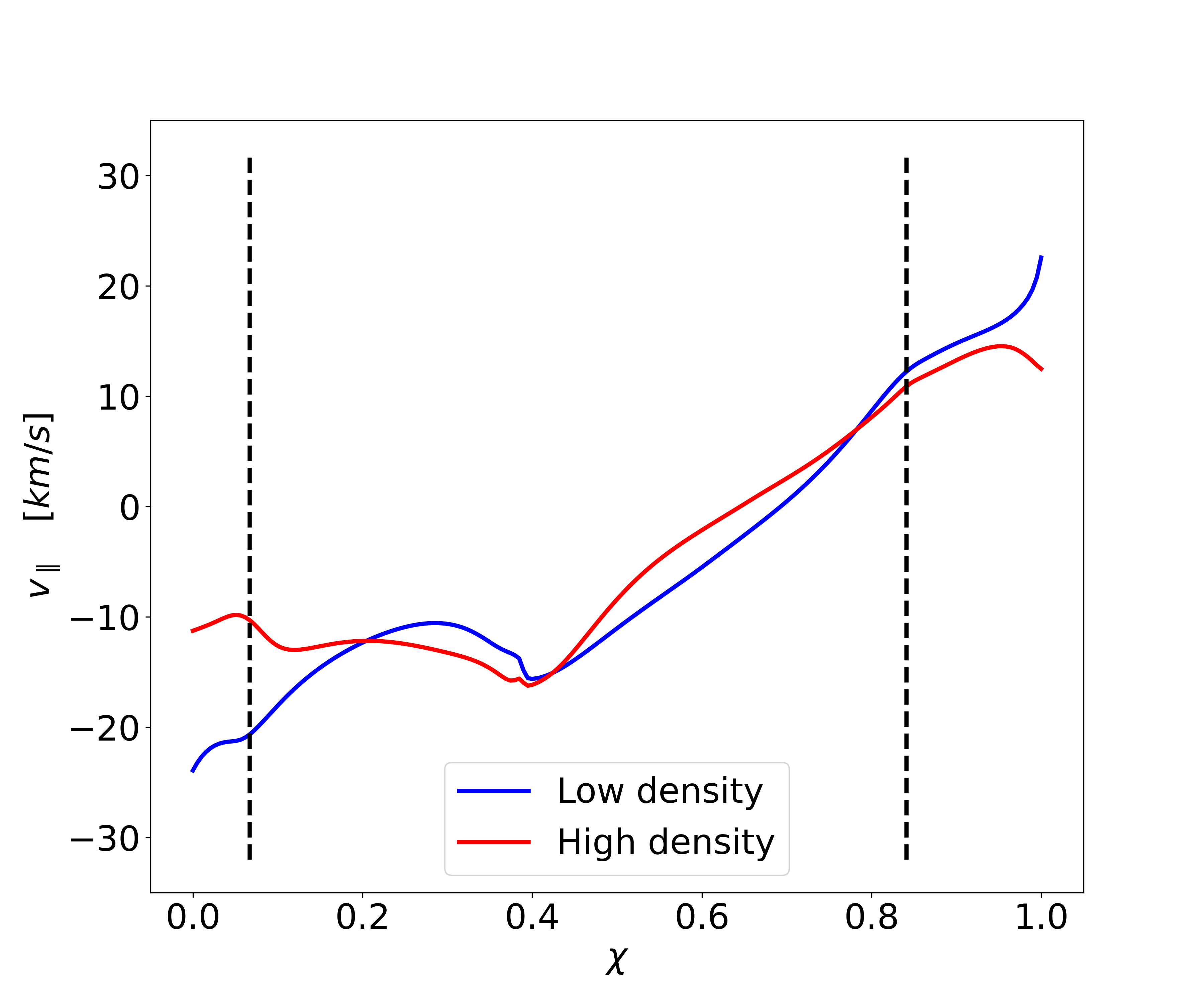}
    \end{subfigure}
    \begin{subfigure}[b]{0.49\textwidth}
        \includegraphics[width=\textwidth]{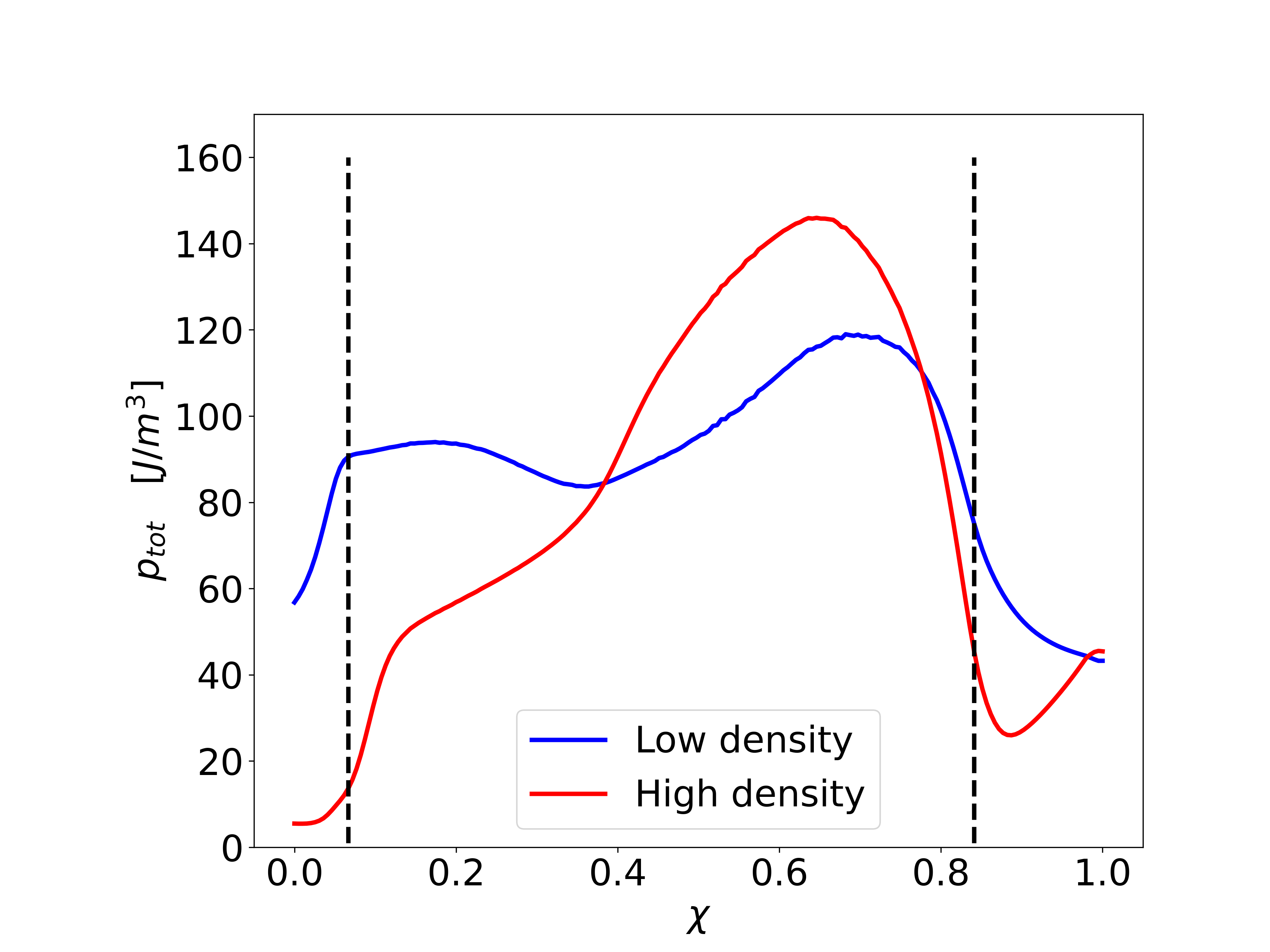}
    \end{subfigure}
    \caption{Averaged ion parallel velocity (left) and total plasma pressure (right) along a flux tube close to the separatrix $1 \leq \rho_{\psi} \leq 1.1$, as a function of the poloidal coordinate $\chi$. The black dashed lines denote the X-point coordinates.}
    \label{fig:ptot_sol}
\end{figure*}

We now turn to the pressure drop appearing in our simulations. Figure \ref{fig:ptot_sol} shows the time- and toroidally-averaged total pressure and $\Dp$ ion parallel velocity, along the flux tube as in Fig. \ref{fig:Sptot_sol}. We evaluate the total pressure as
\begin{equation}
    \label{eq:ptot}
    p_{\text{tot}} = n_{\el} T_{\el} + n_{\Dp} \left( T_{\Dp} + m_{\Dp} v_{\|,\Dp}^2 \right)
\end{equation}
since the $n_{\Dtp}$ as well as the electron dynamic pressure contributions are negligible.
In both simulations the $\Dp$ fluid presents a stagnation point close to the OMP, $\chi = 0.7$, and the module of the velocity increases toward the two targets, as observed in previous simulations \cite{Mancini_2021}. However, the high-density simulation presents lower velocity at both divertor targets, as expected from the large number of charge-exchange reactions and the low temperature (see Sec \ref{subsec:temperatures}). Both pressure profiles present a maximum around the OMP, where the density and temperature are higher. In the high-density simulation, the total pressure drop is larger than in the low-density case. Comparing Fig. \ref{fig:Sptot_sol} with Fig. \ref{fig:ptot_sol}, it is possible to observe that the power loss peaks at $\chi = 0.18$ and $\chi = 0.80$ (see Fig. \ref{fig:Sptot_sol}), where a strong pressure drop occurs, indicating that plasma-neutral interactions described above play a crucial role in determining the pressure profile in our simulations.

\subsection{Plasma and neutral temperature}
\label{subsec:temperatures}
In addition to significant power losses, detachment scenarios are characterized by low temperature, in particular in the divertor region \cite{Stangeby_2000}, resulting from a significant rate of neutral-plasma reactions. Indeed, the relative importance of the atomic reactions is mainly determined by the plasma temperature profile \cite{Coroado_2022_1}.

\begin{figure*}
    \centering
    \begin{subfigure}[b]{0.49\textwidth}
        \includegraphics[width=\textwidth]{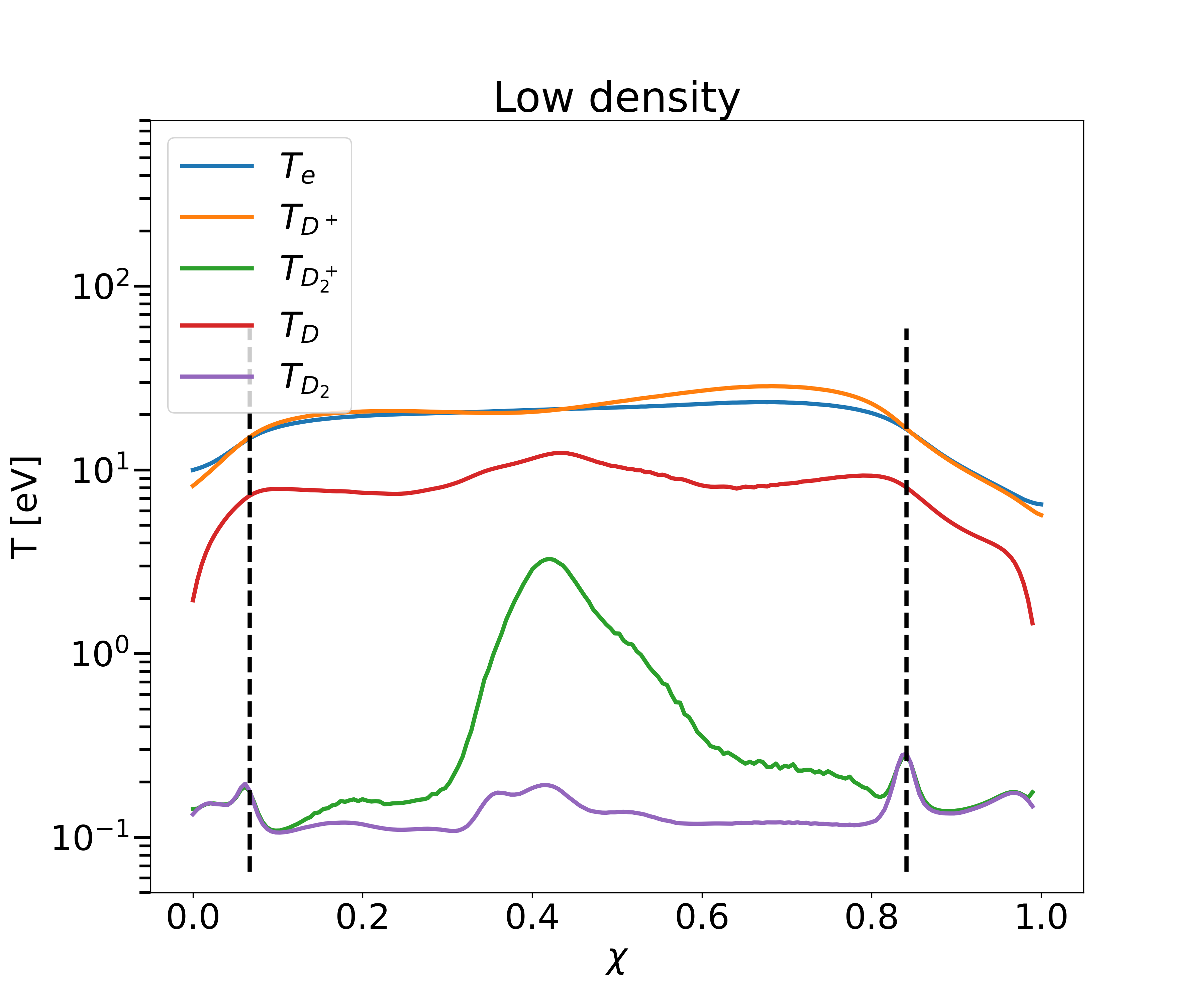}
    \end{subfigure}
    \begin{subfigure}[b]{0.49\textwidth}
        \includegraphics[width=\textwidth]{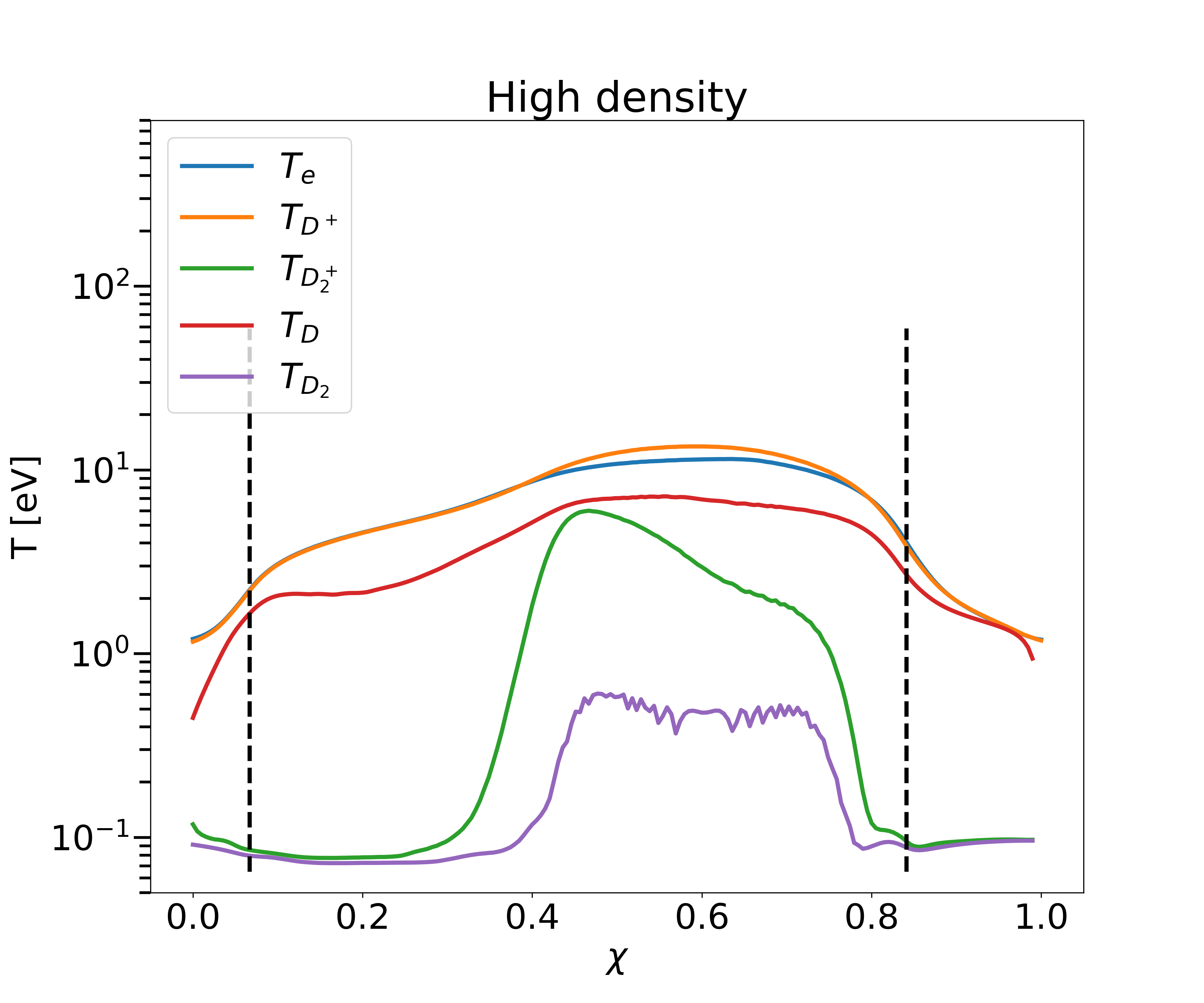}
    \end{subfigure}
    \caption{Time- and toroidally-averaged profiles of all the species temperatures for the low density (left) and high density (right) simulations, in a flux tube close to the separatrix, $1 \leq \rho_{\psi} \leq 1.1$, as a function of the poloidal coordinate $\chi$. The black dashed line denotes the coordinates of the X-point.}
    \label{fig:temp_sol}
\end{figure*}

The temperature of all species present in our simulations are shown in Fig. \ref{fig:temp_sol}. 
In both simulations, the temperature of the molecular species are lower than the atomic species, while the $\Dp$ and electron temperatures are very similar.
At low density, the plasma temperature decreases because of the ionization processes occurring close to the target, causing a steep temperature gradient. Since the temperature remains above $3$~eV, recombination and dissociation reactions are negligible and neutrals are emitted mainly from the wall in this simulation. A fraction $\alpha_{\text{refl}} = 0.2$ are emitted at the incoming ion temperature and the remaining are released at the wall temperature $T_{\text{wall}} = 0.03$~eV, explaining the value of $T_{\D}$. On the other hand, in the high-density simulation, the temperature of the charged species is sufficiently high ($T_{\el} > 3$~eV) only above the X-point for neutrals ionization to occur, while this is not the case closer to the target, a condition that is denoted as power starvation \cite{Pshenov_2017}. In turn, neutrals are produced in the divertor volume through dissociation and recombination processes, since the temperature is lower than $3$~eV (see Table \ref{tab:products}), and not only at the wall, as in the low-density simulation. Due to the asymmetries in the density profiles of the molecules, ultimately determined by the gas puff position, the temperature at the target of the molecular species is asymmetrical between the ISP and the OSP. The plasma energy losses due to charge-exchange are dominant at the targets (see Fig. \ref{fig:Sptot_sol}), lowering the plasma temperature. In particular the presence of the $\Dt$ puff at the outer target leads to higher $\D$ density, resulting in increased charge-exchange processes and in $T_{\Dp} \simeq T_{\D}$.

\subsection{Plasma potential and Ohm's law}
\label{subsec:electric_fields}
Strong electric fields in the divertor volume of high-density discharges are predicted and observed experimentally \cite{Rozhansky_2020, Wensing_2020, Wensing_2021}, leading to $E \times B$ flows in the poloidal plane \cite{Stangeby_1996, Chankin_2001}
Our simulations confirm the presence of these flows.

\begin{figure*}
    \centering
    \begin{subfigure}[b]{0.4\textwidth}
        \includegraphics[width=\textwidth]{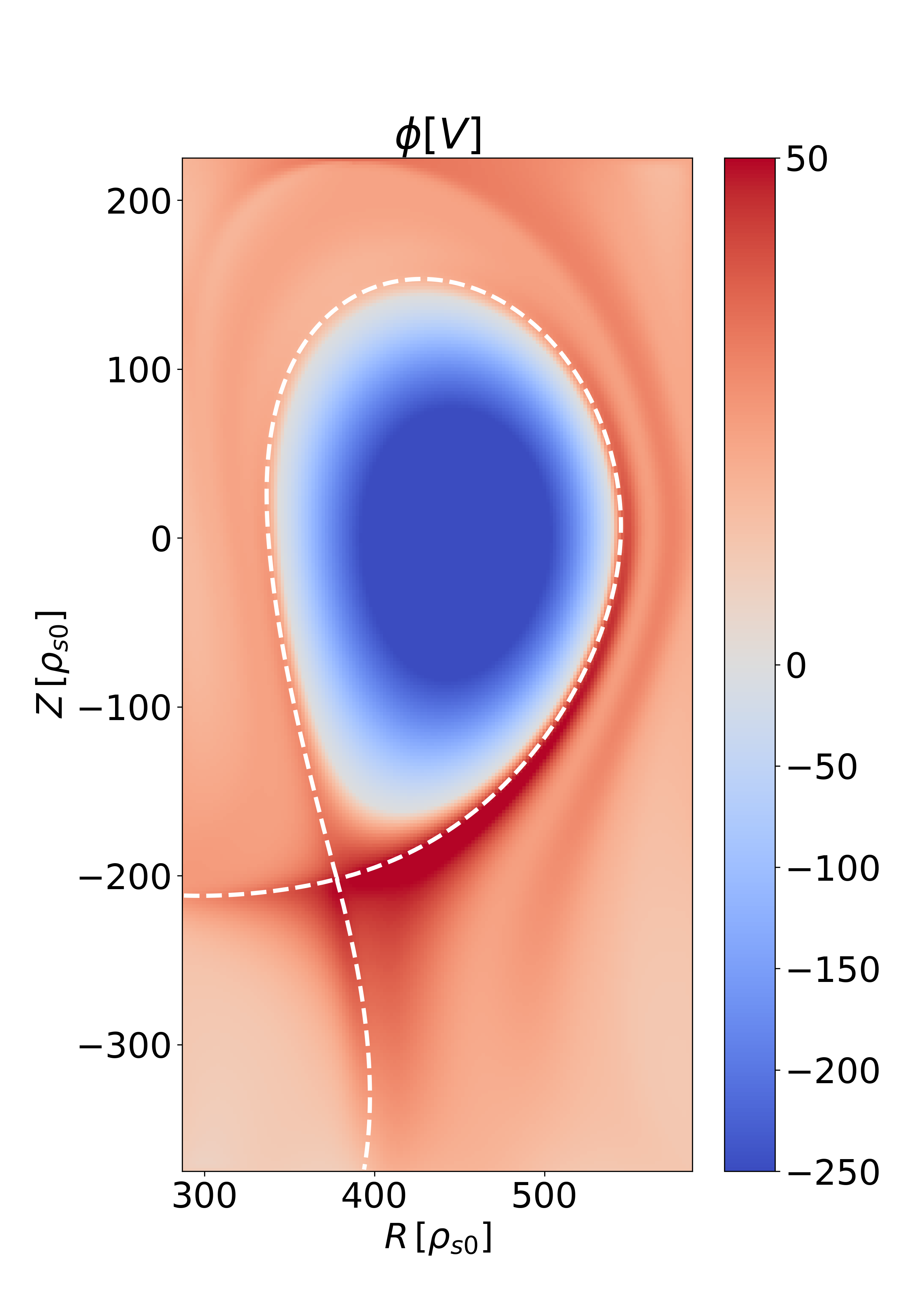}
    \end{subfigure}
    \hfill
    \begin{subfigure}[b]{0.4\textwidth}
        \includegraphics[width=\textwidth]{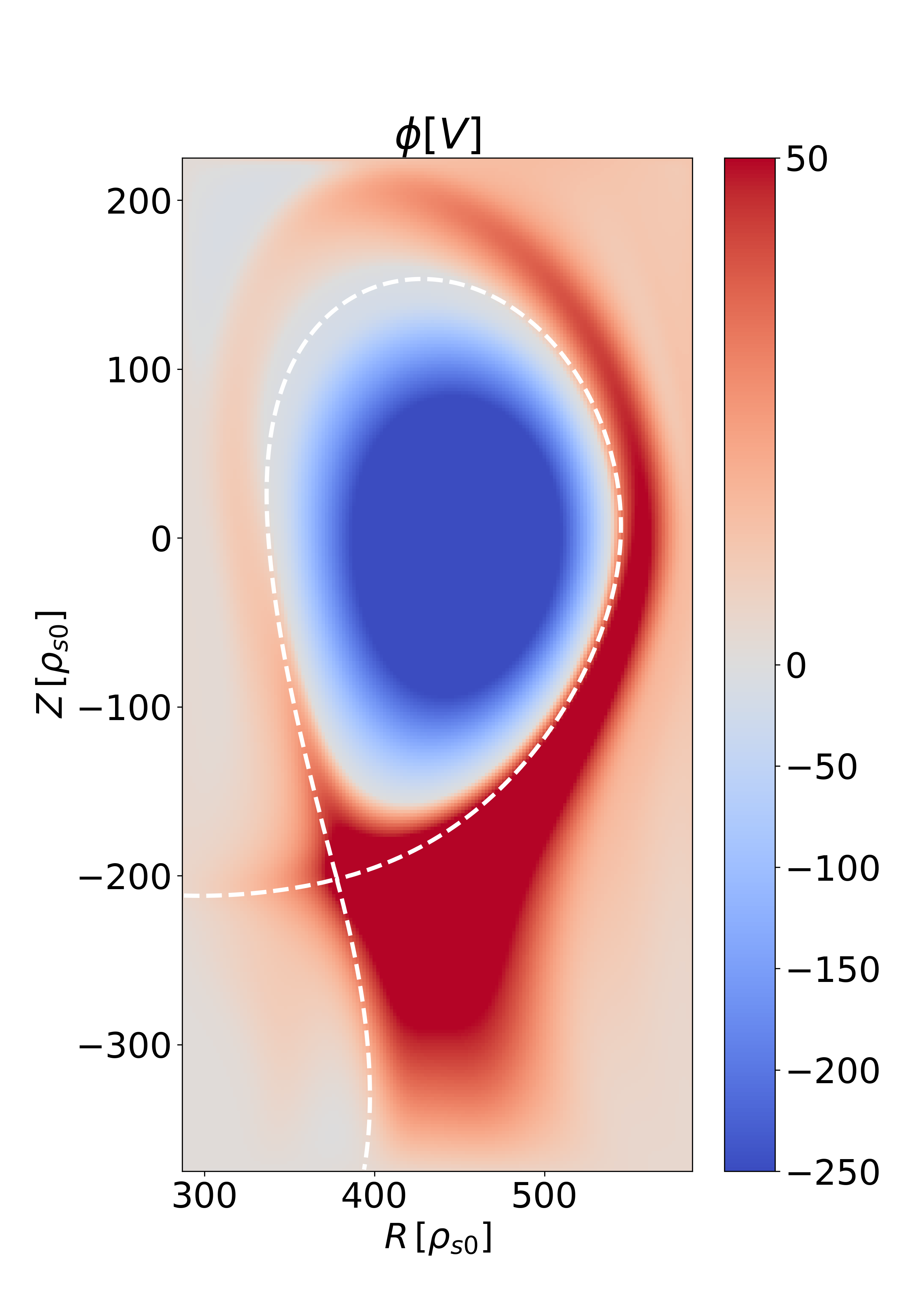}
    \end{subfigure}
    \caption{Time- and toroidally-averaged plasma potential in the low-density (left) and high-density (right) simulation.}
    \label{fig:potential}
\end{figure*}

\begin{figure*}
    \centering
    \begin{subfigure}[b]{0.5\textwidth}
        \includegraphics[width=\textwidth]{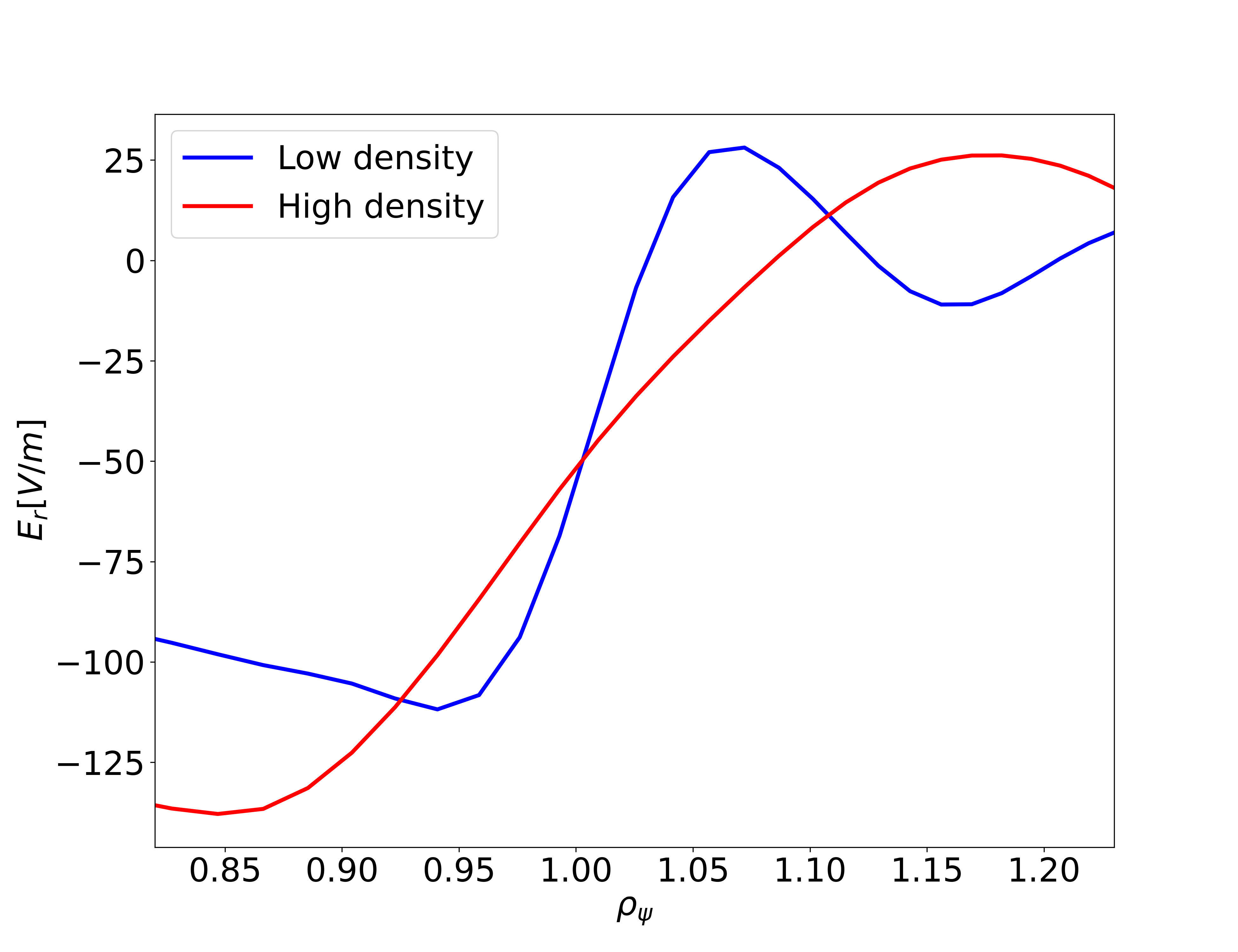}
    \end{subfigure}
    \caption{Time- and toroidally-averaged electric field at OMP.}
    \label{fig:Eromp}
\end{figure*}

The time- and toroidally-averaged electrostatic potential obtained from both simulations is shown in Fig. \ref{fig:potential}.
In both cases the potential has positive values in the SOL, higher at the LFS and around the X-point. A positive value of the plasma potential at the X-point is observed in simulations and experiments with the magnetic field direction corresponding to the one of our simulations \cite{Wensing_2020}. The increase of density leads to higher $\phi$ values in the SOL region at the LFS, except close to the targets, where the potential decreases. This results in the presence of an electric field pointing toward both targets at both strike point regions in the high-density simulation. This is relevant to explain the transport mechanisms at play in the high-density simulation, as described in Sec. \ref{sec:transport}.
The profile of the radial electric field at the outer mid-plane is shown in Fig. \ref{fig:Eromp}.
In both simulations, we observe an increase of the electric field crossing the separatrix, identifying a positive peak outside the separatrix, wider and deeper in the SOL for higher density. Larger electric fields at the OMP are associated with higher effective velocity of the turbulent transport, as observed in Sec. \ref{sec:transport}.

\begin{figure*}
    \centering
    \begin{subfigure}[b]{0.47\textwidth}
        \includegraphics[width=\textwidth]{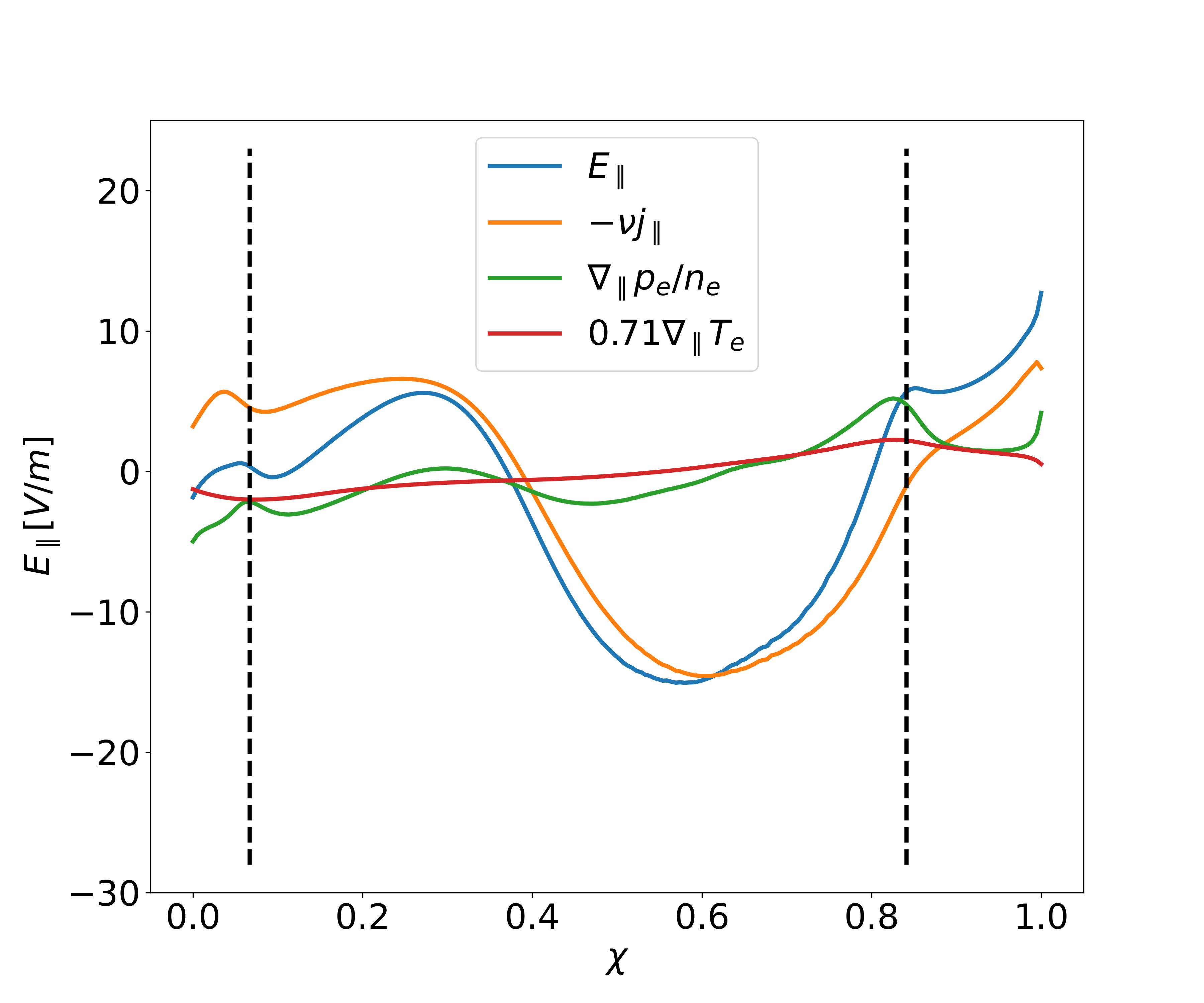}
    \end{subfigure}
    \begin{subfigure}[b]{0.47\textwidth}
        \includegraphics[width=\textwidth]{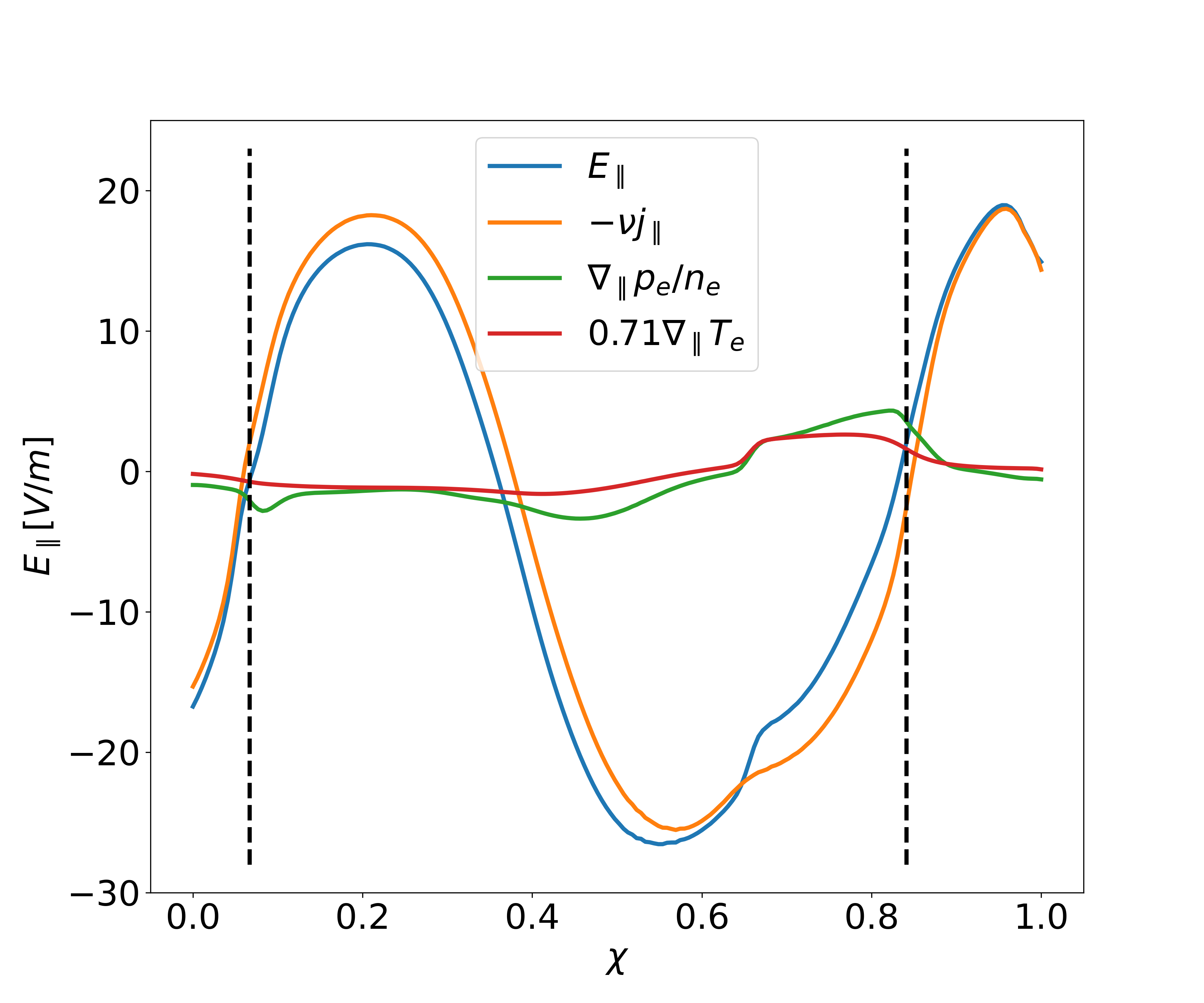}
    \end{subfigure}
    \caption{Time-, toroidally- and radially-averaged parallel electric field and its components appearing in the generalized Ohm's law Eq. \eqref{eq:Ohmlaw}, in a flux tube close to the separatrix, $1 \leq \rho_{\psi} \leq 1.1$, as a function of the poloidal coordinate, in the low-density (left) and high-density (right) simulation. The black dashed line denotes the coordinates of the X-point.}
    \label{fig:Ohmlaw}
\end{figure*}

We study the origin of the electric field by analysing the generalized Ohm's law, Eq. \eqref{eq:vpare}, which defines the relationship between parallel gradients of potential, electron temperature, pressure and parallel current, that is
\begin{equation}
    E_{\parallel} = -\nabla_{\|} \phi = \frac{\nabla_{\parallel} p_e}{n_e e} + 0.71 \frac{\nabla_{\parallel} T_e}{e} - \nu j_{\parallel} \, ,
    \label{eq:Ohmlaw}
\end{equation}
having neglected electron inertia. In Fig. \ref{fig:Ohmlaw} the time- and toroidally-averaged contributes to $E_{\|}$ appearing in Eq. \eqref{eq:Ohmlaw} are shown along the radial direction in a flux tube close to the separatrix.
In both simulations the main contribution to the parallel electric field is given by the term $\nu j_{\|}$. In the high-density case the relative importance of this term is increased, especially close to the target where the temperature is lower, being $\nu \propto T_{\el}^{-3/2}$ (see Eq. \eqref{eq:param}). Experimentally, it is observed that in discharges with relatively high temperature at the target, $T_{\el} \geq 20$~eV, the resistive term do not influence the plasma potential at the OMP \cite{Brida_2022}. Our results show that collisionality determines the potential along the flux tube when the plasma temperature is low.

\subsection{Mid-plane plasma profiles: density shoulder and turbulent transport}
\label{subsec:turbulence}
Experimental operation at high density, achieved through the increase of gas throughput, reveal the tendency to develop flatter density profiles generally associated with an increased level of turbulence, a phenomenon know as \textit{density shoulder} \cite{Vianello_2020,Tsui_2022}. Previous numerical investigation using GBS, which do not include molecular interaction terms, show an increase of turbulence level and the flattening of the pressure profile with the increase of fuelling. In those simulations, the density increase results from the increase of atomic neutrals interactions \cite{Mancini_2021}, mimicked by increasing the plasma resistivity when those interactions are not included \cite{Beadle_2020, Giacomin_2020}.

\begin{figure*}
    \centering
    \begin{subfigure}[b]{0.49\textwidth}
        \includegraphics[width=\textwidth]{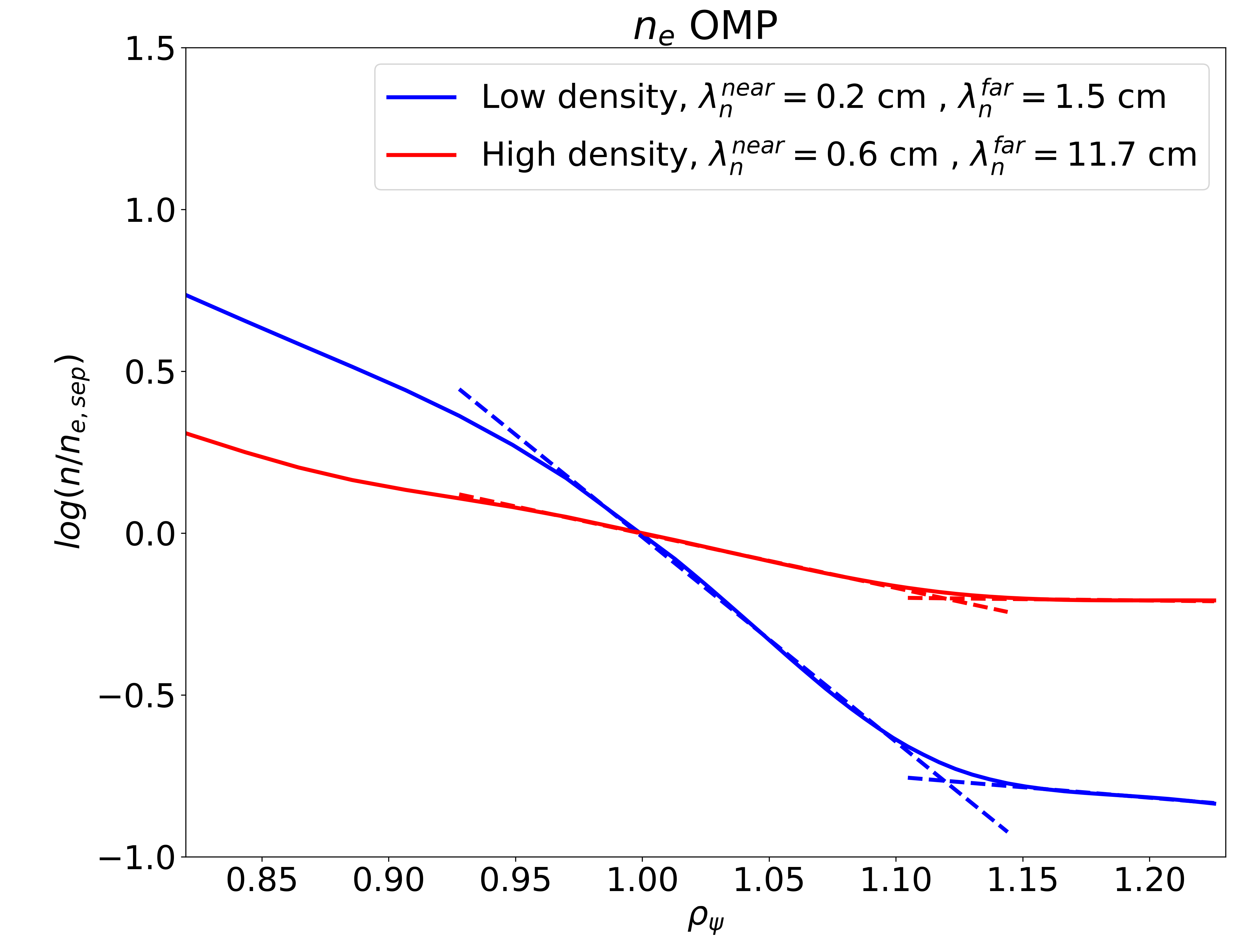}
    \end{subfigure}
    \begin{subfigure}[b]{0.49\textwidth}
        \includegraphics[width=\textwidth]{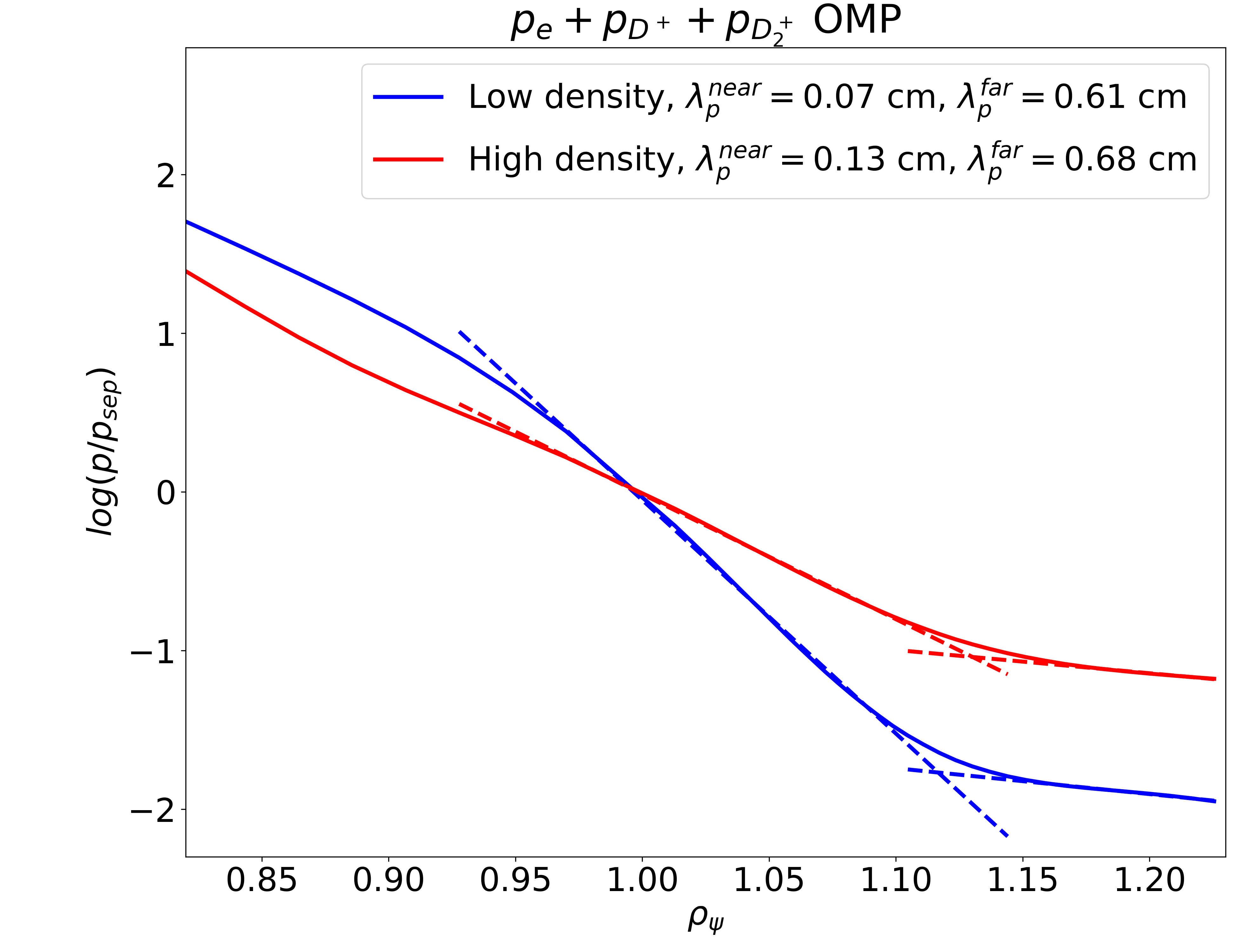}
    \end{subfigure}
    \caption{Time- and toroidally-averaged electron density and total plasma pressure at the OMP. The dashed lines show the linear fits that identify the near SOL the far SOL decay lengths.}
    \label{fig:omp_prof}
\end{figure*}

We investigate the density shoulder formation in the present simulations. Density and pressure profiles for the two simulations considered in this work are shown in Fig. \ref{fig:omp_prof}, normalized to their value at the separatrix.
We identify two decay lengths, one in the near SOL, $1 \leq \rho_{\psi} \leq 1.1 $, and one in the far SOL, in agreement with several experiments \cite{Carralero_2017, Vianello_2020} and simulations \cite{Giacomin_2020}. The increase of near SOL decay length is observed also in the temperature profiles.
The increase of the density yields an increase of the near SOL decay lengths of the density and pressure. In agreement with the simulations that include only atomic contribution \cite{Mancini_2021}, also in the present simulations, the density shoulder appears in combination with a strong reduction of temperature and parallel velocity at the target. The observation of the density shoulder formation with increasing density is in agreement with experimental results that show easier access to it with a open divertor configuration \cite{Tsui_2022}, such as the one used in our simulations.

\begin{figure*}
    \centering
    \begin{subfigure}[b]{0.3\textwidth}
        \includegraphics[width=\textwidth]{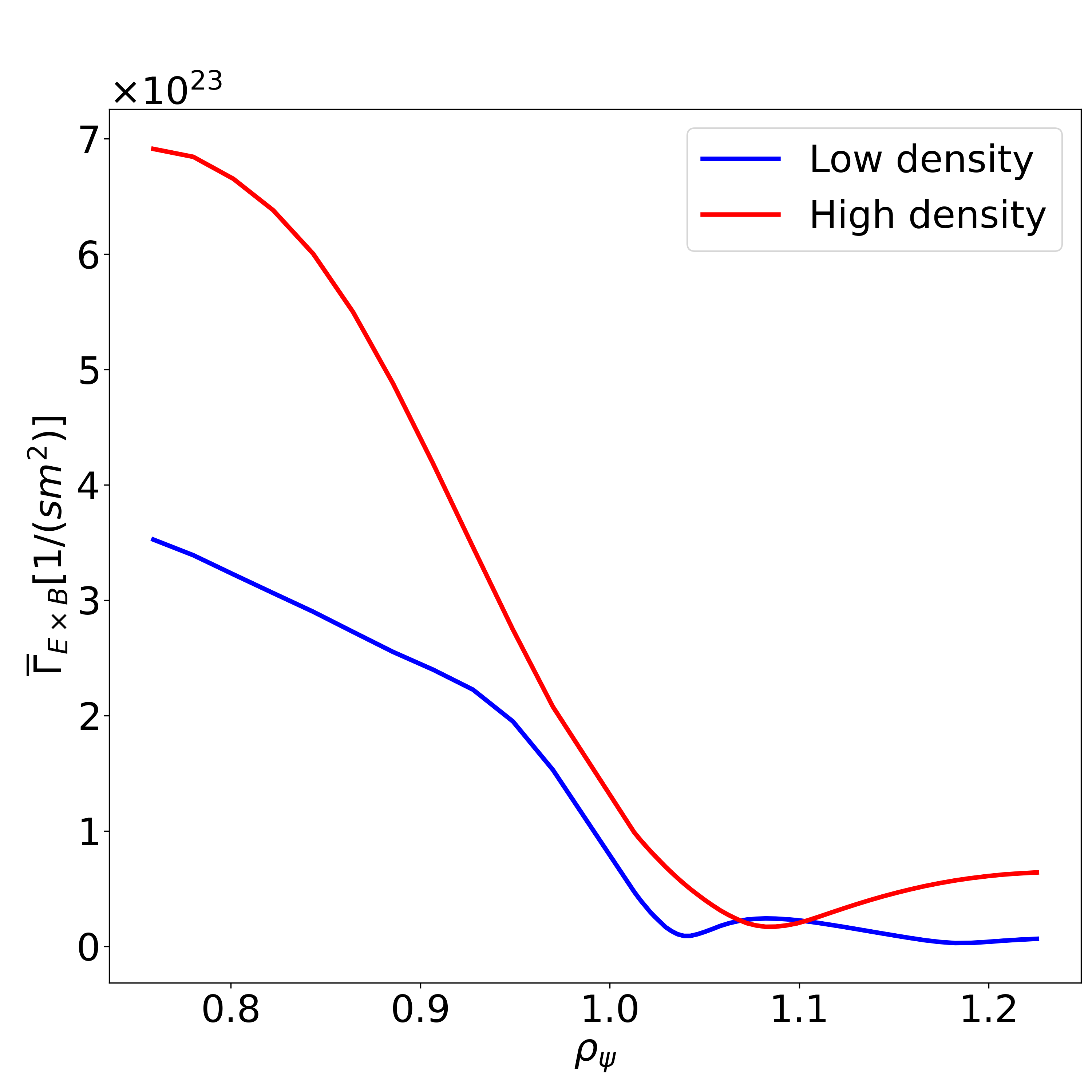}
    \end{subfigure}
    \begin{subfigure}[b]{0.325\textwidth}
        \includegraphics[width=\textwidth]{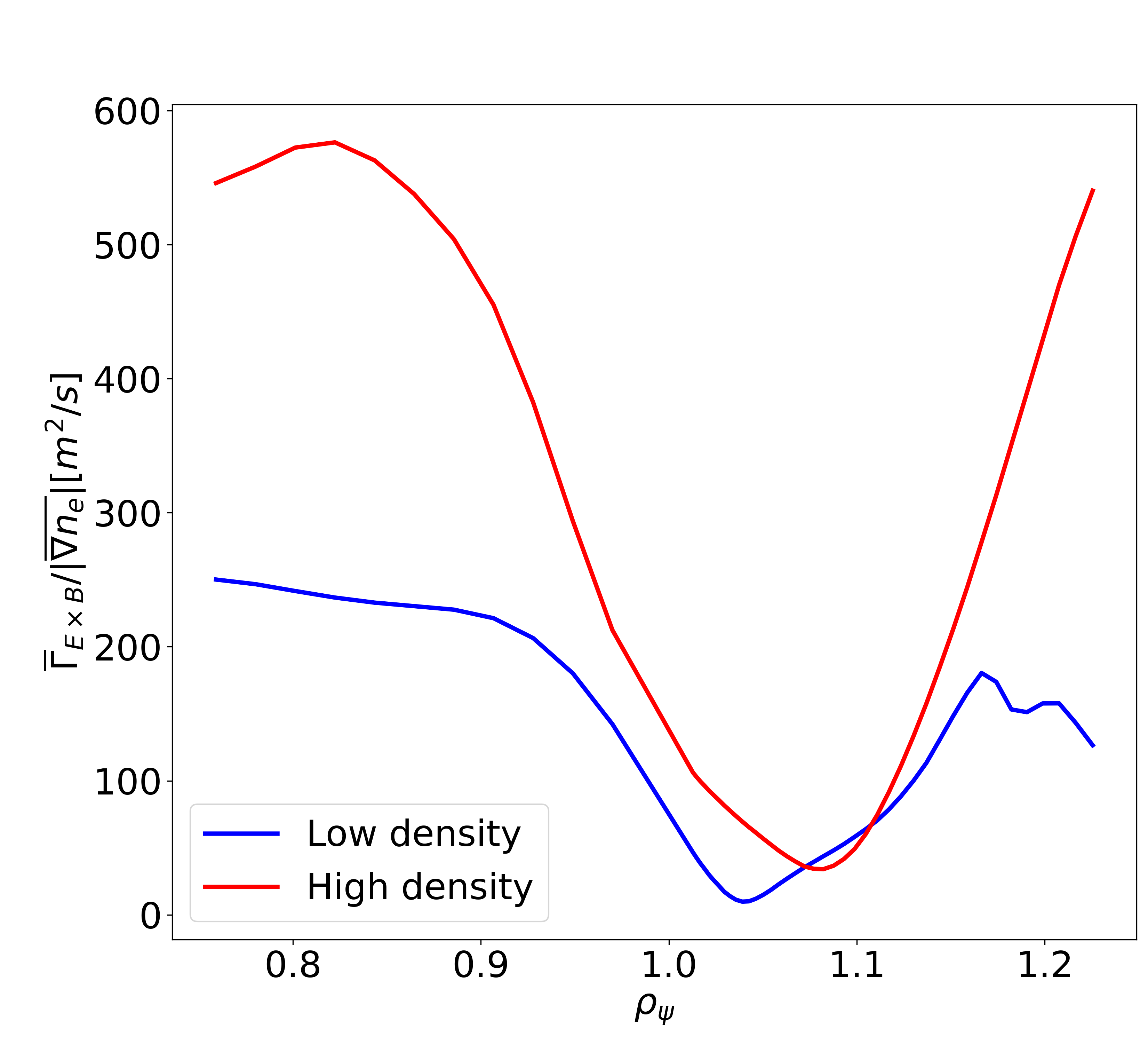}
    \end{subfigure}
    \begin{subfigure}[b]{0.3\textwidth}
        \includegraphics[width=\textwidth]{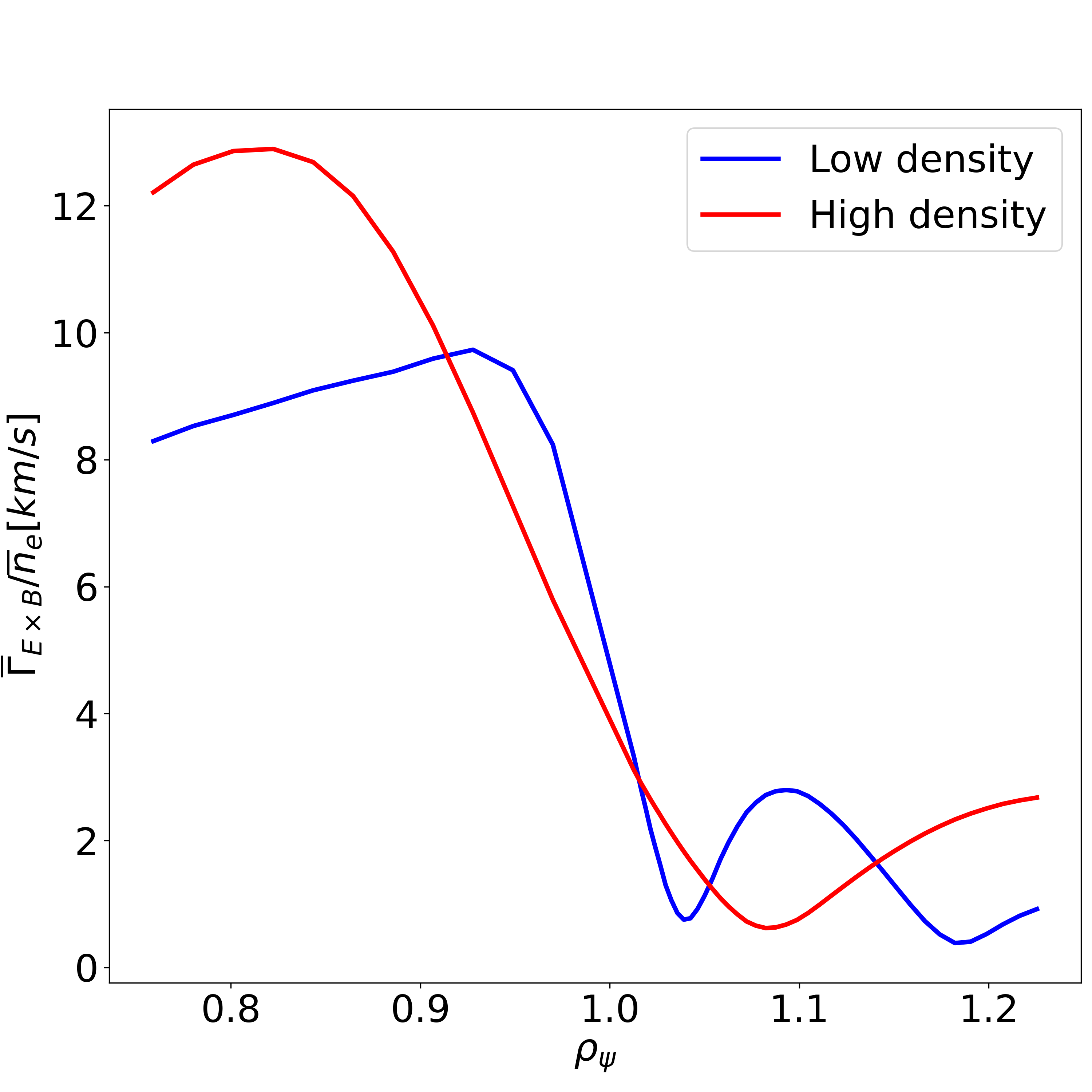}
    \end{subfigure}
    \caption{Time- and toroidally-averaged OMP profiles of $\overline{\Gamma}_{E \times B}$, of the effective coefficient due to $\Gamma_{E \times B}$, $D_{E \times B, \text{eff}} = \overline{\Gamma}_{E \times B}/|\overline{\nabla n}_e|$ and of the effective velocity $v_{E \times B, \text{eff}} = \overline{\Gamma}_{E \times B}/\overline{n}_e $.}
    \label{fig:turb_flux}
\end{figure*}

In order to estimate the perpendicular transport, we consider in Fig. \ref{fig:turb_flux} the radial profiles at the OMP of the averaged $E \times B$ flux, $\overline{\Gamma}_{E \times B}$, of the effective transport coefficient, $D_{E \times B, \text{eff}} = \overline{\Gamma}_{E \times B}/|\overline{\nabla n}_e|$, and of the effective velocity, $v_{E \times B, \text{eff}} = \overline{\Gamma}_{E \times B}/\overline{n}_e$. These include both the time- and toroidally-averaged steady state and the fluctuating flux components of the $E \times B$ flux, $\overline{\Gamma}_{E \times B} = \Gamma_{\overline{E} \times \overline{B}} + \Gamma_{\overline{\tilde{E} \times \tilde{B}}}$, with $\Gamma_{\overline{E} \times \overline{B}} = \overline{n}_{\el} (\overline{E} \times \overline{B})/B^2$ and $\Gamma_{\overline{\tilde{E} \times \tilde{B}}} = \overline{\tilde{n}_{\el} (\tilde{E} \times \tilde{B})}/B^2$. We note that the $E \times B$ flux is significantly larger than the diamagnetic flux, neglected in the present analysis.

Focusing on the SOL, the fluctuating component accounts for half the total flux in the low-density simulation, while its relative importance increases at high density, up to $70 \%$ of the total flux. 
In the SOL of the high-density simulation, not only $\Gamma_{E \times B}$ is larger compared to the low-density case, as expected from the higher density values, but also the effective diffusion coefficient $D_{E \times B, \text{eff}}$ is larger. In addition, the effective velocity, $v_{E \times B, \text{eff}}$, is larger in the high-density simulation for $ \rho_{\psi} > 1.15$. We also note that the high effective velocity values in the SOL, $v_{E \times B, \text{eff}} \simeq 1 \text{km}/s$, in both simulations, are restricted to a narrow region that encompasses the OMP, where the strong turbulence is observed. These high velocity values in the high-density simulation are in contrast with the results of simulations that include only atomic deuterium in a simplified magnetic configuration \cite{Mancini_2021}, where smaller blob velocity with lower temperature at the OMP is observed.


Both experimental results and simulations show that the far SOL turbulent flux in L-mode tokamak discharges is mostly the result of the motion of coherent filamentary structures, denoted as blobs. In GBS simulations, blobs are identified with an algorithm developed and used for the analysis of previous GBS results \cite{Nespoli2017, Beadle_2020}, which was recently extended to detect their three-dimensional structure. The algorithm finds the regions where the density fluctuations are $2.5$ times above the local standard deviation and tracks them in time. A fluctuation is identified as a blob if it is detected over an area of, at least, $20 \rho_{s0}^2$ on a poloidal plane and it has a toroidal extension above $\pi R_0/5$. The blob detection algorithm fits the blob density perturbation in the poloidal plane with a gaussian function.
From the blob center of mass motion, identified as the center of the fitting gaussian function, we retrieve the time-average components of the blob velocity in the poloidal plane, $v_{b, \psi} (R,Z)$ and $v_{b, \chi} (R,Z)$. Our analysis covers a time interval sufficiently large to ensure the statistical convergence of the blob properties. 

\begin{figure*}
    \centering
    \begin{subfigure}[b]{0.47\textwidth}
        \includegraphics[width=\textwidth]{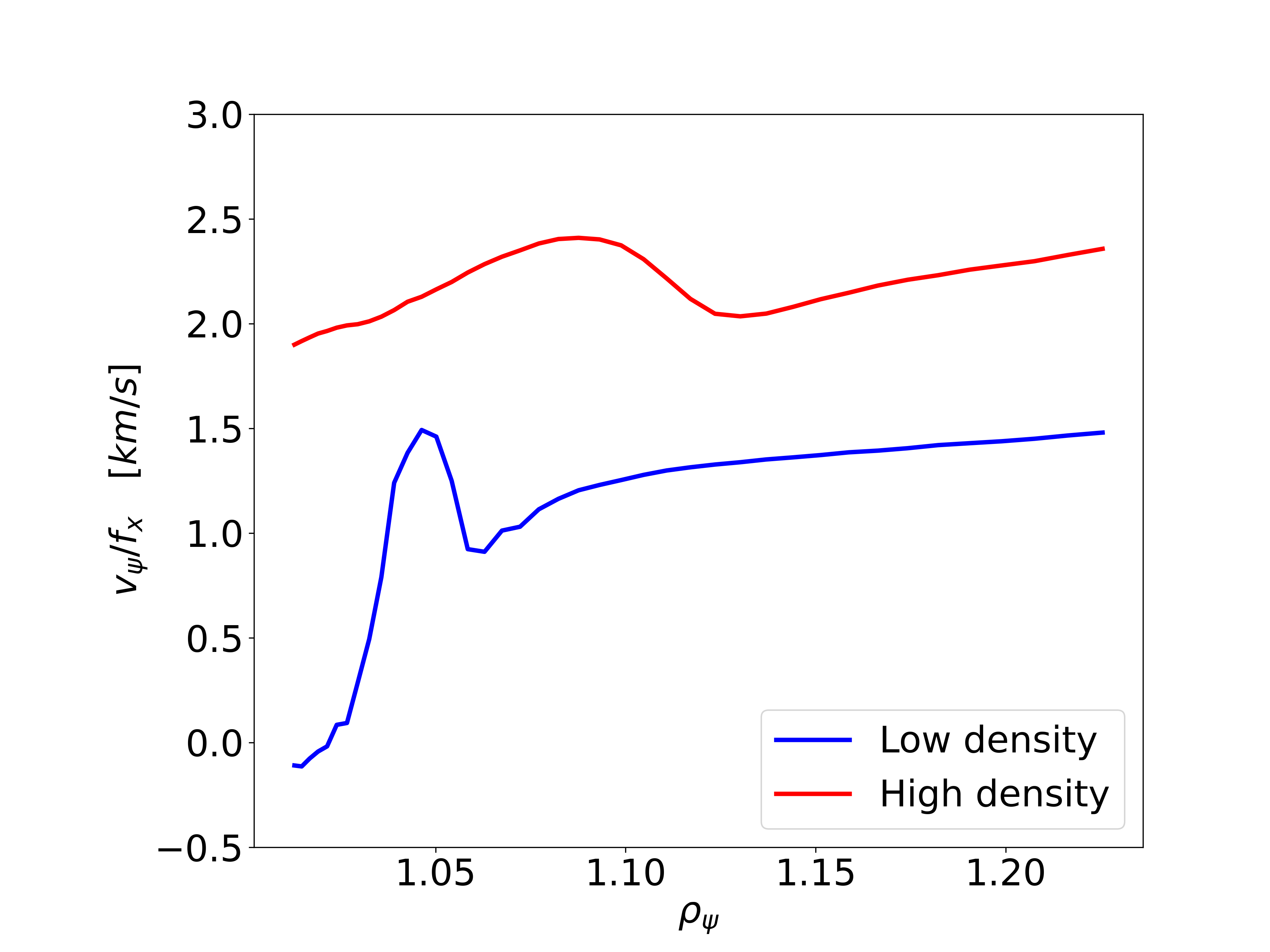}
    \end{subfigure}
    \begin{subfigure}[b]{0.47\textwidth}
        \includegraphics[width=\textwidth]{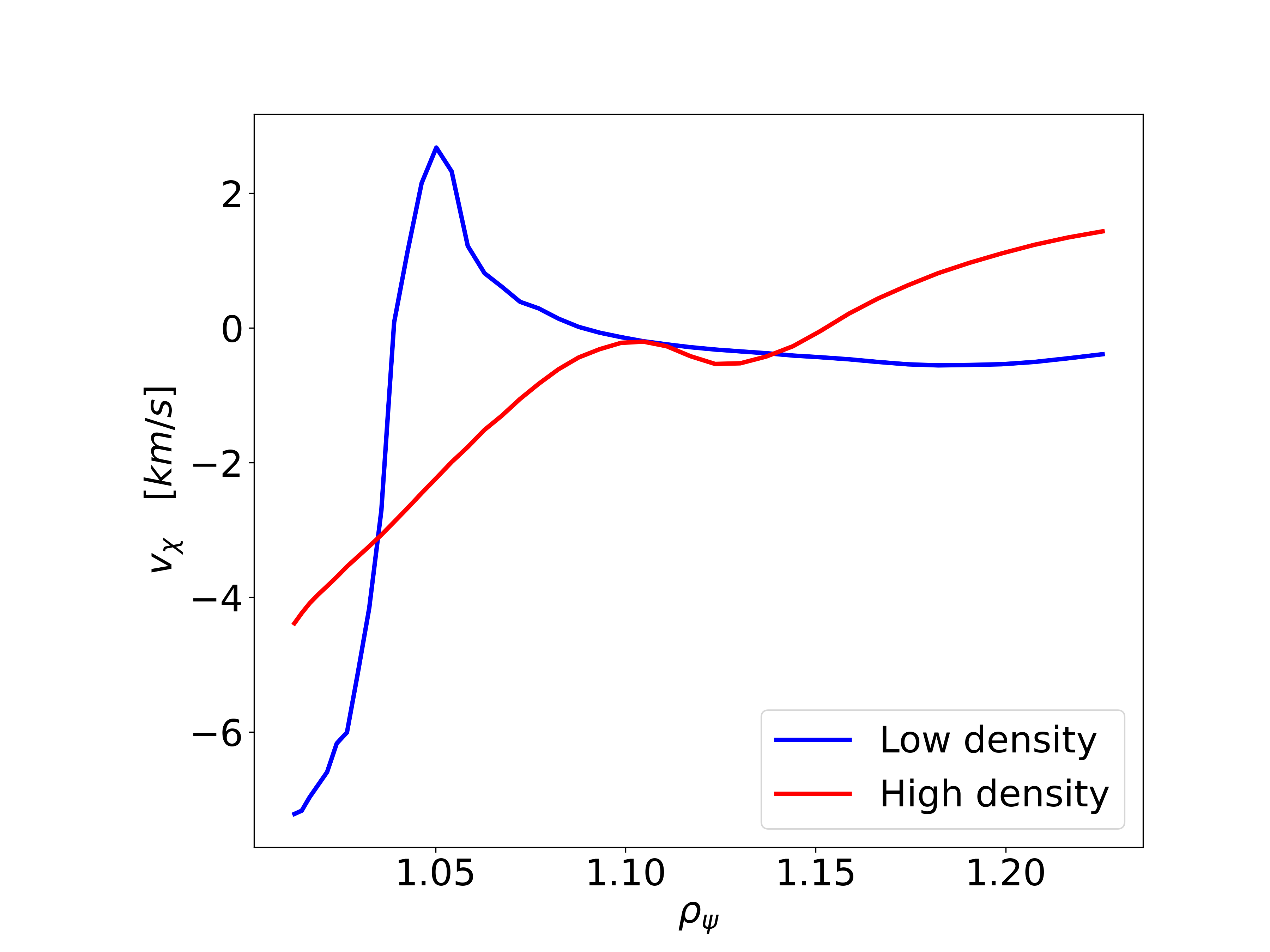}
    \end{subfigure}
    \caption{Radial, $\psi$, (left) and poloidal, $\chi$, (right) components of the blob center of mass velocity. The radial component is divided by the flux-expansion $f_x$. The positive direction of the poloidal coordinate goes from the ISP to the OSP. The velocity is the average over all blobs, at the OMP region. 
    }
    \label{fig:blobs_vel_OMP}
\end{figure*}

In agreement with Ref. \cite{Giacomin_2020} and also in agreement with experimental results \cite{Offeddu_2022}, blobs in our simulations are typically in the resistive ballooning or resistive X-point regimes, with a prevalence for the first one, where blob velocity increases with their size and with the SOL resistivity \cite{Paruta_2019, Mancini_2021}. While it is not common to observe a density shoulder with filaments in resistive ballooning regime, it has been observed in experiments that the high neutral density in the main tokamak chamber facilitates the density shoulder formation \cite{Tsui_2022}. In Fig. \ref{fig:blobs_vel_OMP} the radial and poloidal components of the blob velocity, $v_{\psi}$ and $v_{\chi}$, are plotted at the OMP, as a function of the distance from the separatrix. The velocity $v_{\psi}$ is normalised to the flux expansion $f_{x}$, since the radial velocity of a field-aligned structure is expected to be constant over a flux surface \cite{Wuthrich_2022}.
At low density, both the ratio $v_{\psi}/f_x$ and $v_{\chi}$ increase with $\rho_{\psi}$ in the near SOL and flatten in the far SOL, while in the high-density simulation the blob velocity increases through the SOL. These trends are in qualitative agreement with experimental results, showing also the same order of magnitude \cite{Vianello_2020,Offeddu_2022,Wuthrich_2022}. As expected from Refs. \cite{Wuthrich_2022, Nespoli_2020}, the magnitude of the blob radial velocity is the same as the effective $v_{E \times B,\text{eff}}$ (see Fig. \ref{fig:turb_flux}).

\begin{figure*}
    \centering
    \begin{subfigure}[b]{0.47\textwidth}
        \includegraphics[width=\textwidth]{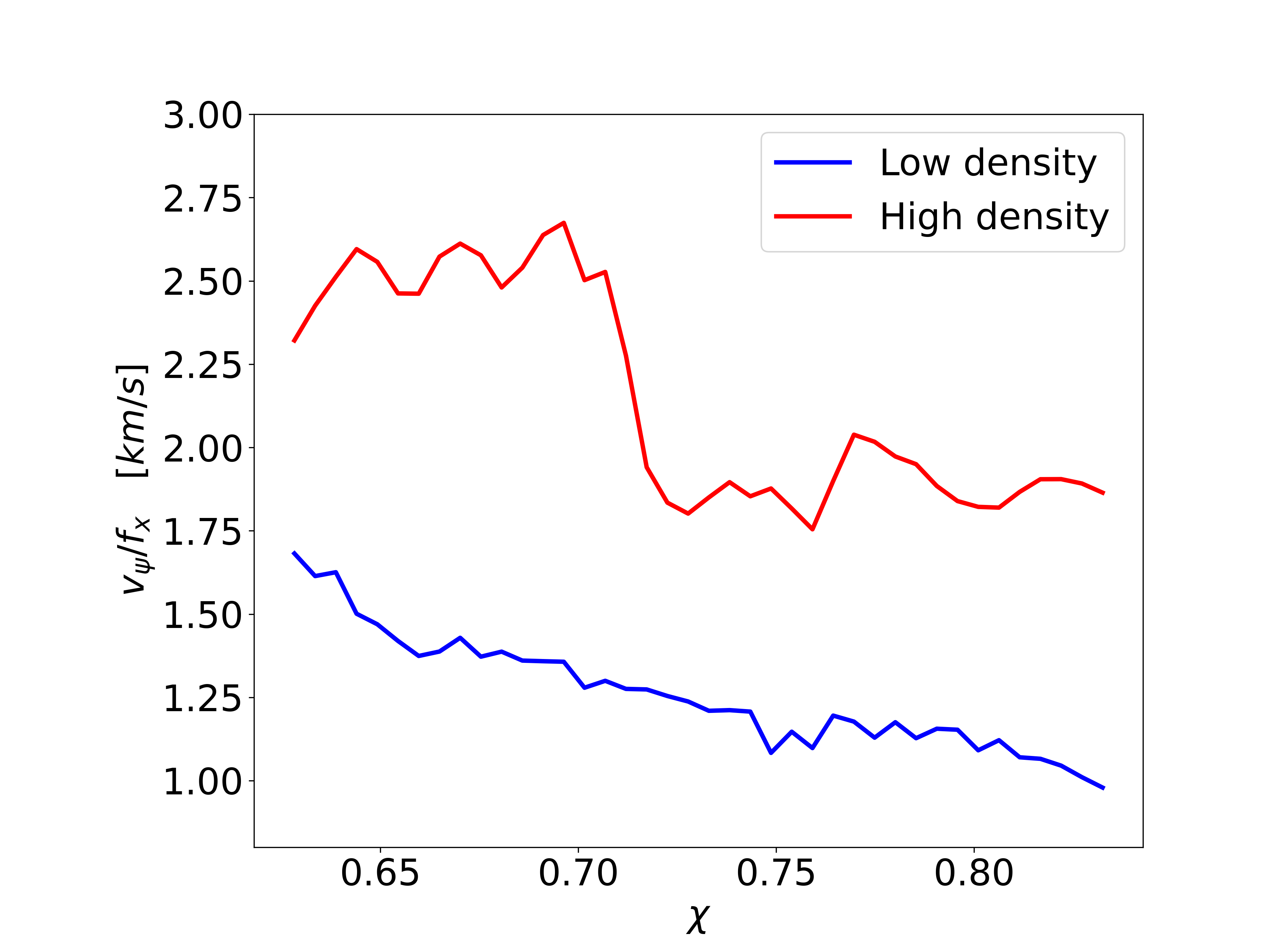}
    \end{subfigure}
    \begin{subfigure}[b]{0.47\textwidth}
        \includegraphics[width=\textwidth]{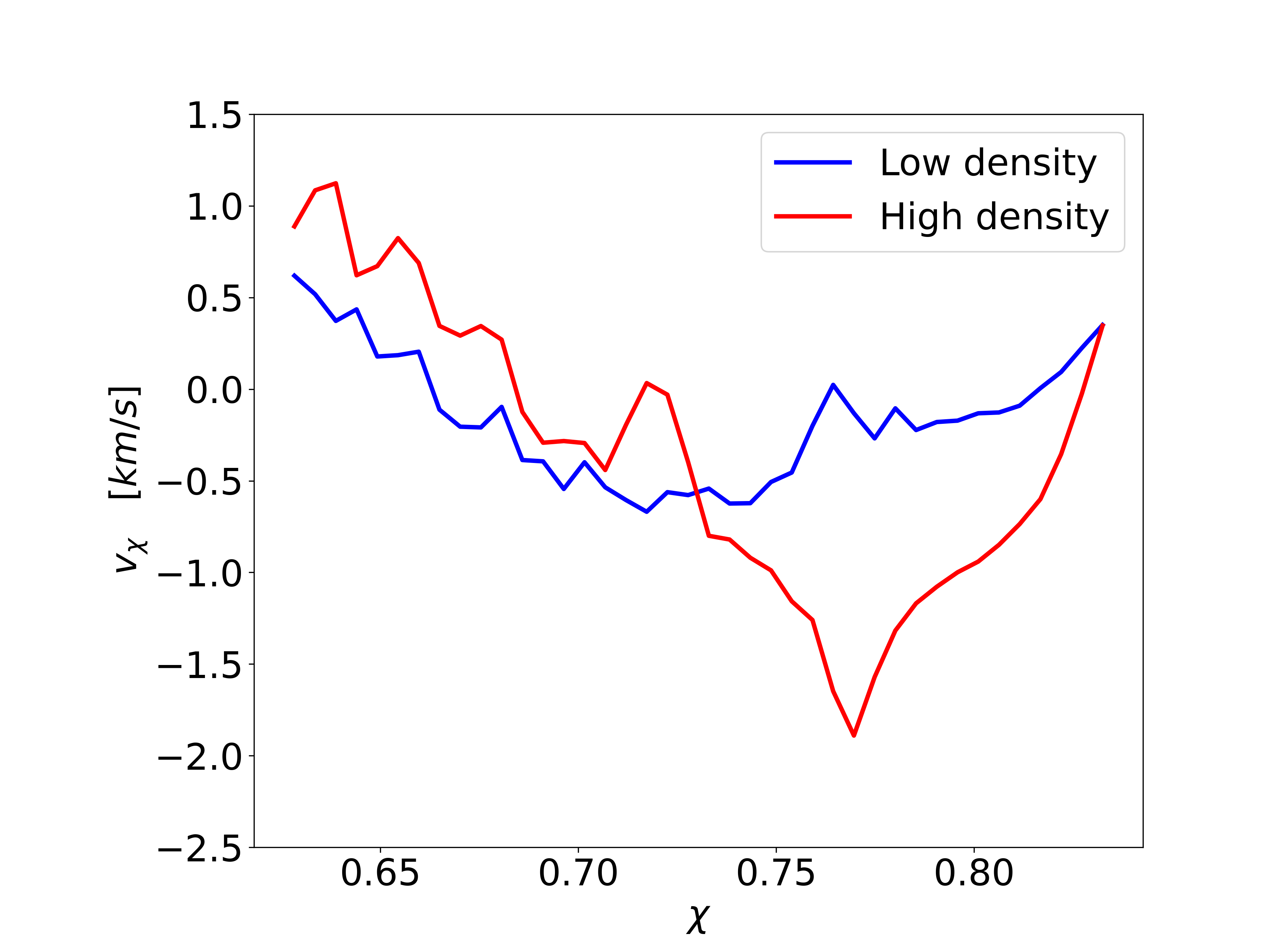}
    \end{subfigure}
    \caption{Radial, $\psi$, (left) and poloidal, $\chi$, (right) components of the blob center of mass velocity, averaged over all blobs, in a flux tube in the far SOL, $\rho_{\psi} \geq 1.1 $. The values are expressed as a function of the poloidal coordinate $\chi$, from the OMP, $\chi_{\text{OMP}} = 0.64$ to the X-point $\chi_{\text{Xpt}} = 0.87$.}
    \label{fig:blobs_vel_sol}
\end{figure*}

To compare the blob velocity dependence on $\chi$, we present the poloidal profile of the blob velocity components in Fig. \ref{fig:blobs_vel_sol}, from the OMP to the X-point, averaged over $\rho_{\psi} \geq 1.1$.
We find higher radial velocity in the entire SOL region for higher density, while poloidal velocities are the same in the two simulations up to the X-point. Both in the low- and high-density simulation, $v_{\psi}$ and $v_{\chi}$ decrease for increasing $\chi$, from the OMP to the X-point. Experimental measurements of the blobs velocities in TCV and Alcator C-mod discharges, made at different distance from the X-point, show the same trends as our simulations \cite{Agostini_2011, Wuthrich_2022}.

To conclude, our blob tracking analysis shows an increase in radial velocity with increasing fuelling, which leads to the larger turbulent $\Gamma_{E \times B}$ and to the larger effective perpendicular transport shown in Fig. \ref{fig:turb_flux}. The increased perpendicular transport, together with the decreased parallel flux, yields a larger $\Gamma_{\perp}/\Gamma_{\parallel}$ ratio, going from $\Gamma_{\perp}/\Gamma_{\parallel} \simeq 0.05$, in the low-density simulation, to $\Gamma_{\perp}/\Gamma_{\parallel} \simeq 0.1$ at higher density. The increase of this ratio is ultimately responsible for the density shoulder formation \cite{Mancini_2021, Militello_2016}.
Compared to single-ion simulations, we observe that the introduction of molecular interactions lowers the SOL plasma temperature, leading to higher resistivity and faster blobs \cite{Beadle_2020}, ultimately increasing the perpendicular transport and the ratio $\Gamma_{\perp}/\Gamma_{\parallel}$.


\section{Fluxes to the divertor targets and detachment}
\label{sec:transport}
In this section we present the analysis of the particle and heat fluxes to the divertor targets, showing that detachment conditions are achieved at the inner target of the high-density simulation. Detachment is characterized by reduced ion and heat fluxes at the divertor targets compared to attached discharges \cite{Loarte_1998}. We start by showing the ion flux profiles at the target and in the divertor volume in Sec. \ref{subsec:ionflux}, followed by the evaluation of the Degree of Detachment (DOD) in Sec. \ref{subsec:DoD} and by the analysis of the target heat flux in Sec. \ref{subsec:heatflux}.

\subsection{Ion particle flux at the target and particle balance}
\label{subsec:ionflux}
When the density is ramped up in a tokamak discharge, the divertor moves across different recycling conditions, from attached to high-recycling and then to partial and, finally, full detachment conditions \cite{Loarte_1998,Potzel_2014}. The saturation of the target ion flux identifies the onset of detachment. This occurs when, as the plasma density increases, the peak ion flux at the target no longer increases and the reduction of the ion flux integrated over the target area is observed. The onset of detachment at the ISP and the OSP in lower-single null discharges often occurs at a different level of core density \cite{Park_2018}, with the differences between the two legs depending on the toroidal magnetic field direction, pointing out to a possible role of the $E \times B$ drift \cite{Potzel_2014}. In fact, simulations of the different phases of detachment, carried out with SOLPS-ITER, show that the introduction of drifts improves the comparison with experimental measurements \cite{Wu_2021}.

\begin{figure*}
    \centering
    \begin{subfigure}[b]{0.45\textwidth}
    \includegraphics[width=\textwidth]{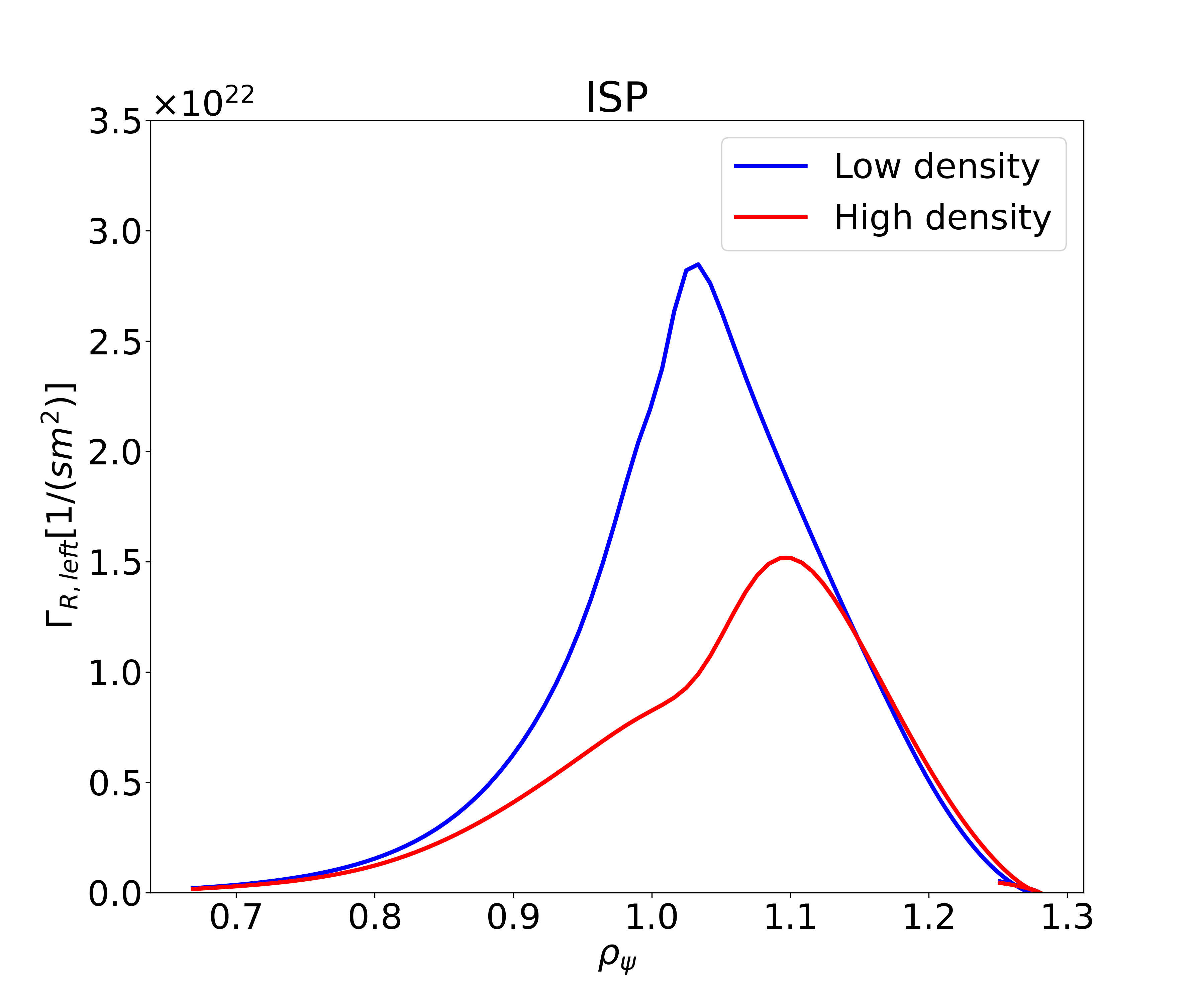}
    \end{subfigure}
    \hfill
    \begin{subfigure}[b]{0.45\textwidth}
    \includegraphics[width=\textwidth]{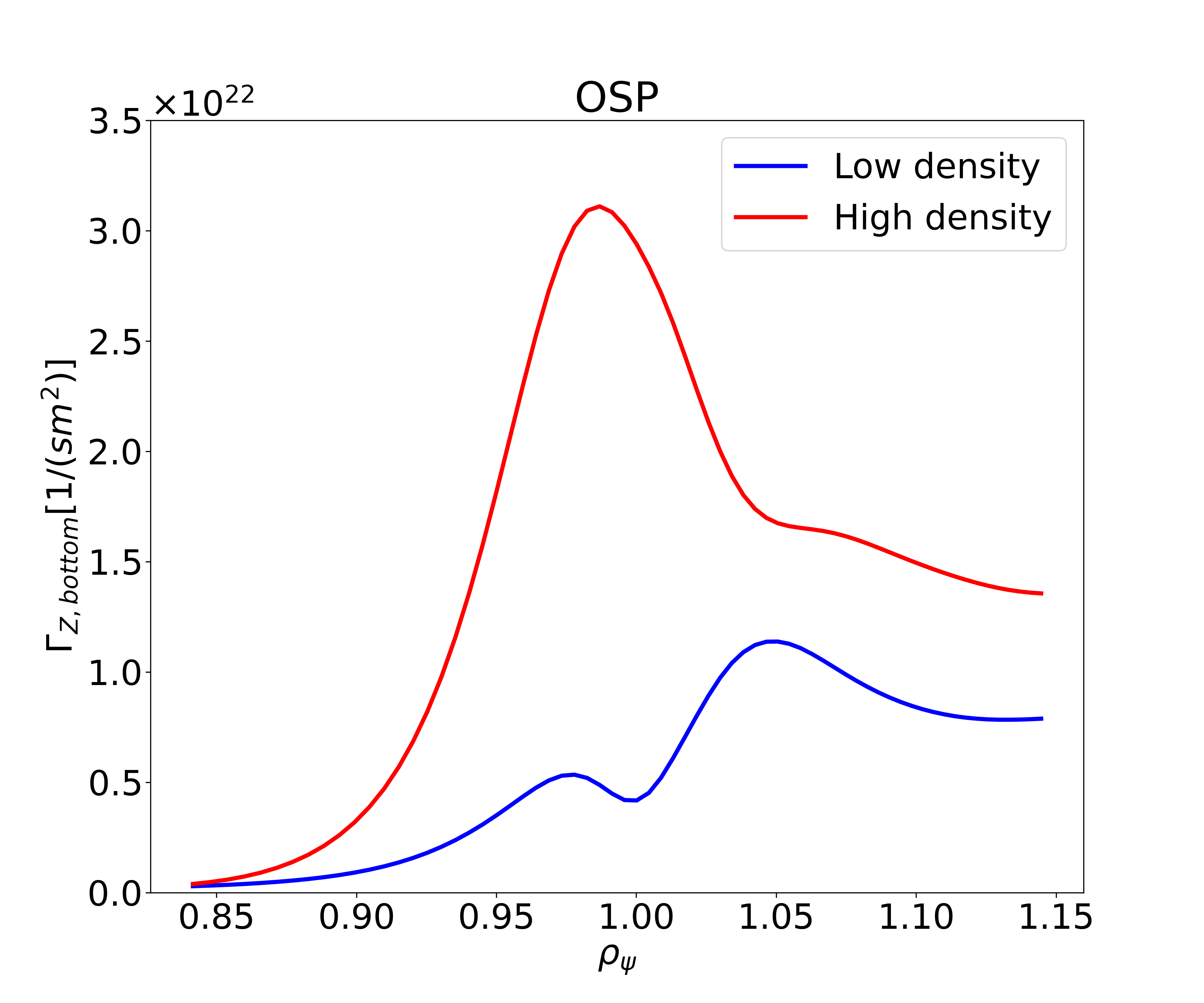}
    \end{subfigure}
    \caption{Time- and toroidally-averagd ion particle flux at the target, for both strike points, as a function of the normalized poloidal flux function.}
    \label{fig:ionflux_target}
\end{figure*}

In Fig. \ref{fig:ionflux_target}, we show the profile of the particle flux to the wall in our simulations. The region that surrounds the ISP at the left wall and, similarly, a region of the bottom wall around the OSP are considered.
The ion flux at the target is evaluated as the sum of the parallel flow and the drift motion in the direction perpendicular to the target, that is
\begin{equation}
    \Gamma_{\Dp,j} = n_{\Dp} (v_{\parallel \Dp , j}  + v_{\perp \Dp ,j}) = n_{\Dp} (v_{\parallel \Dp} b_j + v_{E \times B, \Dp ,j} + v_{di,j}) \, ,
    \label{eq:gammaD_target}
\end{equation}
where $b_j$ is the $j$ component of the unit vector along the direction of the magnetic field, with $j = R$ or $Z$, for the ISP and OSP, respectively.

Starting the analysis from the low-density simulation, we note that the larger contribution to the flux is given by the parallel flux at both targets. The ion flux at the ISP is approximately symmetric around its maximum, which is located in the SOL ($\rho_{\psi} > 1$). On the other hand, it is possible to identify two peaks of the ion flux at the OSP, one in the SOL and a second one in the private flux region ($\rho_{\psi} < 1$). The parallel flux is responsible for the peak in the SOL, while the $E \times B$ flux is dominant for $\rho_{\psi} < 1$, reducing the flow towards the OSP close to the separatrix and increasing the density in the private flux region. This is consistent with SOLPS-ITER simulations showing that the inclusion of $E \times B$ drift in the plasma dynamics can lead to the formation of a hollow profile of the ion flux to the target, as well as a higher density at the ISP than at the OSP in forward field configuration and vice-versa \cite{Wensing_2021_2,Jia_2022}.

The density increase affects the ion particle flux differently at the two targets. At the ISP, the integral of the flux decreases by, approximately, $50 \%$ with respect to the low-density simulation, and the peak of the flux is located further from the separatrix, deeper into the SOL. The decrease of the ion particle flux at the ISP is the consequence of the reduction of the parallel velocity, caused by the increase of charge-exchange reactions and the decrease of the electron and ion temperatures. This is also observed in previous single-component GBS simulations with the increase of the fuelling rate \cite{Mancini_2021}. In contrast, the ion particle flux increases with the density at the OSP, both in the SOL and in the private flux regions, showing a single peak located inside the separatrix. This is due to a lower reduction of the parallel velocity at the OSP than at the ISP, with respect to the low-density simulation. In fact, the parallel velocity at the target is set to be proportional to the electron temperature, which is lowered by an increase of the density due to the ionization reactions occuring at the LFS, as shown in Fig. \ref{fig:neutrals_density}.

\begin{figure*}
    \centering
    \begin{subfigure}[b]{\textwidth}
        \includegraphics[width=\textwidth]{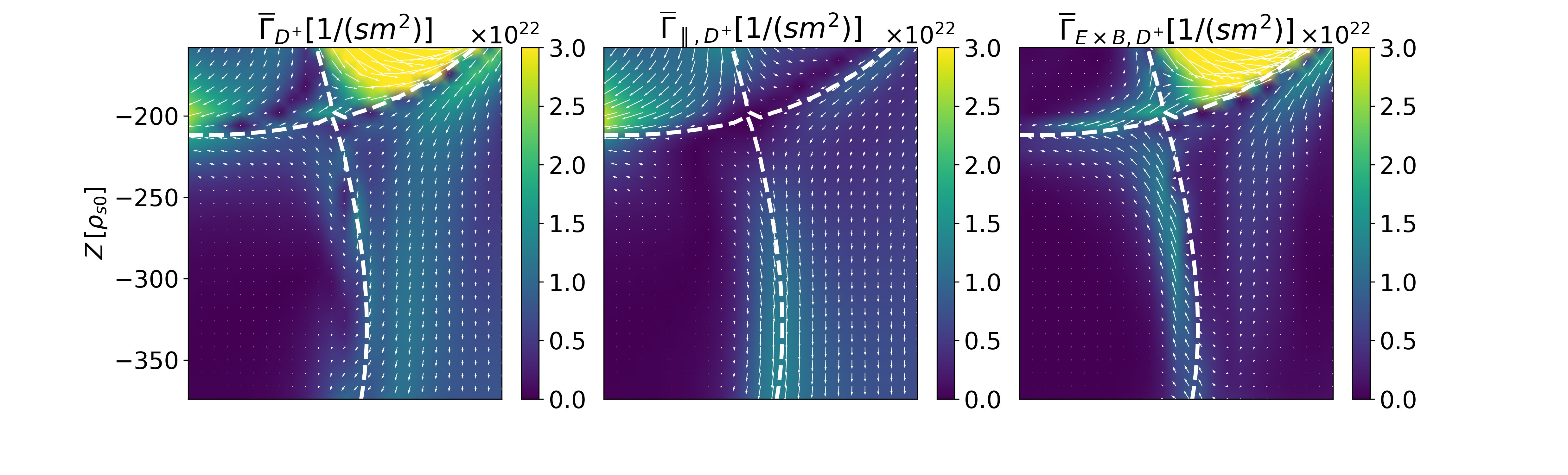}
    \end{subfigure}
    \begin{subfigure}[b]{\textwidth}
        \includegraphics[width=\textwidth]{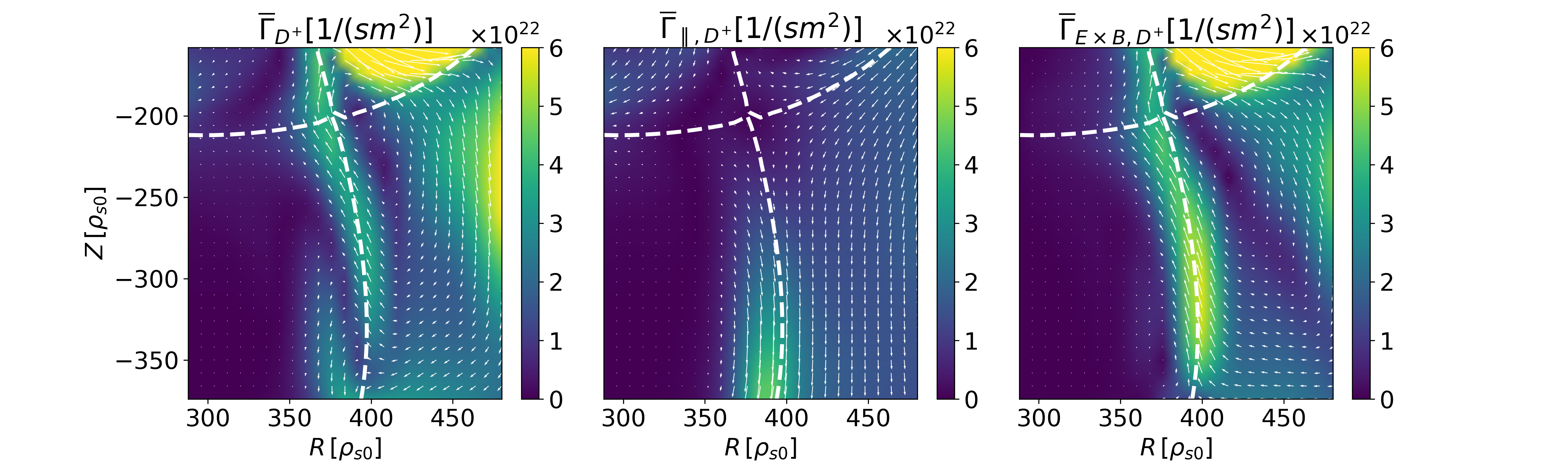}
    \end{subfigure}
    \caption{Vector plot of the time- and toroidally-averaged total ion flux (left) and its two main components, projected on the poloidal plane, in the divertor volume, for the low-density (top row) and high-density (bottom row) simulation. The colormap represents the module of the flux.}
    \label{fig:divflux}
\end{figure*}

The differences in the location of the peak of the ion particle flux between the low- and high-density simulations is explained by the different $E \times B$ drift present in the two simulations.
In Fig. \ref{fig:divflux} we present the vector plot of the main contributions to the ion particle flux on the poloidal plane in the two simulations, with the colormap representing the flux module. 
In the low-density simulation, the flux is dominated in the SOL by the parallel flow and the $E \times B$ drift is comparable to it only inside the private flux region close to the OSP. While the parallel flux peaks close to the strike point, the $E \times B$ drift transports plasma from the OSP to the ISP. In the high-density simulation, the contribution of the $E \times B$ drift increases, as expected from Fig. \ref{fig:potential}. The drift creates a convective cell of circulating plasma, transporting ions from above the X-point to the far SOL, and from the OSP to the X-point. The larger value of the $\Gamma_{E \times B, \Dp}$ flux, with respect to the parallel flux in the high-density simulation, increases the density inside the separatrix at the OSP ($\rho_{\psi} = 0.98 $), while at the ISP the $E \times B$ drift moves the peak to $\rho_{\psi} = 1.12$ (see Fig. \ref{fig:ionflux_target}).

The decrease of the particle flux at the ISP in the high-density simulation is mainly caused by the reduction of its parallel component and by the decrease of the plasma sources, while the role of the $E \times B$ drift is small. The asymmetries between the ion fluxes at the targets are, ultimately, generated by the asymmetries in the plasma-neutral interactions (see Fig. \ref{fig:Sptot_sol}) and are strengthened by the effect of the $E \times B$ drift \cite{Chankin_2001, Potzel_2014,Park_2018}, which is a consequence of the temperature profile set by molecular reactions.

\subsection{Degree of Detachment}
\label{subsec:DoD}
\begin{figure*}
    \centering
    \includegraphics[width=0.55\textwidth]{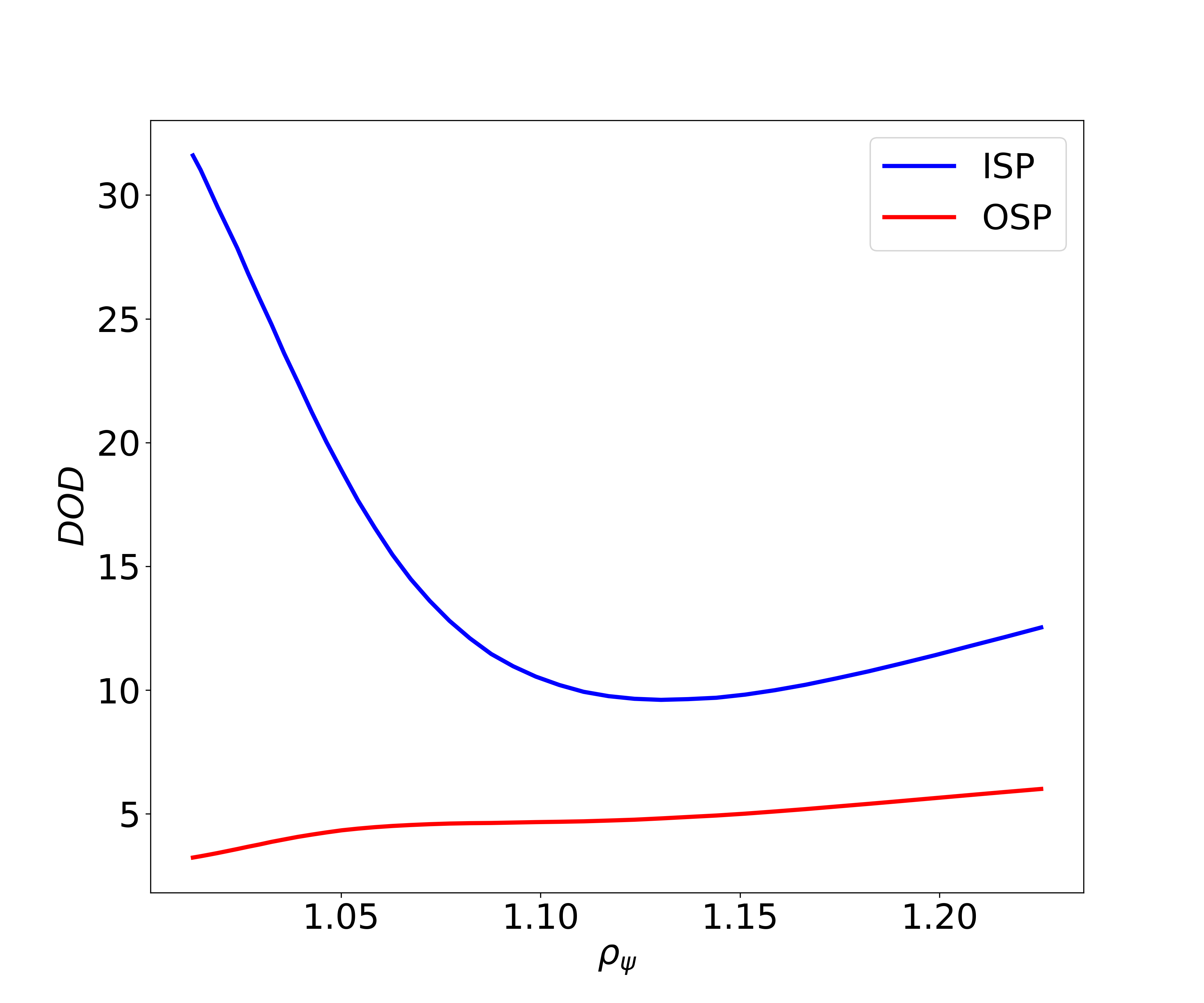}
    \caption{Degree of Detachment, $\text{DOD} = C n^2_{\Dp,\text{up}}/\Gamma_{\Dp}$, for flux tube at different locations, in the high-density simulation.}
    \label{fig:dod_high}
\end{figure*}

The two point model, derived to relate the evolution of the upstream profiles to the target density and temperature, establish a quadratic dependence of the ion flux at the target from plasma density upstream, $\Gamma_{\Dp} \simeq C n_{\text{up} \Dp}^2$ \cite{Stangeby_2000}. This model, validated in several devices (e.g. JET \cite{Loarte_1998} or ASDEX \cite{Potzel_2014}) is valid for density values below the detachment onset.
Based on the two-point model result, the divertor recycling state is often characterized by the Degree Of Detachment, defined as $\text{DOD} = C \frac{n_{\text{up} \Dp}^2}{\Gamma_{\Dp}}$ \cite{Loarte_1998}. If $\text{DOD} > 1$, the measured flux at the target is lower than expected from the two-point model and the plasma is said to be compatible with a detached scenario. While there can be other reasons to observe a weaker scaling of the particle flux at the target compared to the prediction of the two-point model, the DOD is often used as a guideline to help in the characterization of detachment conditions \cite{Potzel_2014}. We also note that the calibration constant $C$ depends on the power crossing the separatrix and the connection length of the specific flux tube.
In Fig. \ref{fig:dod_high} we show the DOD profile of the high-density simulation, at both targets, considering flux tubes at different $\rho_{\psi}$.
The two simulations we consider have the same input power in the SOL and the same magnetic geometry, therefore we evaluate $C$ from the low-density simulation and use this value to determine if the high-density simulation is in a detached state. This methodology is equivalent to considering that the simulation at lower density is not detached, an hypothesis that gives us a lower limit on the DOD value. The density upstream is evaluated as the average density at the separatrix at $Z = 0$.
At the ISP we observe $\text{DOD} > 1 $, across the entire SOL region, as expected from the observed decrease of the ion flux. At the OSP we also observe $\text{DOD} > 1$, in particular at increasing distance from the separatrix, even if the ion flux is larger than in the low-density simulation.
We can conclude that the high-density simulation presents a detached ISP with features that are compatible with the experimental conditions observed at density values larger than the ones necessary for the detachment onset. On the other hand, the OSP is in a 
partially detached state, where the particle flux reduction is not observed \cite{krasheninnikov_2017}.

\subsection{Target heat flux}
\label{subsec:heatflux}
We evaluate the sum of the $\Dp$ ion and electron heat flux considering the contribution from conduction,  convection due the parallel and drift fluxes and recombination energy of $\Dp$ ions at the target, that is
\begin{equation}
    \begin{split}
        q_{\text{tot},j} =& \frac{3}{2} T_{\Dp} \Gamma_{\Dp, j} - (\chi_{\perp, \Dp} \nabla_{\perp, j} T_{\Dp} + b_{j} \, \chi_{\|, \Dp} \nabla_{\|} T_{\Dp}) \\
            &+ \frac{3}{2} T_{\el} \Gamma_{\el, j} - (\chi_{\perp, \el} \nabla_{\perp, j} T_{\el} + b_{j} \, \chi_{\|, \el} \nabla_{\|} T_{\el}) + E_{\text{iz},\D} \Gamma_{\Dp, j}\, ,
    \end{split}
    \label{eq:q_target}
\end{equation}
where $\Gamma_{\Dp, j}$ is defined in Eq. \eqref{eq:gammaD_target} and an analogous definition is used for $\Gamma_{\el, j}$, the diffusion coefficients $\chi_{\perp}$ and $\chi_{\|}$ are those appearing in Eqs. (\ref{eq:temperature_e}-\ref{eq:temperature_D}) and $b_j$ is the component of the magnetic field unit vector, with $j = R$ or $Z$. The $\Dtp$ contribution to the flux is neglected, since $n_{\Dtp} \ll n_{\Dp}$.

\begin{figure*}
    \centering
    \begin{subfigure}[b]{0.45\textwidth}
    \includegraphics[width=\textwidth]{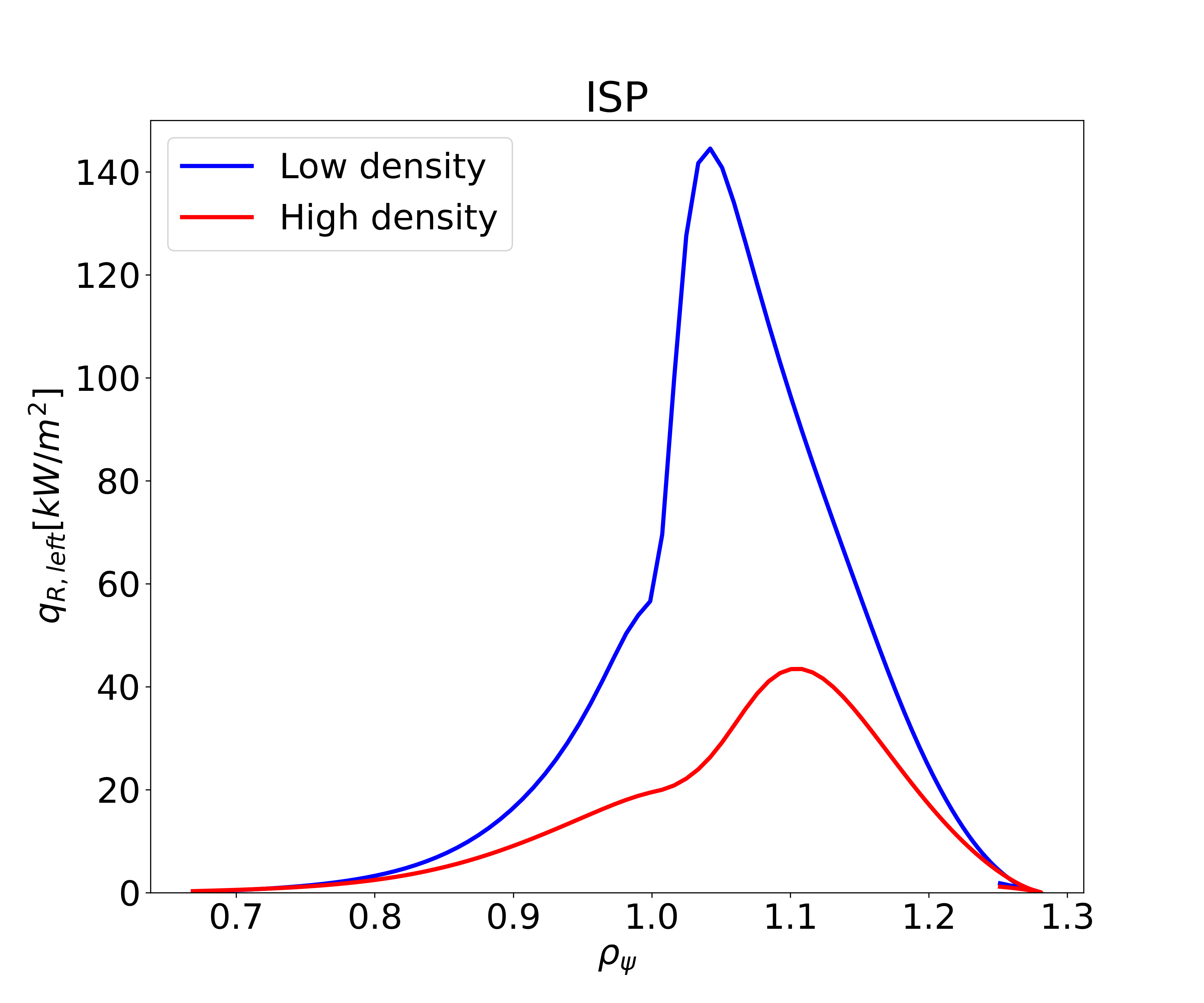}
    \end{subfigure}
    \hfill
    \begin{subfigure}[b]{0.45\textwidth}
    \includegraphics[width=\textwidth]{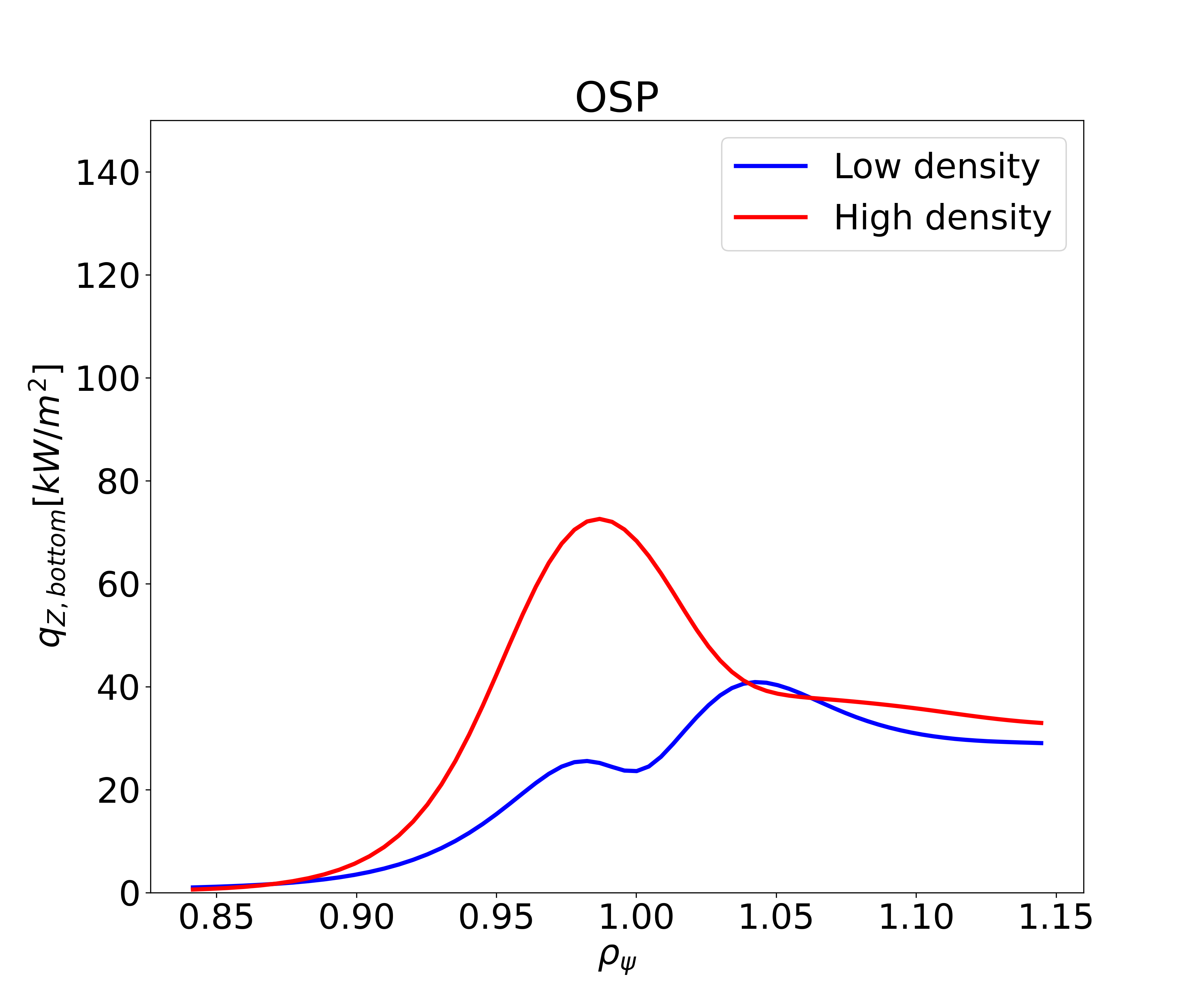}
    \end{subfigure}
    \caption{Time- and toroidally-averaged heat flux on the target, for both strike points, as a function of the normalized poloidal flux function.}
    \label{fig:heatflux_target}
\end{figure*}

The time-averaged profiles of the heat flux for the two targets are shown in Fig. \ref{fig:heatflux_target}.
In the low-density case, the heat flux is mainly determined by conduction and parallel electron convection, equally contributing with their sum and accounting for 80\% of the total heat flux. As a consequence, the heat flux peak is located at the strike points where the parallel ion particle flux peaks, both at the ISP and at the OSP.

The heat flux shows a lower and wider peak at the ISP in the high-density simulation, compared to the low-density one, in agreement with the observed lower pressure values observed (see Fig. \ref{fig:ptot_sol}). The contribution of $\Dp$ ions, largely dominated by convection, increases up to $40 \%$ of the total heat flux in the high-density simulation. As observed for the particle flux, the heat flux decrease is strong at the ISP, due to the sum of the effect of strong reduction of parallel convection toward the target and strong reduction of plasma temperature. We point out that our simulations retrieve the experimental observations of heat flux decrease with the simultaneous increase of upstream radiative and momentum losses (see Fig. \ref{fig:Sptot_sol}) at the onset of detachment \cite{Reimerdes_2017,Stangeby_2000}.

\section{Conclusions}
\label{sec:conclusions}
In this work the first multi-component simulations of plasma turbulence coupled to kinetic neutral dynamics are presented in a diverted tokamak configuration. The simulations are performed by exploiting the multi-component model described in Ref. \cite{Coroado_2022_1} and considering the magnetic equilibrium of a realistic TCV discharge, that is the TCV-X21 configuration \cite{Oliveira_2022}. The self-consistent treatment of the interactions between five species (electrons, $\Dp$, $\Dtp$, $\D$ and $\Dt$) is simulated. The model takes into account the main collisional processes between plasma and neutrals, including ionization, recombination, elastic collisions, charge-exchange and molecular dissociation.

We present the results from two simulations performed at different fuelling rates, obtained by changing the strength of a $\Dt$ gas puff. While GBS simulations that do not include the molecular dynamics retrieve important features associated with increased fuelling \cite{Mancini_2021}, the introduction of molecular dynamics improves the understanding of the processes at play in high-density L-mode tokamak discharges. 
For instance, the relevant density of $\Dt$ creates new channels of $\D$ production, mainly MAR, producing atomic deuterium at a relatively high temperature, $T_{\D} \simeq 3$~eV. The increase of the neutral density yields an increased radiated power through ionization reactions, leading to power starvation in the divertor region and moving the ionization front from the targets to the region above the X-point. The high neutral density is also responsible for the increase of momentum losses, identified by the increase of charge-exchange reactions along both legs of our high-density simulations, which leads to the decrease of the ion parallel velocity. For sufficiently low plasma temperature, $T_e < 3$~eV, molecular dissociations and charge-exchange reactions between $\Dp$ and $\D$ become the main plasma energy sink, leading the ion and electron temperatures to values close to the neutral temperature, $0.03$~eV$< T_{\D} < 3$~eV.

Because of the low temperature values and associated momentum losses, the parallel ion velocity is reduced in the high-density with respect to the low-density simulation, as observed in previous GBS simulations \cite{Mancini_2021}. This yields a reduction of total heat flux at the targets, lowering both heat convection and heat conduction.
In particular, the increase in fuelling causes a strong reduction of particle flux at the ISP, compatible to typical experimental observations in the detachment regime. Indeed, to our knowledge, the simulations presented in this work are the first simulations of plasma detachment that include a self-consistent treatment of plasma turbulence and neutral interactions.
The analysis of the fluxes shows that the decrease of the particle flux to the wall in the high-density simulation is associated with a decreased ionization source, due to a reduced plasma temperature in the SOL. The asymmetries between the ISP and OSP are explained by the different local plasma temperature and molecular neutral density, in turn determined by the magnetic configuration and by the position of the gas puff, combined with the effect of the $E \times B$ flux.

In addition, the reduced plasma temperature leads to an increase of the plasma resistivity. This generates strong electric field in the SOL. Indeed, the equilibrium profile of the parallel electric field $E_{\|}$ follows the generalized Ohm's law, Eq. \eqref{eq:Ohmlaw}, where the contributions of the electron temperature and pressure gradients are negligible compared to the resistive term, $\nu j_{\|}$.

The profiles of both density and pressure at the OMP show that the increase in fuelling leads to the formation of a density shoulder and an increase of the near SOL decay length for both quantities. We observe that the density shoulder is the result of an increase of the perpendicular transport together with the decrease of parallel transport. Leveraging blob tracking routines developed in past studies and recently improved, we perform a detailed investigation of filamentary transport in the SOL, comparing blob velocities with the effective velocity $\overline{\Gamma}_{E \times B}/\overline{n}_e$ . We observe an increase of blob radial velocity with increased fuelling, well reproduced by the increase of the radial $v_{E \times B}$ due to stronger electric field in the far SOL. It is interesting to note that the increase in radial velocity was not observed in simulations where fuelling was increased without including the role of molecules and the plasma temperature in the far SOL was higher \cite{Mancini_2021}.

From the analysis of our simulations we retrieve several qualitative similarities with experimental results. For instance, the increase of deuterium puffing leads to a decrease in the heat flux at the targets and a decrease in the particle flux in the inner target, up to detachment conditions \cite{Stangeby_2000,Loarte_1998}. The asymmetry between the two targets results from the combination of local $\Dt$ and $\Dtp$ density \cite{Park_2018} and stronger $E \times B$ drift \cite{Potzel_2014}.
Our results of filamentary transport reproduce the increase of radial velocity at higher density \cite{Offeddu_2022,Wuthrich_2022}, where faster blobs appear in correspondence of a strong electric field in the SOL, determined by plasma resistivity at low temperature.


\section*{Acknowledgments}
The simulations presented herein were partially carried out on the CINECA Marconi supercomputer under the neutralGBS project, partially at the Swiss National Supercomputing Center (CSCS) under the project IDs s1028 and s1170 and partially at SuperMUC-NG thanks to a PRACE awards. This work, supported in part by the Swiss National Science Foundation, has been carried out within the framework of the EUROfusion Consortium, partially funded by the European Union via the Euratom Research and Training Programme (Grant Agreement No 101052200 — EUROfusion). The Swiss contribution to this work has been funded by the Swiss State Secretariat for Education, Research and Innovation (SERI). Views and opinions expressed are however those of the author(s) only and do not necessarily reflect those of the European Union, the European Commission or SERI. Neither the European Union nor the European Commission nor SERI can be held responsible for them. We thank Davide Galassi for the several useful discussions.

\normalem
\printbibliography

\end{document}